%% file: wns3_2019_quic_arxiv.tex
\documentclass[sigconf]{acmart}

\usepackage[T1]{fontenc}
\usepackage{enumitem}
\usepackage{afterpage}
\usepackage{listings}
\usepackage{xcolor}
\usepackage{graphicx}
\usepackage{float}
\usepackage[font=small]{caption}
\usepackage[font=footnotesize]{subcaption}
\usepackage{placeins}
\usepackage{url}
\usepackage{epstopdf}
\usepackage{booktabs}
\usepackage{footnote}
\makesavenoteenv{table}
\makesavenoteenv{tabular}

\usepackage[acronyms,nonumberlist,nopostdot,nomain,nogroupskip]{glossaries}

\usepackage{tabularx}

\usepackage{paralist}

\input{acronyms.tex}

\usepackage{tikz}
\usepackage{pgfplots}
\pgfplotsset{compat=newest} 
\pgfplotsset{plot coordinates/math parser=false} 
\newlength\fheight
\newlength\fwidth
\usetikzlibrary{plotmarks,patterns,decorations.pathreplacing,backgrounds,calc,arrows,arrows.meta,spy,matrix}
\usepgfplotslibrary{patchplots,groupplots}
\usepackage{tikzscale}

\begin{document}


\flushbottom
\setlength{\parskip}{0ex plus0.1ex}
\addtolength{\skip\footins}{-0.2pc plus 40pt}

\title{A QUIC Implementation for ns-3}

\author{\vspace{-.5cm}\texorpdfstring{Alvise De Biasio, Federico Chiariotti, Michele Polese, Andrea Zanella, Michele Zorzi\\
\small Department of Information Engineering, University of Padova, Padova, Italy \\
\small e-mail: \{debiasio, chiariot, polesemi, zanella, zorzi\}@dei.unipd.it}{}}

\copyrightyear{2019}
\acmYear{2019}
\setcopyright{none}
\acmConference[WNS3 '19]{2018 Workshop on ns-3}{June 19--20, 2019}{Firenze, Italy}
\acmBooktitle{WNS3 '19: 2019 Workshop on ns-3, June 19--20, 2019, Firenze, Italy}
\acmPrice{xxx}
\acmDOI{xxxxxxxxxxxxxxxxxxxxxxx}
\acmISBN{xxxxxxxxxxxxxxxxxxxxxxxxxx}

\pagestyle{empty}

\begin{abstract}

\gls{quic} is a recently proposed transport protocol, currently being standardized by the \gls{ietf}. It aims at overcoming some of the shortcomings of TCP, while maintaining the logic related to flow and congestion control, retransmissions and acknowledgments. It supports multiplexing of multiple application layer streams in the same connection, a more refined selective acknowledgment scheme, and low-latency connection establishment. It also integrates cryptographic functionalities in the protocol design. Moreover, \gls{quic} is deployed at the application layer, and encapsulates its packets in UDP datagrams. Given the widespread interest in the new \gls{quic} features, we believe that it is important to provide to the networking community an implementation in a controllable and isolated environment, i.e., a network simulator such as ns-3, in which it is possible to test \gls{quic}'s performance and understand design choices and possible limitations. Therefore, in this paper we present a native implementation of \gls{quic} for ns-3, describing the features we implemented, the main assumptions and differences with respect to the \gls{quic} Internet Drafts, and a set of examples.

\end{abstract}



\maketitle

\section{Introduction}\label{sec:intro}
\glsresetall

\begin{picture}(0,0)(0,-500)
\put(0,0){
\put(0,0){\small This paper has been submitted to WNS3 2019. Copyright may be transferred without notice.}}
\end{picture}

The infrastructures, technologies and protocols that support communications over the Internet have significantly evolved over the years to support the needs of the connected society and the increase in mobile,  desktop and machine-generated traffic~\cite{ciscoVni}. In particular, the research community has recently shown a renewed interest in topics related to the transport layer, with novel transport protocols and extensions to TCP such as, e.g., \gls{sctp}, \gls{quic}, and \gls{mptcp}~\cite{polese2019survey}. Furthermore, novel congestion control algorithms have also been proposed to address novel communication challenges, such as those introduced by new cellular \glspl{ran} (e.g., 3GPP LTE, NR)~\cite{8613277}. 

One of the most important novelties is \gls{quic}, a transport protocol implemented at the application layer, originally proposed by Google~\cite{Langley:2017:QTP:3098822.3098842} and currently considered for standardization by the \gls{ietf}~\cite{draftquicktr}. \gls{quic} addresses some of the issues that currently affect transport protocols, and TCP in particular. First of all, it is designed to be deployed on top of UDP to avoid any issue with middleboxes in the network that do not forward packets from protocols other than TCP and/or UDP~\cite{7738442}. Moreover, unlike TCP, \gls{quic} is not integrated in the kernel of the \glspl{os}, but resides in user space, so that changes in the protocol implementation will not require \gls{os} updates. Finally, \gls{quic} provides a number of improvements with respect to TCP, such as the integration of the security of the connection, and the reduction of the initial handshake delay.

Despite being still under discussion by the \gls{ietf}, \gls{quic} has been deployed in a number of commercial products (e.g., the Google Chrome browser). According to the estimates provided in~\cite{Langley:2017:QTP:3098822.3098842}, 7\% of the overall Internet traffic and 30\% of Google's egress traffic is currently generated by \gls{quic} connections. The performance of different versions of \gls{quic} has also been studied in a number of recent papers~\cite{DBLP:journals/corr/abs-1801-05168,kakhki2017taking,Carlucci-2015-HOU}, with experiments in real networks based on different open source implementations of \gls{quic}. Nonetheless, given the important role that this protocol will play in the evolution of the networking stack in the near future, we believe that it is fundamental for researchers and network engineers to have the possibility of studying its design and performance in a controlled environment, i.e., in a network simulator.  Therefore, in this paper we present a native implementation of \gls{quic} for ns-3, which is compliant with version 13 of the \gls{ietf} \gls{quic} draft, and is integrated with the existing applications and TCP/IP stack of ns-3.\footnote{The code is publicly available at \url{https://github.com/signetlabdei/quic-ns-3}} Our implementation is based on the design of the ns-3 TCP implementation~\cite{CASONI201681,Patriciello:2017:SCL:3067665.3067666}, and heavily extends it to account for the novelties of \gls{quic}.

Notice that, while there exist multiple real implementations of \gls{quic}~\cite{QGo,LSQ,proto-quic,libquic}, their integration with ns-3 has several issues.\footnote{An integration of the \gls{quic} implementation in~\cite{proto-quic} can be found in~\cite{nsquic}.} For example, \gls{quic} also implements the \gls{tls} functionalities and thus requires the generation and setup of security certificates. Installing and integrating the \gls{quic} implementations and ns-3 can also be complex and cause compatibility issues on some operating systems. Furthermore, a direct integration would not allow researchers to easily change the parameters of the protocol or test new features, requiring them to alter the code-base directly. Finally, ns-3 applications and existing implementations of congestion control algorithms are based on the structure of the ns-3 TCP socket, and cannot be reused without a native implementation. Therefore, we advocate that a \gls{quic} implementation for ns-3 can benefit the networking research community due to its ease of use, despite introducing the burden of re-implementing and updating the protocol. Moreover, we designed the \gls{quic} stack so that it is possible to re-use the various implementations of TCP congestion control in \gls{quic}. This makes it possible to directly compare different congestion control designs for \gls{quic}, while real implementations only offer adaptations of the NewReno and CUBIC schemes~\cite{kakhki2017taking}.

In the remainder of the paper, we will first provide an overview of the main features of \gls{quic} in Sec.~\ref{sec:quic}. Then, we will describe the ns-3 implementation in Sec.~\ref{sec:implementation}, with details on the code structure, the compatibility with TCP and the missing elements with respect to the \gls{quic} Internet Drafts. In Sec.~\ref{sec:examples} we will illustrate some examples of \gls{quic}'s performance evaluation. Finally, Sec.~\ref{sec:conclusions} will conclude the paper. 

\section{QUIC: protocol Description}
\label{sec:quic}

\begin{figure}[t]
	\centering
	\includegraphics[width=\columnwidth]{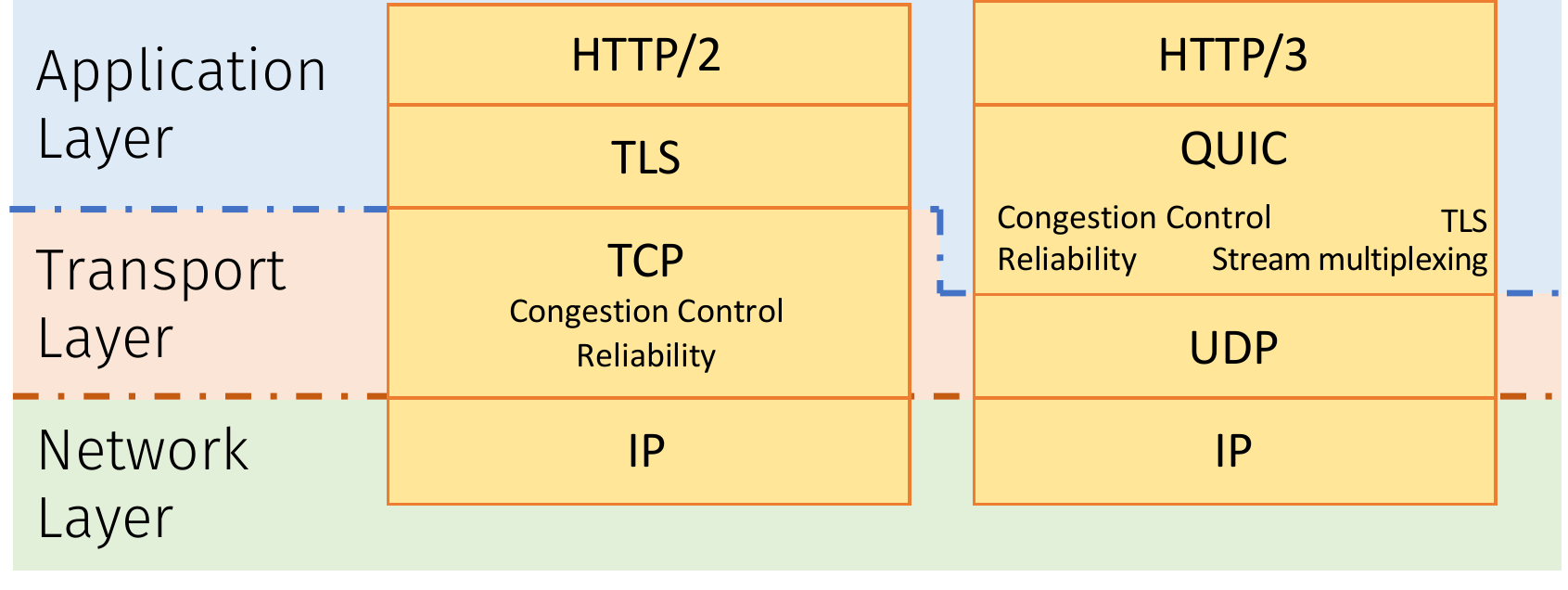}
	\caption{Protocol stack when considering an application over \gls{quic} or TCP.}
	\label{fig:stack}
\end{figure}

As mentioned in Sec.~\ref{sec:intro}, \gls{quic} is a transport protocol implemented on top of UDP, initially proposed by Google in 2013. Since 2016, the \gls{ietf} has started a process to standardize the protocol into multiple Internet Drafts, e.g., for the main transport functionalities~\cite{draftquicktr}, recovery and congestion control~\cite{draftquickrec}, security~\cite{draftquictls} and the bindings to the protocols at the application layer, e.g., the \gls{http}~\cite{draftquichttp}. 

Fig.~\ref{fig:stack} reports the configuration of the protocol stack (above the network layer) when considering a secure \gls{http} connection over \gls{quic} or TCP. \gls{quic}'s design indeed builds on that of TCP, but the two protocols also present significant differences. Like TCP, \gls{quic} uses acknowledgments, duplicate acknowledgments, and the lack of acknowledgments to infer the status of the end-to-end connection and react by updating the congestion window or performing the necessary retransmissions. However, loss recovery is improved with respect to TCP: acknowledgments explicitly differentiate between out-of-order and lost packets, and intrinsically support \gls{sack} with a higher number of blocks than in TCP~\cite{draftquicktr}. Moreover, \gls{quic} can provide more information to the congestion control algorithms compared to TCP, consequently allowing better estimation of connection \glspl{rtt} (e.g., by explicitly signaling the delay between the reception of a packet and its ACK) and more efficient loss detection and recovery~\cite{polese2019survey}.

\gls{quic} is also designed to support the multiplexing of several data (and control) streams over a single connection, with flow control implemented at both the stream and the connection level. The multiplexing is based on a novel packet structure, in which multiple frames (belonging to different streams, or carrying control information such as acknowledgments) are combined to create a \gls{quic} packet. 
Thanks to the stream support, \gls{quic} naturally combines with the streaming interface of \gls{http}/2,\footnote{Notice that the integration of \gls{http} on top of \gls{quic} will be defined as \gls{http}/3.} so that different application-layer streams can be mapped to different and independent streams that do not require in-order delivery, while still being part of the same transport layer flow. Therefore, even if one of the streams is affected by packet loss, the others can still reliably deliver their packets to the application at the receiver, preventing the so-called \gls{hol} blocking issue that affects TCP. With TCP, indeed, the multiple streams that are generated in an HTTP/2 connection are bundled together at the transport layer. If a single packet is lost, TCP cannot release any packet to the receiver's application layer, consequently blocking all the streams.

Moreover, \gls{quic} integrates the functionalities of \gls{tls} 1.3, making it possible to authenticate and encrypt both the payload and (at least part of) the header. Thanks to this integration, \gls{quic} makes it harder to perform injection attacks, eavesdropping or traffic analysis in middleboxes~\cite{draftquictls}. Additionally, this allows \gls{quic} to reduce the latency of the initial connection establishment. Indeed, while TCP needs to perform a protocol handshake first, and then the \gls{tls} cryptographic handshake, with \gls{quic} it is possible to embed the relevant cryptographic parameters into the first message sent between the two endpoints of the connection , thus reducing the initial handshake latency to a single \gls{rtt}. If two endpoints have already connected in some previous interactions, it is also possible to perform a 0-\gls{rtt} handshake, in which the encrypted data can already be sent in the first exchanged packet.

Finally, \gls{quic} supports a connection-level identifier that provides robustness against updates in the underlying layers of the protocol stack, e.g., IP address updates caused by \gls{nat} and/or mobility in cellular and Wi-Fi networks~\cite{draftquicktr}.


\section{Implementation of QUIC in ns-3}
\label{sec:implementation}

\begin{figure*}[t]
	\includegraphics[width=.9\textwidth]{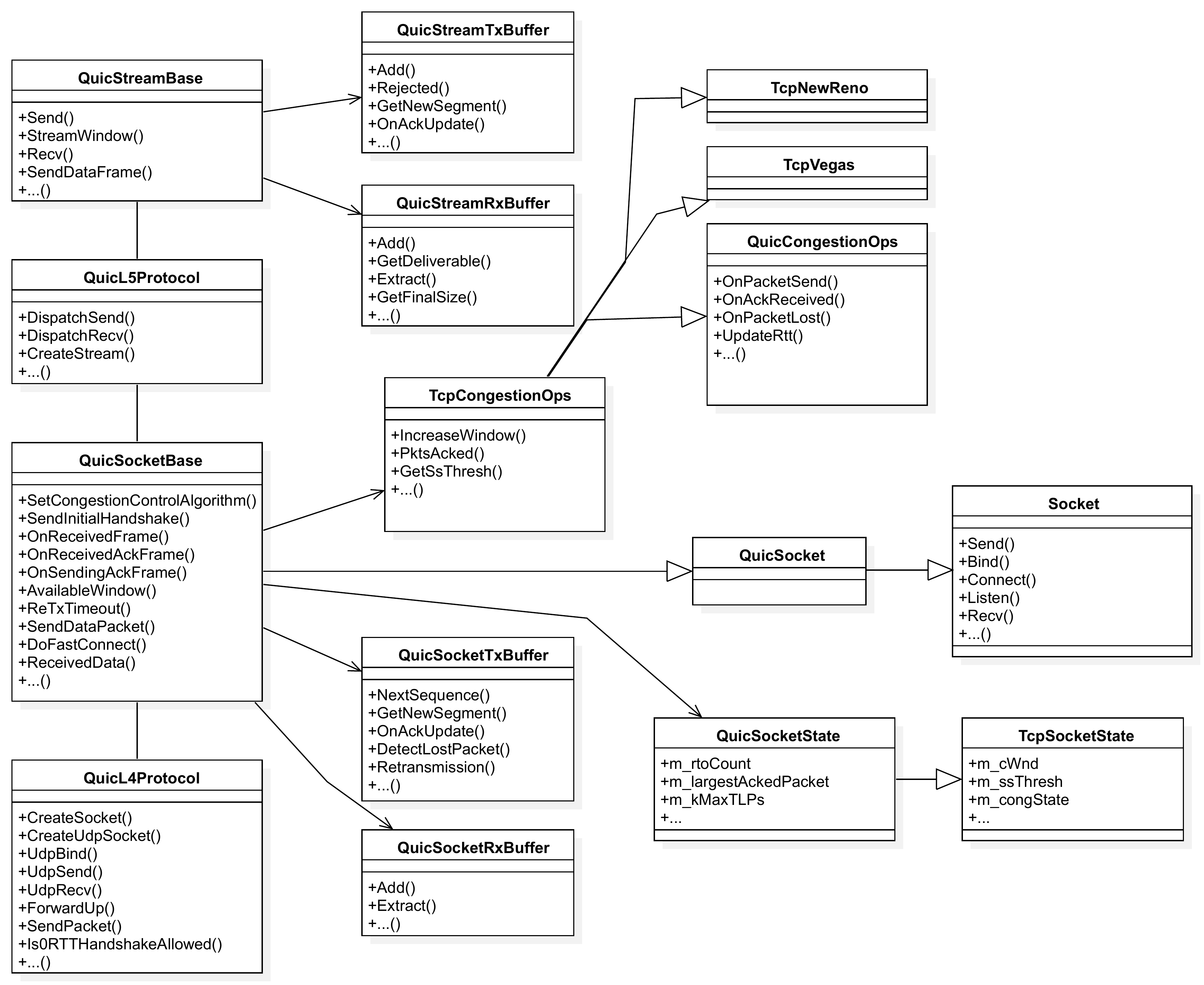}
	\caption{Simplified \gls{uml} diagram for the ns-3 \gls{quic} implementation. Notice that not all the classes, methods and relations are shown.}
	\label{fig:uml}
\end{figure*}

This Section describes the main features of the native \gls{quic} implementation for ns-3. Fig.~\ref{fig:uml} reports the \gls{uml} diagram for the \gls{quic} socket, congestion control, headers and buffers code. As the diagram shows, we developed the main classes in the ns-3 \texttt{internet} module, which already models other transport protocols such as UDP and TCP. Moreover, the structure of the code resembles that of the TCP implementation, replicating the separation of congestion control from the main socket logic and the presence of stand alone buffer classes~\cite{CASONI201681}, but at the same time introduces new elements to account for the novelties of \gls{quic} with respect to TCP.

\subsection{Code Structure}
\label{sub:code}

\texttt{QuicSocket\-Base} (which extends \texttt{QuicSocket}), the main class of the \gls{quic} implementation, models the basic functionalities of a \gls{quic} socket, like in the TCP implementation. Each client will instantiate a single \texttt{QuicSocketBase} object, while the server will fork a new socket for each incoming connection. 
A \texttt{Quic\-SocketBase} object receives and transmits \gls{quic} packets and acknowledgments, accounts for retransmissions, performs flow and congestion control at a connection level, takes care of the initial handshake and exchange of transport parameters, and handles the life cycle and the state machine of a \gls{quic} connection.
An instance of the \texttt{QuicSocketBase} class holds pointers to multiple other relevant items, including:
\begin{compactitem}
	\item the socket transmission and reception buffers, implemented by \texttt{QuicSocketTxBuffer} and \texttt{QuicSocketRxBuffer}, respectively;
	\item the \texttt{QuicSocketState} object, which extends \texttt{TcpSocket\-State}~\cite{CASONI201681} with additional variables that are used by the \gls{quic} state machine and congestion control;
	\item an object extending the \texttt{TcpCongestionOps} class, which performs the congestion control operations and provides a basic compatibility with the TCP congestion control implementations, as we will discuss later.
\end{compactitem}

The \gls{quic} socket is bound to an underlying UDP socket through a \texttt{QuicL4Protocol} object, which also handles the initial creation of the \texttt{QuicSocketBase} object, triggers the UDP socket to bind and connect, and handles the delivery of packets between the UDP socket and \texttt{QuicSocketBase}.

A stream is instead modeled by the \texttt{QuicStreamBase} class, which extends the basic \texttt{QuicStream} class. It buffers application data, performs stream-level flow control, and delivers the received data to the application. Similarly to the full socket, a \texttt{QuicStream\-Base} object also has pointers to transmission and reception buffers, which are implemented by the \texttt{QuicStreamTxBuffer} and \texttt{Quic\-StreamRxBuffer} classes.

Multiple \texttt{QuicStreamBase} objects are connected to a single \texttt{Quic\-SocketBase} through an object belonging to the \texttt{QuicL5Protocol} class. A \texttt{QuicSocketBase} holds a pointer to a \texttt{QuicL5Protocol} object, and the latter contains a vector of pointers to multiple \texttt{QuicStreamBase} instances. The \texttt{QuicL5Protocol} class creates and configures the streams, and takes care of delivering packets or frames to be transmitted and received across the streams and the socket.

\subsection{QUIC Packets, Frames and Headers} 
\label{sub:quic_packets_frames_and_headers}

According to the \gls{quic} Internet Drafts~\cite{draftquicktr,draftquickrec}, a \gls{quic} packet is composed by a header and a payload. The \gls{quic} packet is then encapsulated into a UDP datagram to be transmitted over the wire. 

The header of a \gls{quic} packet is modeled by the \texttt{QuicHeader} class, which extends the ns-3 \texttt{Header} class. It takes care of serializing and deserializing the header when attached to an ns-3 packet, and represents the information that a \gls{quic} header carries. 
One of the main novelties of the \gls{quic} header design with respect to the traditional TCP headers is that it can have two different formats (i.e., \textit{long} or \textit{short}) according to the amount of information that needs to be exchanged. Moreover, for short headers, the packet number entry in the header can have a variable length, i.e., one, two or four bytes according to how many bytes are actually necessary to represent the packet number. 
Long headers (17 bytes) are used for the message exchanges which are relevant to the connection life cycle (e.g., the initial handshake, version negotiation) and are sent prior to the establishment of the \gls{tls} keys. 
Short headers (from 2 to 13 bytes) are instead used for data packets and acknowledgments. The \texttt{QuicHeader} class also provides various static methods that can be used to create different kinds of headers (short, long, for handshake messages) and set the relevant parameters.

The payload of a \gls{quic} packet, instead, is composed of multiple frames, with each frame mapped to a stream and/or a control operation. Each frame carries an additional subheader, which is implemented by the \texttt{QuicSubheader} class, and specifies the type of frame. Data frames are explicitly associated to a stream, through a stream identifier, and their subheader may carry the size of the frame, the offset with respect to the first stream byte, and a flag that signals the end of the stream. Control frames, instead, are used to perform specific actions and have custom formats. The most relevant control frames for the ns-3 \gls{quic} code are the ACK frames, implemented according to the format specified in~\cite{draftquicktr,draftquickrec}, and used to acknowledge the packets received by an endpoint of the connection. Each ACK specifies the largest acknowledged sequence number and can carry multiple selective acknowledgment blocks if gaps in the received packet sequence are identified. Moreover, it is possible to explicitly specify in the ACK the delay between the reception of the acknowledged packet with the highest sequence number and the time at which its ACK was sent.

\subsection{QUIC Buffers} 
\label{sub:quic_buffers}
The current implementation of \gls{quic} for ns-3 features 4 different kinds of buffers, as mentioned in Sec.~\ref{sub:code}. We decided to keep separate buffer implementations between the socket and stream level because the functionalities of the various buffers are slightly different, and the buffers contain different items.

At the socket level, the transmission buffer, implemented in the \texttt{QuicSocketTxBuffer} class, stores stream or control frames to be delivered by \texttt{QuicSocketBase} and returns packets (composed by one or multiple frames) of the desired size to the socket. This buffer holds two different lists of packets, one for those not yet transmitted, and one for those transmitted (until they are acknowledged), in case retransmissions are needed. 
Each item stored in these lists is encapsulated in a \texttt{QuicSocketTxItem}, which associates the smart pointer to the packet with a possible sequence number, information on the timing of transmission and ACK reception, and a number of flags that identify the packet as lost, retransmitted, acknowledged or released. The \texttt{QuicSocketTxBuffer} implements a prioritization mechanism for the special frames of stream 0, which should be transmitted first whenever there is a transmission opportunity. Moreover, it is capable of splitting and assembling frames according to the actual packet size that the socket requires. Once a packet is delivered to the socket for the first transmission, the associated frames are removed from the unsent list, and the whole packet is added to the sent list. In this way, it is possible to correctly handle the reception of ACKs, release correctly received packets, handle retransmissions, and compute the number of bytes in flight.

At the stream level, the sender-side buffering is handled by \texttt{Quic\-StreamTxBuffer} instead. The transmission buffer stores application packets until they are acknowledged, but does not perform retransmissions, which, according to the \gls{quic} Internet Draft, are only handled at the socket level for full \gls{quic} packets. 
The packets in the stream buffer are identified by an offset with respect to the beginning of the stream. 
Moreover, the stream transmission buffer needs to gracefully handle the case in which it tries to deliver a packet to the underlying socket buffer, but the socket buffer is full and rejects the packet. In this case, the frame is re-added to the unsent list of the stream buffer, with the correct offset. 

Receiver-side buffering follows the opposite paradigm: packets received by the socket are read, disgregated into individual subframes and stored in the \texttt{QuicStreamRxBuffer}. If the received bytes are out of order, they are stored in the stream buffer until the packet containing the missing bytes is received correctly. If the stream buffer gets some bytes in the correct order, it transfers them to the \texttt{QuicSocketRxBuffer}, which holds the received application packets released by the streams until the application requests them. Moreover, in case the stream has received the FIN bit (i.e., the transmission has ended), the receiver stream-level buffer can return the total amount of received bytes in the stream.

Given that the buffers are always initialized in either \texttt{Quic\-SocketBase} or \texttt{QuicStreamBase}, and following the ns-3 TCP implementation approach, it is possible to set the size of the buffer (i.e., the number of bytes they can buffer before rejecting packets) using attributes of these classes for the socket (\texttt{SocketSndBufSize} and \texttt{SocketRcvBuf\-Size}) and the stream buffers (\texttt{StreamSndBufSize} and \texttt{Stream\-RcvBufSize}), respectively.

\subsection{Data Flow in a QUIC Connection} 
\label{sub:data_flow_in_a_quic_connection}
In the following, we will describe the chain of events and calls that are triggered whenever an application sends a packet to the socket, and when the UDP socket at the other endpoint receives a datagram with a \gls{quic} payload.

\texttt{QuicSocketBase} implements the ns-3 \texttt{Socket} \glspl{api}, thus the application calls the \texttt{Send} method to deliver a packet to the socket. Notice that possible extensions of the classic BSD socket interface are being discussed for the interfacing between \gls{http}/3 streams and \gls{quic}~\cite{draftquichttp}, e.g., to explicitly signal the mapping between a packet and a stream to the socket. In order to make this possible in ns-3 without modifying the \texttt{Socket} implementation, we exploit the piggybacking paradigm and embed the information in the \texttt{flag} integer parameter that can be set when using the \texttt{Send} call. \texttt{QuicSocketBase} will then use this parameter as the identifier of the stream on which the packet should be sent, and the packet will be delivered to that stream through the \texttt{QuicL5Protocol} object. If it is the first packet of that stream, the corresponding \texttt{QuicStreamBase} instance is also initialized. Otherwise, if the application does not express a preference, then \texttt{QuicL5Protocol} will assign the packet to the first stream. The maximum number of streams can be set through attributes of \texttt{QuicSocketBase}.

If the stream flow control allows the transmission of the packet, then \texttt{QuicStreamBase} will create the \texttt{QuicSubheader} and deliver the frame to \texttt{QuicL5Protocol}, which will relay it to \texttt{QuicSocket\-Base}. If the socket buffer has enough space, the packet is added to the \texttt{QuicSocketTxBuffer}. The frame is then possibly aggregated to other frames and sent to the underlying UDP socket through \texttt{QuicL4\-Protocol}, according to the possible transmission opportunities enabled by the socket flow and congestion controls.

At the other endpoint, when the UDP socket receives the packet, it will trigger a callback to the \texttt{ForwardUp} method of the associated \texttt{QuicL4Protocol}. This method handles the possible cloning of the server socket if the packet is received in a listening state, and will eventually trigger an additional callback to deliver the packet to the \texttt{ReceivedData} method of \texttt{QuicSocketBase}. This method checks if the packet is a control packet (e.g., for the initial handshake and configuration) or a packet with a mixed payload with one or more data and ACK frames. If this is the case, \texttt{QuicSocketBase} will deliver the packet to the \texttt{DispatchRecv} method of the associated \texttt{QuicL5Procotol} object. This method cycles through the frames of the \gls{quic} packet, triggering the relevant methods either in \texttt{QuicSocketBase} or in \texttt{QuicStreamBase}. The first is called whenever a control frame that needs to be handled by the socket is received (e.g., an ACK), while the second is called for stream-related control or data frames. The latter are buffered in the \texttt{QuicStreamRx\-Buffer} and released in the correct order to \texttt{QuicSocketBase}, which will notify the application through the \texttt{NotifyDataReceived} callback. The application can then call the \textit{RecvFrom} method to extract the packets from the socket and consume them.

\subsection{Retransmission Process} 
\label{sub:retransmission_process}
The retransmission process of \gls{quic} works at the socket level, i.e., it retransmits complete packets (composed of one or multiple frames) that were lost, by leveraging the ACK transmission and reception flow that is embedded in the \gls{quic} operations. In this implementation, the \texttt{QuicSocketBase} class handles the reception of ACKs through the \texttt{OnReceivedAckFrame} method, which is triggered from \texttt{QuicL5Protocol} when a received packet contains an ACK frame. 

As mentioned in Sec.~\ref{sub:quic_packets_frames_and_headers}, an ACK frame always specifies the largest acknowledged packet number, and possibly contains several additional ACK blocks, which also specify missing packets. The socket relays this information to the associated \texttt{QuicSocketTx\-Buffer} object, which iterates over the list of sent packets, marks them as acknowledged or lost, following the procedure specified in~\cite[Sec. 4.2.1]{draftquickrec}, and returns the list of the acknowledged packets to the socket. The socket then performs the relevant congestion control operations (e.g., by updating the socket state in case of losses, or increasing the congestion window), and triggers the retransmission process.

In \gls{quic}, the retransmissions are always associated with new, monotonically increasing sequence numbers. Therefore, for each retransmission, \texttt{QuicSocketBase} increases the sequence number by one, and notifies the socket buffer. The latter moves the sent packets marked as lost back to the buffer of not-yet-transmitted packets, updating their packet number. Finally, \texttt{QuicSocketBase} retrieves the packets to be retransmitted and forwards them to \texttt{QuicL4Protocol}.


\subsection{Compatibility with TCP Congestion Control implementations} 
\label{sub:compatibility_with_tcp_congestion_control_implementations}

One of the main features of the ns-3 TCP redesign described in~\cite{CASONI201681} is the modularity of the congestion control algorithm with respect to the main socket code. This is achieved by using a separate object, which extends a class with the basic congestion control functionalities, i.e., \texttt{TcpCongestionOps}. A number of congestion control algorithms have been implemented and added to the \texttt{internet} module, including, for example, TCP NewReno, HighSpeed, BIC, Illinois, Vegas~\cite{Nguyen:2016:ISV:2915371.2915386}.
Therefore, \texttt{QuicSocketBase} can be used in legacy mode, with which it is possible to use the TCP congestion control algorithms to drive the congestion window of the \gls{quic} connection as well. Besides, it is also possible to use more refined algorithms, which exploit the additional information that the \gls{quic} socket provides, and develop congestion control methods for \gls{quic} only.

The compatibility with legacy TCP congestion control algorithms is achieved by having a \texttt{TcpCongestionOps} instance as the basic congestion control object in \texttt{QuicSocketBase}. Then, we introduced a new class \texttt{QuicCongestionControl}, which extends \texttt{TcpNewReno}.\footnote{As expected, \texttt{TcpNewReno} extends \texttt{TcpCongestion\-Ops} itself.} \texttt{QuicCongestionControl} features additional methods that inject additional information specified by the \gls{quic} Internet Draft in the congestion control algorithm (e.g., the transmission of a packet, more refined information on the \gls{rtt}) if needed. When the congestion control algorithm is set in \texttt{QuicSocketBase} (i.e., after the socket is created by \texttt{QuicL4Protocol}), the \texttt{SetCongestion\-ControlAlgorithm} checks if the specific algorithm extends \texttt{Quic\-CongestionControl}, and in this case sets the \texttt{m\_quicCongestion\-ControlLegacy} flag to false. Otherwise, the latter is set to true and the legacy mode is activated. Then, every time the socket needs to trigger the methods of the congestion control algorithm, e.g., when an acknowledgment is received, or a \gls{rto} expires, it will check whether the operations are in legacy mode to call the relevant methods.

\subsection{Connection Establishment} 
\label{sub:connection_establishment}

\begin{figure*}
\centering
	\begin{subfigure}[t]{0.4\textwidth}	
		\centering
		\includegraphics[width=.9\textwidth]{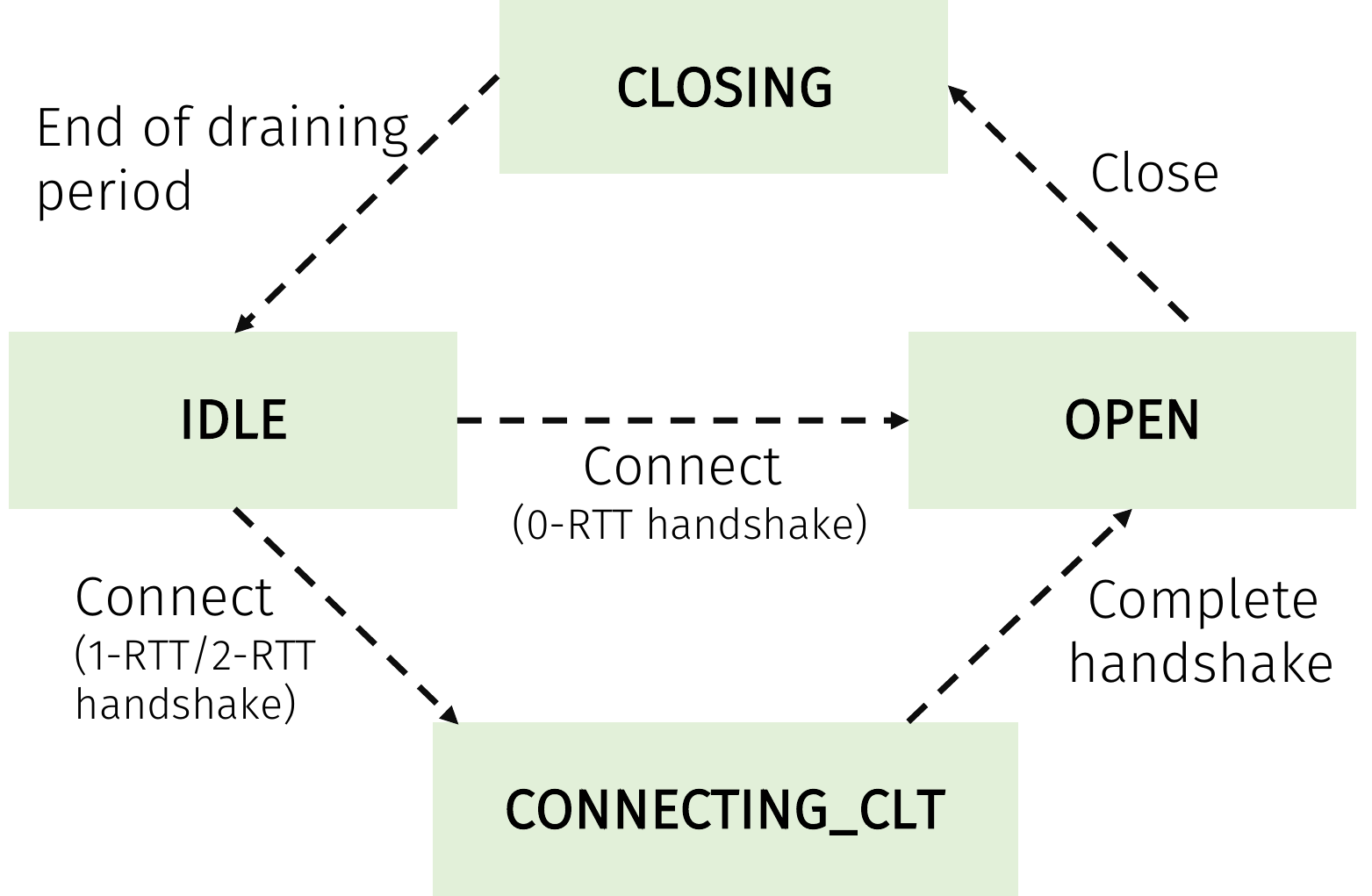}
		\caption{\gls{quic} client state machine}
		\label{fig:clt_state}
	\end{subfigure}%
	\begin{subfigure}[t]{0.6\textwidth}	
		\centering
		\includegraphics[width=.9\textwidth]{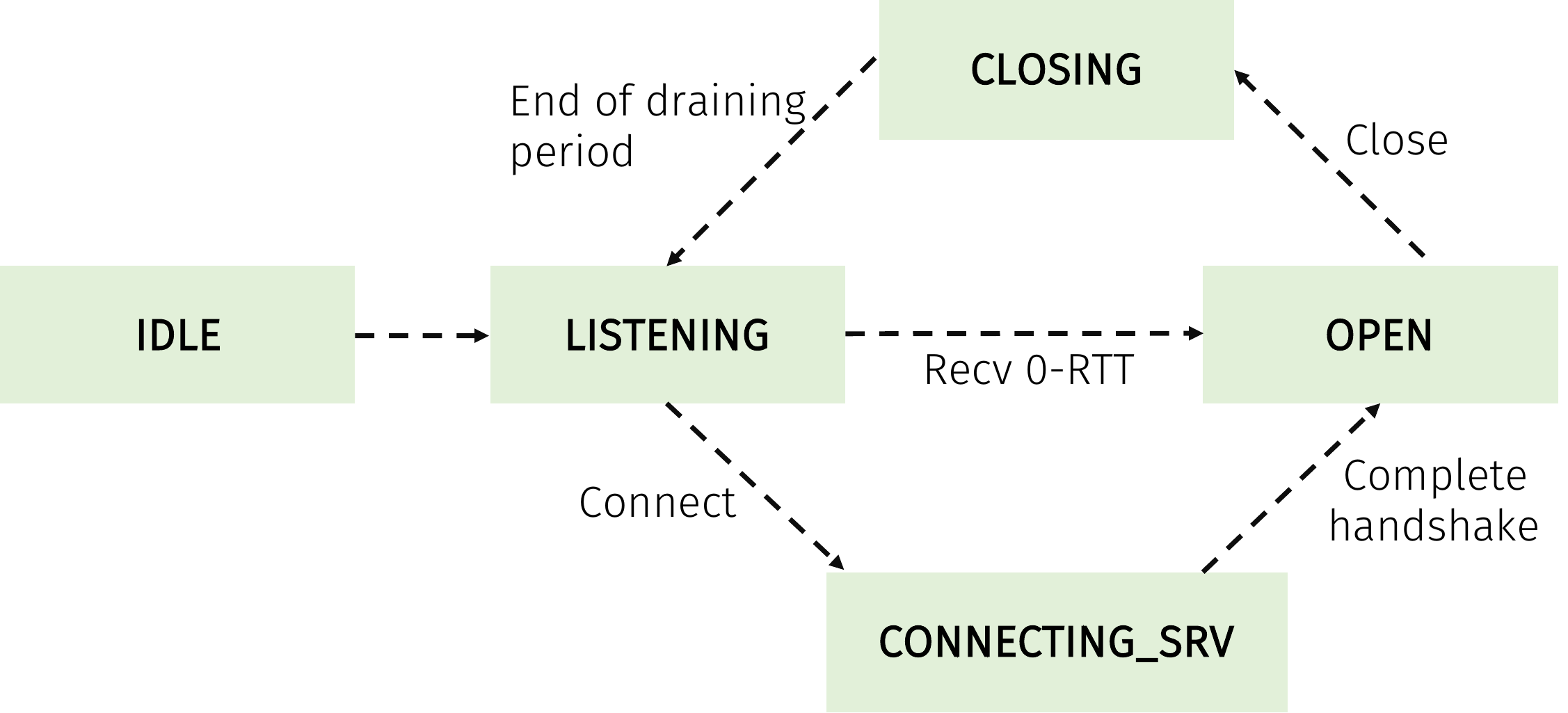}
		\caption{\gls{quic} server state machine}
		\label{fig:srv_state}
	\end{subfigure}
	\caption{State machine of the \gls{quic} implementation (i.e., of a \texttt{QuicSocketBase} object).}
	\label{fig:state}
\end{figure*}

A \gls{quic} connection is a single conversation between two QUIC endpoints~\cite{draftquicktr}, each unequivocally identified by the tuple of IP addresses and UDP ports. We implemented a connection opening mechanism that follows \gls{quic}'s Internet Draft, and performs the version negotiation and the transport handshake between the client and the server. Moreover, the procedure we implemented models the cryptographic handshake, but does not perform actual encryption, as we will discuss later. Fig.~\ref{fig:state} reports a diagram of the \gls{quic} state machine, with the transitions that happen during connection establishment and termination.

Notice that \gls{quic} can execute the handshake in two different ways, according to the shared history of the two endpoints. If the endpoints have recently authenticated, then a 0-\gls{rtt} handshake can be used. We model this by adding to \texttt{QuicL4Protocol} a list of authenticated endpoints, which is checked before sending the first packet and when it is received at the other endpoint. Moreover, the 0-\gls{rtt} can be forced using the \texttt{0RTT-Handshake} attribute in \texttt{QuicL4Protocol}. If 0-\gls{rtt} is not allowed, the socket falls back to the 1-RTT handshake using the \textit{client initial} message. Finally, if there is a version mismatch\footnote{Given the availability of early \gls{quic} implementations, the Internet Draft defines a version field in the initial handshake header.} between the two endpoints, then a 2-\gls{rtt} handshake is performed to settle to a common \gls{quic} version between endpoints.

\textbf{2-RTT handshake - }
In the ns-3 \gls{quic} implementation, the 2-RTT handshake is forced by setting the 0-\gls{rtt} attribute to false, and by configuring the \texttt{InitialVersion} attribute of \texttt{QuicSocketBase} to \texttt{QUIC\_VERSION\_NEGOTIATION}.\footnote{Other supported versions are defined in \texttt{quic-socket.h}. By setting this attribute to another version, the client and the server will assume to use the same version and avoid the 2-\gls{rtt} handshake.} The client (unaware of the fact that the server does not support the same \gls{quic} version) sends a \textit{client hello} initial message using the \texttt{SendInitialHandshake} method of \texttt{Quic\-SocketBase}. The server then rejects the connection and replies with a packet carrying a version negotiation frame, whose payload contains the list of all the versions the server supports encoded in 32 bits. The client then matches the list with one of the versions it supports and updates its version; finally, the handshake continues as a 1-\gls{rtt} handshake.

\textbf{1-RTT handshake - }
This handshake is very similar to TCP's, but it also embeds the setup of the \gls{tls} parameters. As in the 2-\gls{rtt} handshake, the client sends a \textit{client hello} initial message, i.e., a packet with a long header of type \textit{client initial}, a certain \gls{quic} version (which was possibly just negotiated) and a first packet number.\footnote{While according to the \gls{quic} specification the initial packet number should be random, in the ns-3 implementation it is systematically set to zero in order to simplify the debugging of the protocol stack. } Upon reception of the \textit{client initial} packet, the server replies with a packet carrying a \textit{handshake} header containing the relevant transport and cryptographic parameters and the connection identifier chosen by the server. 
Once the client receives this packet, it can conclude the transport and cryptographic handshake sending a new \textit{client hello}. Its connection ID field will contain the server-selected connection ID, while its packet number has to be one higher than the last packet that was sent. From now on the packet number will be monotonically increased for each subsequent packet. The payload of this packet can contain stream frames, padding and ACK frames.

\textbf{0-RTT handshake - }
As mentioned in Sec.~\ref{sec:quic}, this handshake is one of \gls{quic}'s novelties with respect to TCP, which allows the setup of a new connection with just a one-way packet exchange, from the client to the server. Once the application triggers the socket, and if the aforementioned conditions are met, then the client will send a packet with a \texttt{QuicHeader} of the \texttt{ZRTT\_PROTECTED} kind, carrying the previous connection identifier, the version and the cryptographic parameters.

\subsection{Connection Termination} 
\label{sub:connection_termination}

Besides the initial handshake, the implementation of \gls{quic} for ns-3 also features a simplified connection termination mechanism, that triggers a draining period with an idle timeout or an immediate close message.

The draining period is a specific time interval in which the two endpoints synchronize on the closing of the connection and handle all the packets that could be received in the meantime. In fact, even if a socket has entered the \textit{CLOSING} state, it could still receive some packets, because of imprecise synchronization between the idle timers of the two endpoints, or retransmissions and delayed packets still in flight in the network.
The immediate close is the standard mechanism through which an application can close a connection. For this purpose, the \texttt{QuicSocketBase} has a specific \texttt{Close} method which, once called by the application, triggers the transmission of a specific frame signaling the end of the connection to the other endpoint. Once the closing packet is sent, the socket moves to the \textit{CLOSING} state and enters the draining period. The same is done by the other endpoint upon receiving this packet. Additionally, both the client and the server keep an idle timeout timer, which is reset at every packet transmission or reception. The timer is initialized at a fixed value (300 s) and keeps running during all the connection lifetime. If it expires, both endpoints enter the draining period and move to the \textit{CLOSING} state. 

The draining period makes it possible to receive and acknowledge packets which were in flight after the closing of the connection, while avoiding new transmissions, except for those with application or connection close frames. Once the draining timer expires, the connection is definitively closed and all incoming packets are discarded. Both the client and server sockets are destroyed.

In our specific implementation, the \texttt{Close} method in \texttt{QuicSocket\-Base} is invoked either by the application to terminate the connection or by the socket itself when the idle timeout expires. This method moves the connection to the \textit{CLOSING} state and starts the draining period timer.
When the draining period timeout expires, the \texttt{DoClose} method is called. The socket callbacks are revoked, and then the socket is removed from the socket list of \texttt{QuicL4Protocol} calling the \texttt{RemoveSocket} method. If the closed socket was a server's listening socket, then any other forked sockets are closed at the same time. In any case, when the socket list is empty, the UDP socket is closed as well.

%


\subsection{Missing Features With Respect to QUIC Internet Drafts} 
\label{sub:missing_features_with_respect_to_quic_rfcs}

This implementation is aligned with version 13 of the \gls{quic} Internet Drafts, and it will be possible to update it in future releases to add new features and align it with the final version of the \gls{rfc}. Moreover, we list here the features that are missing from this implementation with respect to draft 13, and discuss the main assumptions that were made. 

First of all, we decided not to implement the \gls{tls} stack or rely on external cryptographic libraries. While this represents a significant difference with respect to the Draft-compliant \gls{quic} design, it will not affect the evaluation of the end-to-end performance of \gls{quic} in terms of congestion control behavior and retransmission processes. Besides, we implemented the connection establishment, as described in Section~\ref{sub:connection_establishment}, so that the need for the cryptographic handshake is accounted for, even though not actually performed. Moreover, given that this implementation targets a network simulator, there is no explicit need for the privacy and security features that \gls{tls} is expected to provide in a real \gls{quic} implementation. Finally, researchers who want to use the ns-3 \gls{quic} implementation will not need to install additional external libraries and security certificates.

In terms of transport capabilities, we do not allow the deletion of a stream, and some special frames (e.g., \texttt{STREAM\_BLOCK} frames) are not currently supported, but can be easily added in future releases. Finally, there is a difference with the Draft with respect to how the server and client exchange the transport parameters, which specify a number of properties for the connection. According to the \gls{quic} Internet Draft, the server sends its transport parameters to the client during the initial handshake, and the client applies them. While this exchange is actually simulated, it has no relevant effect on the configuration of the \gls{quic} stack in ns-3, whose parameters are already shared by the server and the client in the implementation of \texttt{QuicSocketBase}. The reason is that \texttt{QuicSocketBase} itself is a class that represents both a client and a server, and has a single set of attributes for both. Therefore, by setting the attributes for the server, the user also sets those for the client.

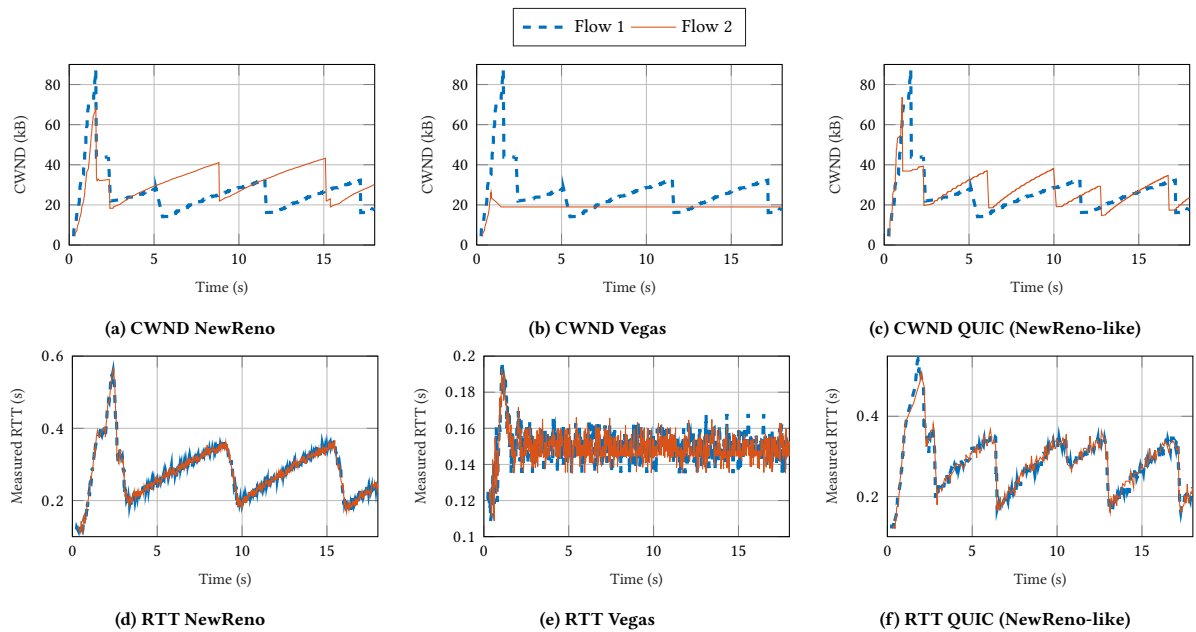
\begin{figure*}[t]
	\begin{subfigure}[t]{.3\textwidth}
		\centering
		\setlength\fwidth{.8\textwidth}
		\setlength\fheight{.45\textwidth}
		\input{figures/newreno-cwnd-2.tex}
		\caption{CWND NewReno}
		\label{fig:nr-cw}
	\end{subfigure}
	\begin{subfigure}[t]{.3\textwidth}
		\centering
		\setlength\fwidth{.8\textwidth}
		\setlength\fheight{.45\textwidth}
		\input{figures/vegas-cwnd-2.tex}
		\caption{CWND Vegas}
		\label{fig:vegas-cw}
	\end{subfigure}
	\begin{subfigure}[t]{.3\textwidth}
		\centering
		\setlength\fwidth{.8\textwidth}
		\setlength\fheight{.45\textwidth}
		\input{figures/quic-cwnd-2.tex}
		\caption{CWND \gls{quic} (NewReno-like)}
		\label{fig:quic-cw}
	\end{subfigure}
	\begin{subfigure}[t]{.3\textwidth}
		\centering
		\setlength\fwidth{.8\textwidth}
		\setlength\fheight{.45\textwidth}
		\input{figures/newreno-rtt-2.tex}
		\caption{RTT NewReno}
		\label{fig:nr-rtt}
	\end{subfigure}
	\begin{subfigure}[t]{.3\textwidth}
		\centering
		\setlength\fwidth{.8\textwidth}
		\setlength\fheight{.45\textwidth}
		\input{figures/vegas-rtt-2.tex}
		\caption{RTT Vegas}
		\label{fig:vegas-rtt}
	\end{subfigure}
	\begin{subfigure}[t]{.3\textwidth}
		\centering
		\setlength\fwidth{.8\textwidth}
		\setlength\fheight{.45\textwidth}
		\input{figures/quic-rtt-2.tex}
		\caption{RTT \gls{quic} (NewReno-like)}
		\label{fig:quic-rtt}
	\end{subfigure}
	\caption{Congestion window (CWND) and \gls{rtt} for different congestion control algorithms, with two flows over a shared bottleneck.}
	\label{fig:cwrtt}
\end{figure*}

\subsection{Testing and Applications} 
\label{sub:testing}
Finally, we implemented a basic set of unit tests, which will be extended in future releases. The first test suite (i.e., \texttt{quic-tx-buffer}) checks the correctness of the behavior of the socket and stream transmission buffers, implemented by \texttt{QuicSocketTxBuffer} and \texttt{QuicStreamTxBuffer}, respectively. For both the socket and the stream, the test checks different insertion and removal events from the buffer, including test cases in which the insertion is not successful due to a full socket buffer and the packet needs to be re-inserted in the stream buffer. Moreover, the socket test verifies the behavior in case of retransmissions. Similarly, the second test suite (i.e., \texttt{quic-rx-buffer}) performs add and remove operations on the reception buffers of the socket and stream. Finally, the third test suite (i.e., \texttt{quic-header}) verifies the correctness of the \texttt{QuicHeader} and \texttt{QuicSubheader} implementations.

Moreover, in order to explicitly use the stream multiplexing capabilities, we implemented an adaptation of the UDP client/server applications (i.e., \texttt{QuicClient} and \texttt{QuicServer}) which call the \texttt{Send} \gls{api} of the socket using a stream identifier. In this first implementation, the data is scheduled over the available streams (whose number can be set with an attribute) in a round robin fashion.

\section{Examples}
\label{sec:examples}


We now provide a set of preliminary results to validate the behavior of the ns-3 \gls{quic} implementation when using different congestion control algorithms, both in legacy and non-legacy modes. We consider a classic dumb-bell topology, often used to evaluate TCP performance~\cite{mishra2016tcp}, with two pairs of clients and servers sharing a common bottleneck of 2 Mbps. The minimum \gls{rtt} is 100 ms, and the example we used to run this experiment is \texttt{quic-variants-compari\-son}, adapted from the respective TCP example.

Fig.~\ref{fig:cwrtt} reports the evolution of the congestion window and the \gls{rtt} for the two flows, for two legacy congestion control algorithms (i.e., NewReno and Vegas) and the non-legacy option, which adapts NewReno with a slightly different congestion window increase in the congestion avoidance phase~\cite{draftquickrec}. The evolution of the congestion windows matches the expected behavior of the algorithms, with the two flows that settle in steady state to the \gls{bdp} (i.e., 25 kB). Moreover, when comparing the behaviors of delay-based (i.e., Vegas) and loss-based (i.e., NewReno and \gls{quic}) congestion control, it can be seen that, as expected, Vegas manages to maintain a smaller \gls{rtt}. Finally, NewReno and \gls{quic} congestion control exhibit similar trends, with the latter being characterized by a faster window ramp-up during the congestion avoidance phase.

\section{Conclusions}
\label{sec:conclusions}

In this paper, we presented a native ns-3 implementation of \gls{quic}, a new transport protocol that is being standardized by the \gls{ietf} and already accounts for a significant portion of the Internet traffic. The implementation of \gls{quic} described in this paper extends the code structure of the TCP ns-3 code base with the features that characterize \gls{quic}, e.g., stream multiplexing, low-latency initial handshake, improved \gls{sack} through ACK frames. Moreover, we designed the \gls{quic} socket implementation so that it is possible to plug both the legacy TCP and new \gls{quic}-only congestion control algorithms. We validated the performance of our implementation through simulations on a dumb-bell topology, which showed the typical behavior of the legacy congestion control algorithms and compared their performance to a non-legacy, Internet-Draft-based \gls{quic} congestion control. 
The code is publicly available, and is compatible with the latest version of ns-3 at the time of writing. 

\vspace{-.1cm}
\begin{acks}
	This work was partially supported  by the U.S. Commerce Department/NIST through the project ``An End-to-End Research Platform for Public Safety Communications above 6 GHz.'' The authors would also like to acknowledge Davide Marcato and Stefano Ravazzolo for their initial contribution to the project.
\end{acks}

\bibliographystyle{ACM-Reference-Format.bst}
\bibliography{bibl.bib}

\end{document}

%% file: acronyms.tex
\newacronym{quic}{QUIC}{Quick UDP Internet Connections}
\newacronym{3gpp}{3GPP}{3rd Generation Partnership Project}
\newacronym{adc}{ADC}{Analog to Digital Converter}
\newacronym{5g}{5G}{5th generation}
\newacronym{aimd}{AIMD}{Additive Increase Multiplicative Decrease}
\newacronym{am}{AM}{Acknowledged Mode}
\newacronym{amc}{AMC}{Adaptive Modulation and Coding}
\newacronym{aqm}{AQM}{Active Queue Management}
\newacronym{awgn}{AGWN}{Additive White Gaussian Noise}
\newacronym{balia}{BALIA}{Balanced Link Adaptation}
\newacronym{bdp}{BDP}{Bandwidth-Delay Product}
\newacronym{bf}{BF}{Beamforming}
\newacronym{cc}{CC}{Congestion Control}
\newacronym{cdf}{CDF}{Cumulative Distribution Function}
\newacronym{cn}{CN}{Core Network}
\newacronym{cqi}{CQI}{Channel Quality Information}
\newacronym{cp}{CP}{Control Plane}
\newacronym{csirs}{CSI-RS}{Channel State Information - Reference Signal}
\newacronym{dc}{DC}{Dual Connectivity}
\newacronym{dce}{DCE}{Direct Code Execution}
\newacronym{dci}{DCI}{Downlink Control Information}
\newacronym{dl}{DL}{Downlink}
\newacronym{dmr}{DMR}{Deadline Miss Ratio}
\newacronym{dmrs}{DMRS}{DeModulation Reference Signal}
\newacronym{e2e}{E2E}{End-to-End}
\newacronym{ecn}{ECN}{Explicit Congestion Notification}
\newacronym{edf}{EDF}{Earliest Deadline First}
\newacronym{enb}{eNB}{evolved Node Base}
\newacronym{epc}{EPC}{Evolved Packet Core}
\newacronym{es}{ES}{Edge Server}
\newacronym{fdma}{FDMA}{Frequency Division Multiple Access}
\newacronym{fdd}{FDD}{Frequency Division Duplexing}
\newacronym[firstplural=Radio Access Technologies (RATs)]{rat}{RAT}{Radio Access Technology}
\newacronym{fs}{FS}{Fast Switching}
\newacronym{ftp}{FTP}{File Transfer Protocol}
\newacronym{gnb}{gNB}{Next Generation Node Base}
\newacronym{harq}{HARQ}{Hybrid Automatic Repeat reQuest}
\newacronym{hetnet}{HetNet}{Heterogeneous Network}
\newacronym{hh}{HH}{Hard Handover}
\newacronym{hol}{HOL}{Head-of-Line}
\newacronym{ia}{IA}{Initial Access}
\newacronym{imt}{IMT}{International Mobile Telecommunication}
\newacronym{iot}{IoT}{Internet of Things}
\newacronym{los}{LOS}{Line of Sight}
\newacronym{lte}{LTE}{Long Term Evolution}
\newacronym{m2m}{M2M}{Machine to Machine}
\newacronym{mac}{MAC}{Medium Access Control}
\newacronym{mc}{MC}{Multi-Connectivity}
\newacronym{mcs}{MCS}{Modulation and Coding Scheme}
\newacronym{mec}{MEC}{Mobile Edge Cloud}
\newacronym{mi}{MI}{Mutual Information}
\newacronym{mimo}{MIMO}{Multiple Input, Multiple Output}
\newacronym{mmwave}{mmWave}{millimeter wave}
\newacronym{mr}{MR}{Maximum Rate}
\newacronym{mss}{MSS}{Maximum Segment Size}
\newacronym{mtd}{MTD}{Machine-Type Device}
\newacronym{mtu}{MTU}{Maximum Transmission Unit}
\newacronym{nfv}{NFV}{Network Function Virtualization}
\newacronym{nlos}{NLOS}{Non Line of Sight}
\newacronym{nr}{NR}{New Radio}
\newacronym{ofdm}{OFDM}{Orthogonal Frequency Division Multiplexing}
\newacronym{pdcch}{PDCCH}{Physical Downlonk Control Channel}
\newacronym{pdcp}{PDCP}{Packet Data Convergence Protocol}
\newacronym{pdsch}{PDSCH}{Physical Downlink Shared Channel}
\newacronym{pdu}{PDU}{Packet Data Unit}
\newacronym{pf}{PF}{Proportional Fair}
\newacronym{pgw}{PGW}{Packet Gateway}
\newacronym{phy}{PHY}{Physical}
\newacronym{pbch}{PBCH}{Physical Broadcast Channel}
\newacronym[plural=\gls{mme}s,firstplural=Mobility Management Entities (MMEs)]{mme}{MME}{Mobility Management Entity}
\newacronym{prb}{PRB}{Physical Resource Block}
\newacronym{pss}{PSS}{Primary Synchronization Signal}
\newacronym{pucch}{PUCCH}{Physical Uplink Control Channel}
\newacronym{pusch}{PUSCH}{Physical Uplink Shared Channel}
\newacronym{rach}{RACH}{Random Access Channel}
\newacronym{ran}{RAN}{Radio Access Network}
\newacronym{red}{RED}{Random Early Detection}
\newacronym{rf}{RF}{Radio Frequency}
\newacronym{rlc}{RLC}{Radio Link Control}
\newacronym{rlf}{RLF}{Radio Link Failure}
\newacronym{rrc}{RRC}{Radio Resource Control}
\newacronym{rrm}{RRM}{Radio Resource Management}
\newacronym{rr}{RR}{Round Robin}
\newacronym{rs}{RS}{Remote Server}
\newacronym{rsrp}{RSRP}{Reference Signal Received Power}
\newacronym{rss}{RSS}{Received Signal Strength}
\newacronym{rtt}{RTT}{Round Trip Time}
\newacronym{rw}{RW}{Receive Window}
\newacronym{rx}{RX}{Receiver}
\newacronym{sa}{SA}{standalone}
\newacronym{sack}{SACK}{Selective Acknowledgment}
\newacronym{sap}{SAP}{Service Access Point}
\newacronym{sch}{SCH}{Secondary Cell Handover}
\newacronym{scoot}{SCOOT}{Split Cycle Offset Optimization Technique}
\newacronym{sdma}{SDMA}{Spatial Division Multiple Access}
\newacronym{sinr}{SINR}{Signal to Interference plus Noise Ratio}
\newacronym{sm}{SM}{Saturation Mode}
\newacronym{snr}{SNR}{Signal to Noise Ratio}
\newacronym{son}{SON}{Self-Organizing Network}
\newacronym{ss}{SS}{Synchronization Signal}
\newacronym{srs}{SRS}{Sounding Reference Signal}
\newacronym{sss}{SSS}{Secondary Synchronization Signal}
\newacronym{tb}{TB}{Transport Block}
\newacronym{tcp}{TCP}{Transmission Control Protocol}
\newacronym{tdd}{TDD}{Time Division Duplexing}
\newacronym{tdma}{TDMA}{Time Division Multiple Access}
\newacronym{tfl}{TfL}{Transport for London}
\newacronym{tm}{TM}{Transparent Mode}
\newacronym{trp}{TRP}{Transmitter Receiver Pair}
\newacronym{tti}{TTI}{Transmission Time Interval}
\newacronym{ttt}{TTT}{Time-to-Trigger}
\newacronym{tx}{TX}{Transmitter}
\newacronym{ue}{UE}{User Equipment}
\newacronym{ul}{UL}{Uplink}
\newacronym{uml}{UML}{Unified Modeling Language}
\newacronym{um}{UM}{Unacknowledged Mode}
\newacronym{utc}{UTC}{Urban Traffic Control}
\newacronym{vm}{VM}{Virtual Machine}
\newacronym{rsrq}{RSRQ}{Reference Signal Received Quality}
\newacronym{rssi}{RSSI}{Received Signal Strength Indicator}
\newacronym{crs}{CRS}{Cell Reference Signal}
\newacronym{comp}{CoMP}{Coordinated Multi-Point}
\newacronym{cran}{C-RAN}{Cloud \acrlong{ran}}
\newacronym{ca}{CA}{Carrier Aggregation}
\newacronym{cco}{CC}{Carrier Component}
\newacronym{nsa}{NSA}{Non Stand Alone}
\newacronym{embb}{eMBB}{Enhanced Mobility Broadband}
\newacronym{bsr}{BSR}{Buffer Status Report}
\newacronym{srb}{SRB}{Service Radio Bearer}
\newacronym{scm}{SCM}{Spatial Channel Model}
\newacronym{sctp}{SCTP}{Stream Control Transmission Protocol}
\newacronym{mptcp}{MPTCP}{Multi-path TCP}
\newacronym{ietf}{IETF}{Internet Engineering Task Force}
\newacronym{os}{OS}{Operating System}
\newacronym{tls}{TLS}{Transport Layer Security}
\newacronym{rfc}{RFC}{Request for Comments}
\newacronym{http}{HTTP}{HyperText Transfer Protocol}
\newacronym{nat}{NAT}{Network Address Translation}
\newacronym{api}{API}{Application Programming Interface}
\newacronym{rto}{RTO}{Retransmission Timeout}

%% file: figures/newreno-cwnd-2.tex
%
%
\definecolor{mycolor1}{rgb}{0.00000,0.44700,0.74100}%
\definecolor{mycolor2}{rgb}{0.85000,0.32500,0.09800}%
\begin{tikzpicture}
\pgfplotsset{every tick label/.append style={font=\scriptsize}}

\begin{axis}[%
width=0.951\fwidth,
height=\fheight,
at={(0\fwidth,0\fheight)},
scale only axis,
xmajorgrids,
ymajorgrids,
xmin=0,
xmax=18,
xlabel style={font=\scriptsize\color{white!15!black}},
xlabel={Time (s)},
ymin=0,
ymax=90,
ylabel style={font=\scriptsize\color{white!15!black}},
ylabel={CWND (kB)},
axis background/.style={fill=white},
legend style={legend cell align=left, font=\scriptsize,  align=left, draw=white!15!black}
]
\addplot [color=mycolor1,dashed,line width=1.2pt]
  table[row sep=crcr]{%
0.25657	4.38\\
0.27159	5.84\\
0.27159	7.3\\
0.30782	8.76\\
0.40544	10.22\\
0.41802	11.68\\
0.42123	13.14\\
0.44252	14.6\\
0.45509	16.059\\
0.45797	17.519\\
0.57329	18.979\\
0.57329	20.439\\
0.599	21.899\\
0.599	23.359\\
0.599	24.819\\
0.6357	26.279\\
0.6357	27.739\\
0.6357	29.199\\
0.67348	30.659\\
0.67348	32.119\\
0.67348	33.579\\
0.84773	35.039\\
0.84773	36.499\\
0.84773	37.959\\
0.88431	39.419\\
0.88431	40.879\\
0.91002	42.339\\
0.91002	43.799\\
0.91002	45.259\\
0.94704	46.719\\
0.94704	48.179\\
0.94704	49.639\\
0.98419	51.099\\
0.98419	52.558\\
0.98419	54.018\\
1.02093	55.478\\
1.02093	56.938\\
1.02093	58.398\\
1.0592	59.858\\
1.0592	61.318\\
1.0592	62.778\\
1.09537	64.238\\
1.09537	65.69\\
1.09537	67.15\\
1.13212	68.61\\
1.13212	70.07\\
1.44216	71.53\\
1.44216	72.99\\
1.44216	74.45\\
1.47875	75.91\\
1.47875	77.37\\
1.50418	78.83\\
1.50418	80.29\\
1.50418	81.75\\
1.54078	83.21\\
1.54078	84.67\\
1.5661	86.13\\
1.5661	87.59\\
1.60328	43.795\\
2.31211	43.843\\
2.31211	43.891\\
2.43227	21.945\\
3.34434	23.013\\
3.34434	23.105\\
3.34434	23.197\\
3.45268	23.84\\
3.79063	23.929\\
3.79063	24.018\\
3.79063	24.106\\
3.82726	24.194\\
3.82726	24.282\\
3.85534	24.369\\
3.85534	24.456\\
3.85534	24.543\\
3.89202	24.629\\
3.89202	24.715\\
3.91931	24.801\\
3.91931	24.886\\
3.91931	24.971\\
3.95572	25.056\\
3.95572	25.141\\
3.95572	25.225\\
4.31951	25.309\\
4.31951	25.393\\
4.31951	25.476\\
4.35614	25.559\\
4.35614	25.642\\
4.38266	25.725\\
4.38266	25.807\\
4.38266	25.889\\
4.41895	25.971\\
4.41895	26.053\\
4.44547	26.134\\
4.44547	26.215\\
4.44547	26.296\\
4.48176	26.377\\
4.48176	26.457\\
4.50759	26.537\\
4.50759	26.617\\
4.50759	26.697\\
4.54547	26.776\\
4.89685	26.855\\
4.89685	26.934\\
4.92216	27.013\\
4.92216	27.091\\
4.94794	27.169\\
4.94794	27.247\\
4.94794	27.325\\
4.98463	27.403\\
4.98463	27.48\\
5.01081	27.557\\
5.01081	27.634\\
5.01081	27.711\\
5.04804	27.787\\
5.04804	27.863\\
5.04804	27.939\\
5.08508	28.015\\
5.08508	28.091\\
5.11131	28.166\\
5.11131	28.241\\
5.49953	14.12\\
5.95844	14.27\\
5.95844	14.419\\
5.95844	14.566\\
5.99521	14.712\\
5.99521	14.856\\
5.99521	14.999\\
6.03259	15.141\\
6.03259	15.281\\
6.03259	15.42\\
6.0697	15.558\\
6.256	15.695\\
6.256	15.83\\
6.256	15.964\\
6.29277	16.097\\
6.29277	16.229\\
6.29277	16.36\\
6.33027	16.49\\
6.33027	16.619\\
6.33027	16.747\\
6.36658	16.874\\
6.36658	17\\
6.58178	17.125\\
6.58178	17.249\\
6.58178	17.372\\
6.61808	17.494\\
6.61808	17.615\\
6.6434	17.736\\
6.6434	17.856\\
6.66956	17.975\\
6.66956	18.093\\
6.66956	18.21\\
6.70587	18.327\\
6.70587	18.443\\
6.73244	18.558\\
6.94552	18.672\\
6.94552	18.786\\
6.97135	18.899\\
6.97135	19.011\\
6.97135	19.123\\
7.0084	19.234\\
7.0084	19.344\\
7.03417	19.454\\
7.03417	19.563\\
7.03417	19.671\\
7.07174	19.779\\
7.07174	19.886\\
7.07174	19.993\\
7.34776	20.099\\
7.34776	20.205\\
7.38499	20.31\\
7.38499	20.414\\
7.38499	20.518\\
7.42255	20.621\\
7.42255	20.724\\
7.42255	20.826\\
7.4596	20.928\\
7.4596	21.029\\
7.48577	21.13\\
7.48577	21.23\\
7.48577	21.33\\
7.52208	21.429\\
7.52208	21.528\\
7.76264	21.627\\
7.76264	21.725\\
7.8246	21.823\\
7.8246	21.92\\
7.85037	22.017\\
7.85037	22.113\\
7.85037	22.209\\
7.88668	22.304\\
7.88668	22.399\\
7.91325	22.494\\
7.91325	22.588\\
7.91325	22.682\\
7.95036	22.775\\
7.95036	22.868\\
8.2024	22.961\\
8.2024	23.053\\
8.2024	23.145\\
8.239	23.237\\
8.2779	23.328\\
8.2779	23.419\\
8.315	23.51\\
8.315	23.6\\
8.3403	23.69\\
8.3403	23.779\\
8.3661	23.868\\
8.3661	23.957\\
8.3661	24.045\\
8.4028	24.133\\
8.4028	24.221\\
8.4281	24.309\\
8.4281	24.396\\
8.68	24.483\\
8.7053	24.57\\
8.7053	24.656\\
8.7306	24.742\\
8.7306	24.828\\
8.7931	24.913\\
8.7931	24.998\\
8.8184	25.083\\
8.8184	25.167\\
8.8438	25.251\\
8.8438	25.335\\
8.8438	25.419\\
8.8805	25.502\\
8.8805	25.585\\
8.9062	25.668\\
8.9062	25.751\\
9.1464	25.833\\
9.1464	25.915\\
9.183	25.997\\
9.183	26.078\\
9.2084	26.159\\
9.2084	26.24\\
9.2084	26.321\\
9.246	26.401\\
9.3206	26.481\\
9.3206	26.561\\
9.3464	26.641\\
9.3464	26.72\\
9.3464	26.799\\
9.3839	26.878\\
9.3839	26.957\\
9.4096	27.036\\
9.4096	27.114\\
9.6604	27.192\\
9.6604	27.27\\
9.6604	27.348\\
9.6979	27.425\\
9.6979	27.502\\
9.7237	27.579\\
9.7237	27.656\\
9.7237	27.733\\
9.7624	27.809\\
9.7624	27.885\\
9.8115	27.961\\
9.8738	28.037\\
9.8738	28.113\\
9.9	28.188\\
9.9	28.263\\
9.9	28.338\\
9.9376	28.413\\
9.9376	28.488\\
9.9376	28.562\\
9.9747	28.636\\
10.2271	28.71\\
10.2271	28.784\\
10.2271	28.858\\
10.2647	28.931\\
10.2647	29.004\\
10.2647	29.077\\
10.3013	29.15\\
10.3013	29.223\\
10.3266	29.295\\
10.3266	29.367\\
10.3529	29.439\\
10.3904	29.511\\
10.4524	29.583\\
10.4524	29.655\\
10.4777	29.726\\
10.4777	29.797\\
10.5035	29.868\\
10.5035	29.939\\
10.5035	30.01\\
10.5402	30.081\\
10.793	30.151\\
10.793	30.221\\
10.793	30.291\\
10.8297	30.361\\
10.8297	30.431\\
10.8558	30.501\\
10.8558	30.57\\
10.8558	30.639\\
10.8925	30.708\\
10.8925	30.777\\
10.9191	30.846\\
10.9191	30.915\\
10.9554	30.983\\
11.0313	31.051\\
11.0313	31.119\\
11.0572	31.187\\
11.0572	31.255\\
11.0572	31.323\\
11.0942	31.391\\
11.0942	31.458\\
11.3475	31.525\\
11.3475	31.592\\
11.3475	31.659\\
11.3841	31.726\\
11.3841	31.793\\
11.4107	31.86\\
11.4107	31.926\\
11.4107	31.992\\
11.4479	32.058\\
11.4479	32.124\\
11.4479	32.19\\
11.4857	32.256\\
11.4857	32.322\\
11.5233	32.387\\
11.5233	32.452\\
11.6233	16.226\\
12.15	16.357\\
12.2971	16.487\\
12.2971	16.616\\
12.2971	16.744\\
12.3342	16.871\\
12.3342	16.997\\
12.3342	17.122\\
12.3706	17.246\\
12.3706	17.369\\
12.3706	17.491\\
12.4073	17.612\\
12.4707	17.733\\
12.4707	17.853\\
12.5077	17.972\\
12.6461	18.09\\
12.6723	18.207\\
12.6723	18.324\\
12.6723	18.44\\
12.7086	18.555\\
12.7086	18.669\\
12.7339	18.783\\
12.7339	18.896\\
12.7601	19.008\\
12.7601	19.12\\
12.7601	19.231\\
12.8361	19.341\\
12.8361	19.451\\
12.8859	19.56\\
13.012	19.668\\
13.012	19.776\\
13.0745	19.883\\
13.0745	19.99\\
13.0745	20.096\\
13.1112	20.202\\
13.1112	20.307\\
13.137	20.411\\
13.137	20.515\\
13.137	20.618\\
13.2129	20.721\\
13.2129	20.823\\
13.2501	20.925\\
13.2501	21.026\\
13.4138	21.127\\
13.4138	21.227\\
13.4138	21.327\\
13.477	21.426\\
13.5023	21.525\\
13.5023	21.624\\
13.5277	21.722\\
13.5277	21.82\\
13.5277	21.917\\
13.5645	22.014\\
13.5645	22.11\\
13.5645	22.206\\
13.6293	22.301\\
13.6293	22.396\\
13.6293	22.491\\
13.6665	22.585\\
13.842	22.679\\
13.8681	22.772\\
13.9062	22.865\\
13.9062	22.958\\
13.9438	23.05\\
13.9438	23.142\\
13.9438	23.234\\
13.9813	23.325\\
13.9813	23.416\\
13.9813	23.506\\
14.018	23.596\\
14.0569	23.686\\
14.0569	23.775\\
14.0569	23.864\\
14.0935	23.953\\
14.0935	24.041\\
14.2846	24.129\\
14.3462	24.217\\
14.3462	24.304\\
14.3728	24.391\\
14.3728	24.478\\
14.3728	24.565\\
14.4091	24.651\\
14.4091	24.737\\
14.4348	24.823\\
14.4348	24.908\\
14.4602	24.993\\
14.5105	25.078\\
14.5105	25.162\\
14.5105	25.246\\
14.5471	25.33\\
14.5471	25.414\\
14.5724	25.497\\
14.5724	25.58\\
14.7501	25.663\\
14.7991	25.746\\
14.8611	25.828\\
14.8611	25.91\\
14.8868	25.992\\
14.8868	26.074\\
14.9122	26.155\\
14.9122	26.236\\
14.9122	26.317\\
14.9502	26.397\\
14.9885	26.477\\
14.9885	26.557\\
14.9885	26.637\\
15.0248	26.717\\
15.0248	26.796\\
15.0501	26.875\\
15.0501	26.954\\
15.2404	27.033\\
15.2404	27.111\\
15.2767	27.189\\
15.2767	27.267\\
15.3275	27.345\\
15.3653	27.422\\
15.3653	27.499\\
15.3911	27.576\\
15.3911	27.653\\
15.3911	27.73\\
15.4294	27.806\\
15.4674	27.882\\
15.4674	27.958\\
15.4674	28.034\\
15.5037	28.11\\
15.5037	28.185\\
15.5295	28.26\\
15.5295	28.335\\
15.5295	28.41\\
15.5679	28.485\\
15.8063	28.559\\
15.8063	28.633\\
15.8063	28.707\\
15.8426	28.781\\
15.8807	28.855\\
15.8807	28.928\\
15.9304	29.001\\
15.9304	29.074\\
15.957	29.147\\
15.957	29.22\\
16.0069	29.292\\
16.0069	29.364\\
16.0069	29.436\\
16.0435	29.508\\
16.0435	29.58\\
16.0697	29.652\\
16.0697	29.723\\
16.0697	29.794\\
16.1084	29.865\\
16.1084	29.936\\
16.1084	30.007\\
16.3468	30.078\\
16.3468	30.148\\
16.3979	30.218\\
16.3979	30.288\\
16.473	30.358\\
16.473	30.428\\
16.4988	30.498\\
16.4988	30.567\\
16.4988	30.636\\
16.5372	30.705\\
16.5372	30.774\\
16.5372	30.843\\
16.5738	30.912\\
16.5738	30.98\\
16.5992	31.048\\
16.5992	31.116\\
16.6245	31.184\\
16.6245	31.252\\
16.6499	31.32\\
16.6499	31.388\\
16.6499	31.455\\
16.6874	31.522\\
16.9378	31.589\\
16.9378	31.656\\
16.9884	31.723\\
16.9884	31.79\\
17.0644	31.857\\
17.0644	31.923\\
17.0897	31.989\\
17.0897	32.055\\
17.1151	32.121\\
17.1151	32.187\\
17.1151	32.253\\
17.1518	32.319\\
17.1778	16.159\\
17.5534	16.29\\
17.5896	16.42\\
17.5896	16.549\\
17.5896	16.677\\
17.6269	16.804\\
17.6269	16.93\\
17.6269	17.055\\
17.6636	17.179\\
17.8232	17.303\\
17.8232	17.426\\
17.8232	17.548\\
17.86	17.669\\
17.86	17.789\\
17.86	17.908\\
17.8968	18.026\\
17.8968	18.144\\
17.8968	18.261\\
17.9347	18.377\\
17.9347	18.492\\
17.9347	18.607\\
};

\addplot [color=mycolor2]
  table[row sep=crcr]{%
0.33764	4.38\\
0.35345	5.84\\
0.459	7.3\\
0.47482	8.76\\
0.5213	10.22\\
0.57985	11.68\\
0.6086	13.14\\
0.66457	14.6\\
0.70169	16.06\\
0.7267	17.52\\
0.74545	18.98\\
0.79539	20.44\\
0.85455	21.9\\
0.86702	23.36\\
0.90688	24.82\\
0.917	26.28\\
0.92602	27.74\\
0.95449	29.2\\
0.95723	30.66\\
0.98787	32.12\\
0.99799	33.58\\
1.02323	35.04\\
1.03823	36.5\\
1.12878	37.96\\
1.14399	39.42\\
1.16611	40.88\\
1.18133	42.34\\
1.19716	43.8\\
1.20638	45.26\\
1.22853	46.72\\
1.23739	48.18\\
1.25934	49.64\\
1.26823	51.1\\
1.28423	52.56\\
1.29329	54.02\\
1.32088	55.48\\
1.33081	56.94\\
1.34645	58.4\\
1.36183	59.86\\
1.38397	61.32\\
1.39935	62.78\\
1.41498	64.24\\
1.48618	65.7\\
1.54572	67.16\\
1.57142	68.62\\
1.60514	33.579\\
1.642	32.119\\
1.72085	32.848\\
1.77954	32.118\\
1.9446	32.184\\
1.97573	32.25\\
2.02958	32.316\\
2.09549	32.381\\
2.14343	32.446\\
2.18068	32.511\\
2.28148	32.576\\
2.31257	32.641\\
2.3614	32.706\\
2.39804	18.249\\
2.52509	18.365\\
2.54474	18.249\\
2.58211	18.365\\
2.59711	18.481\\
2.60742	18.979\\
2.65286	19.091\\
2.65511	19.202\\
2.69593	19.313\\
2.70455	19.423\\
2.74078	19.532\\
2.80758	19.641\\
2.83743	19.749\\
2.86834	19.856\\
2.89258	19.963\\
2.92349	20.069\\
2.94817	20.175\\
2.99111	20.28\\
3.01941	20.385\\
3.02194	20.489\\
3.04395	20.593\\
3.05892	20.696\\
3.06963	20.798\\
3.08462	20.9\\
3.10047	21.001\\
3.11547	21.102\\
3.14056	21.203\\
3.16228	21.303\\
3.17182	21.403\\
3.18779	21.502\\
3.21993	21.601\\
3.23278	21.699\\
3.24524	21.797\\
3.25786	21.894\\
3.27035	21.991\\
3.28532	22.087\\
3.29566	22.183\\
3.30832	22.279\\
3.32403	22.374\\
3.34605	22.469\\
3.36105	22.563\\
3.39343	22.657\\
3.42774	22.751\\
3.4404	22.844\\
3.45537	22.937\\
3.4718	23.029\\
3.47453	23.121\\
3.49674	23.213\\
3.51171	23.304\\
3.52475	23.395\\
3.54081	23.486\\
3.5533	23.576\\
3.57158	23.666\\
3.59103	23.756\\
3.60603	23.845\\
3.64112	23.934\\
3.65391	24.023\\
3.66909	24.111\\
3.69121	24.199\\
3.70035	24.287\\
3.72604	24.374\\
3.75112	24.461\\
3.77337	24.548\\
3.80501	24.634\\
3.81743	24.72\\
3.85121	24.806\\
3.86805	24.891\\
3.88305	24.976\\
3.89929	25.061\\
3.91429	25.146\\
3.93978	25.23\\
3.95584	25.314\\
3.97084	25.398\\
3.98405	25.481\\
4.00626	25.564\\
4.02126	25.647\\
4.0375	25.73\\
4.0525	25.812\\
4.06894	25.894\\
4.08394	25.976\\
4.10058	26.058\\
4.12569	26.139\\
4.16615	26.22\\
4.19793	26.301\\
4.21367	26.382\\
4.22596	26.462\\
4.25378	26.542\\
4.28291	26.622\\
4.29553	26.702\\
4.31053	26.781\\
4.32697	26.86\\
4.34271	26.939\\
4.37653	27.018\\
4.40465	27.096\\
4.4247	27.174\\
4.45186	27.252\\
4.47171	27.33\\
4.47441	27.407\\
4.50911	27.484\\
4.52157	27.561\\
4.54685	27.638\\
4.57829	27.715\\
4.59713	27.791\\
4.60972	27.867\\
4.62854	27.943\\
4.64351	28.019\\
4.66336	28.095\\
4.68489	28.17\\
4.69463	28.245\\
4.71669	28.32\\
4.75409	28.395\\
4.76675	28.47\\
4.7794	28.544\\
4.80468	28.618\\
4.82327	28.692\\
4.83827	28.766\\
4.87603	28.84\\
4.8925	28.913\\
4.90747	28.986\\
4.92937	29.059\\
4.94562	29.132\\
4.96151	29.205\\
4.97647	29.277\\
4.99897	29.349\\
5.02744	29.421\\
5.04336	29.493\\
5.05599	29.565\\
5.08743	29.637\\
5.11274	29.708\\
5.12792	29.779\\
5.15956	29.85\\
5.191	29.921\\
5.21588	29.992\\
5.24443	30.063\\
5.24764	30.133\\
5.28835	30.203\\
5.30121	30.273\\
5.31383	30.343\\
5.34507	30.413\\
5.37634	30.483\\
5.38896	30.552\\
5.4352	30.621\\
5.44554	30.69\\
5.45488	30.759\\
5.4858	30.828\\
5.51792	30.897\\
5.53963	30.965\\
5.55463	31.033\\
5.5713	31.101\\
5.58627	31.169\\
5.60903	31.237\\
5.62169	31.305\\
5.63669	31.373\\
5.65313	31.44\\
5.66848	31.507\\
5.72522	31.574\\
5.74657	31.641\\
5.75591	31.708\\
5.77235	31.775\\
5.79133	31.842\\
5.8007	31.908\\
5.83835	31.974\\
5.86057	32.04\\
5.87319	32.106\\
5.88568	32.172\\
5.89831	32.238\\
5.92255	32.304\\
5.93209	32.369\\
5.94853	32.434\\
5.96122	32.499\\
5.97997	32.564\\
5.99497	32.629\\
6.01773	32.694\\
6.04305	32.759\\
6.04574	32.824\\
6.07721	32.888\\
6.10882	32.952\\
6.13739	33.016\\
6.15005	33.08\\
6.1686	33.144\\
6.17149	33.208\\
6.21183	33.272\\
6.23674	33.336\\
6.28173	33.399\\
6.29419	33.462\\
6.30681	33.525\\
6.31929	33.588\\
6.32869	33.651\\
6.36319	33.714\\
6.36605	33.777\\
6.40365	33.84\\
6.42587	33.902\\
6.43578	33.964\\
6.47267	34.026\\
6.49544	34.088\\
6.53234	34.15\\
6.55832	34.212\\
6.57084	34.274\\
6.59863	34.336\\
6.61779	34.398\\
6.64631	34.459\\
6.67376	34.52\\
6.68407	34.581\\
6.69926	34.642\\
6.72488	34.703\\
6.74678	34.764\\
6.77139	34.825\\
6.78053	34.886\\
6.79697	34.947\\
6.81233	35.007\\
6.83454	35.067\\
6.84954	35.127\\
6.86638	35.187\\
6.87887	35.247\\
6.88156	35.307\\
6.92273	35.367\\
6.94495	35.427\\
6.96685	35.487\\
6.97948	35.547\\
7.00462	35.606\\
7.02629	35.665\\
7.06047	35.724\\
7.08665	35.783\\
7.10503	35.842\\
7.11772	35.901\\
7.1365	35.96\\
7.15147	36.019\\
7.17427	36.078\\
7.19322	36.137\\
7.2122	36.195\\
7.25589	36.253\\
7.26492	36.311\\
7.28737	36.369\\
7.29982	36.427\\
7.31248	36.485\\
7.3219	36.543\\
7.35334	36.601\\
7.38098	36.659\\
7.39032	36.717\\
7.41329	36.775\\
7.44473	36.832\\
7.46371	36.889\\
7.4762	36.946\\
7.47889	37.003\\
7.50128	37.06\\
7.51377	37.117\\
7.52659	37.174\\
7.54159	37.231\\
7.55783	37.288\\
7.57283	37.345\\
7.59539	37.402\\
7.60445	37.458\\
7.62683	37.514\\
7.64183	37.57\\
7.65484	37.626\\
7.67093	37.682\\
7.68321	37.738\\
7.71485	37.794\\
7.72751	37.85\\
7.74013	37.906\\
7.75871	37.962\\
7.77117	38.018\\
7.79921	38.074\\
7.82142	38.129\\
7.83408	38.184\\
7.84673	38.239\\
7.85935	38.294\\
7.87817	38.349\\
7.89059	38.404\\
7.90938	38.459\\
7.92207	38.514\\
7.94668	38.569\\
7.95562	38.624\\
7.98171	38.679\\
8.01	38.734\\
8.0128	38.789\\
8.0352	38.843\\
8.0478	38.897\\
8.0607	38.951\\
8.0756	39.005\\
8.0921	39.059\\
8.1383	39.113\\
8.1611	39.167\\
8.1736	39.221\\
8.1862	39.275\\
8.2113	39.329\\
8.2397	39.383\\
8.2617	39.437\\
8.2767	39.491\\
8.2923	39.544\\
8.312	39.597\\
8.321	39.65\\
8.3526	39.703\\
8.375	39.756\\
8.3873	39.809\\
8.3902	39.862\\
8.4223	39.915\\
8.4379	39.968\\
8.4529	40.021\\
8.4841	40.074\\
8.5008	40.127\\
8.5132	40.18\\
8.5161	40.233\\
8.5384	40.285\\
8.5637	40.337\\
8.5761	40.389\\
8.6074	40.441\\
8.6166	40.493\\
8.6387	40.545\\
8.6537	40.597\\
8.6795	40.649\\
8.6955	40.701\\
8.7043	40.753\\
8.7265	40.805\\
8.7415	40.857\\
8.7674	40.909\\
8.7832	40.961\\
8.7984	41.013\\
8.8208	41.064\\
8.8523	21.899\\
8.8901	21.996\\
8.9028	22.092\\
8.9154	22.188\\
8.9277	22.284\\
8.9616	22.379\\
8.9839	22.474\\
8.9988	22.568\\
9.0213	22.662\\
9.0363	22.756\\
9.055	22.849\\
9.0774	22.942\\
9.0921	23.034\\
9.1109	23.126\\
9.1334	23.218\\
9.1418	23.309\\
9.1765	23.4\\
9.1789	23.491\\
9.2102	23.581\\
9.2405	23.671\\
9.2776	23.761\\
9.3085	23.85\\
9.3324	23.939\\
9.3695	24.028\\
9.4063	24.116\\
9.4306	24.204\\
9.4649	24.292\\
9.4898	24.379\\
9.499	24.466\\
9.5265	24.553\\
9.5368	24.639\\
9.5519	24.725\\
9.5647	24.811\\
9.5803	24.896\\
9.6092	24.981\\
9.6242	25.066\\
9.6467	25.151\\
9.6748	25.235\\
9.6949	25.319\\
9.7226	25.403\\
9.732	25.486\\
9.7484	25.569\\
9.7582	25.652\\
9.7763	25.735\\
9.8113	25.817\\
9.8337	25.899\\
9.8554	25.981\\
9.8807	26.063\\
9.8957	26.144\\
9.9058	26.225\\
9.9208	26.306\\
9.9375	26.387\\
9.9465	26.467\\
9.9716	26.547\\
10.0004	26.627\\
10.0032	26.707\\
10.0285	26.786\\
10.056	26.865\\
10.0659	26.944\\
10.0818	27.023\\
10.0945	27.101\\
10.1071	27.179\\
10.1221	27.257\\
10.1476	27.335\\
10.1784	27.412\\
10.1916	27.489\\
10.2077	27.566\\
10.2227	27.643\\
10.2392	27.72\\
10.2419	27.796\\
10.2706	27.872\\
10.2833	27.948\\
10.296	28.024\\
10.3084	28.1\\
10.3326	28.175\\
10.342	28.25\\
10.3613	28.325\\
10.3985	28.4\\
10.4148	28.475\\
10.4177	28.549\\
10.4525	28.623\\
10.4552	28.697\\
10.4807	28.771\\
10.5029	28.845\\
10.5154	28.918\\
10.5407	28.991\\
10.5532	29.064\\
10.5682	29.137\\
10.5903	29.21\\
10.6036	29.282\\
10.6159	29.354\\
10.6189	29.426\\
10.6351	29.498\\
10.6602	29.57\\
10.6694	29.642\\
10.6914	29.713\\
10.701	29.784\\
10.7294	29.855\\
10.7418	29.926\\
10.7445	29.997\\
10.7765	30.068\\
10.7985	30.138\\
10.8074	30.208\\
10.8394	30.278\\
10.8613	30.348\\
10.8739	30.418\\
10.8961	30.488\\
10.9241	30.557\\
10.9366	30.626\\
10.949	30.695\\
10.9678	30.764\\
10.9828	30.833\\
10.9994	30.902\\
11.0086	30.97\\
11.0246	31.038\\
11.0395	31.106\\
11.0625	31.174\\
11.0811	31.242\\
11.084	31.31\\
11.1128	31.378\\
11.1157	31.445\\
11.1442	31.512\\
11.1469	31.579\\
11.1695	31.646\\
11.1785	31.713\\
11.2069	31.78\\
11.2544	31.847\\
11.2855	31.913\\
11.3173	31.979\\
11.3487	32.045\\
11.364	32.111\\
11.379	32.177\\
11.4042	32.243\\
11.4235	32.309\\
11.4577	32.374\\
11.467	32.439\\
11.49	32.504\\
11.4988	32.569\\
11.5308	32.634\\
11.5529	32.699\\
11.5619	32.764\\
11.5778	32.829\\
11.5928	32.893\\
11.6121	32.957\\
11.628	33.021\\
11.6407	33.085\\
11.6533	33.149\\
11.6683	33.213\\
11.688	33.277\\
11.7215	33.341\\
11.7505	33.404\\
11.7821	33.467\\
11.8194	33.53\\
11.8355	33.593\\
11.8504	33.656\\
11.8607	33.719\\
11.8698	33.782\\
11.8985	33.845\\
11.9112	33.907\\
11.9262	33.969\\
11.9453	34.031\\
11.9677	34.093\\
11.9829	34.155\\
12.0199	34.217\\
12.0401	34.279\\
12.0617	34.341\\
12.0766	34.403\\
12.087	34.464\\
12.096	34.525\\
12.1182	34.586\\
12.1332	34.647\\
12.1496	34.708\\
12.1523	34.769\\
12.1686	34.83\\
12.1811	34.891\\
12.2087	34.952\\
12.2251	35.012\\
12.2279	35.072\\
12.2564	35.132\\
12.2691	35.192\\
12.2878	35.252\\
12.3028	35.312\\
12.3224	35.372\\
12.3509	35.432\\
12.3636	35.492\\
12.3791	35.552\\
12.414	35.611\\
12.4326	35.67\\
12.4354	35.729\\
12.4516	35.788\\
12.4642	35.847\\
12.4893	35.906\\
12.492	35.965\\
12.5173	36.024\\
12.5585	36.083\\
12.5835	36.142\\
12.5934	36.2\\
12.6085	36.258\\
12.6235	36.316\\
12.6561	36.374\\
12.6714	36.432\\
12.6841	36.49\\
12.6866	36.548\\
12.7188	36.606\\
12.7341	36.664\\
12.747	36.722\\
12.7594	36.78\\
12.7688	36.837\\
12.7907	36.894\\
12.8001	36.951\\
12.8286	37.008\\
12.8411	37.065\\
12.875	37.122\\
12.8852	37.179\\
12.9105	37.236\\
12.9255	37.293\\
12.9481	37.35\\
12.9509	37.407\\
12.9761	37.463\\
13.0134	37.519\\
13.0356	37.575\\
13.0485	37.631\\
13.0701	37.687\\
13.1016	37.743\\
13.1295	37.799\\
13.1388	37.855\\
13.1681	37.911\\
13.1808	37.967\\
13.1934	38.023\\
13.2059	38.079\\
13.2216	38.134\\
13.2371	38.189\\
13.2465	38.244\\
13.2812	38.299\\
13.2836	38.354\\
13.3092	38.409\\
13.3378	38.464\\
13.3504	38.519\\
13.3723	38.574\\
13.4033	38.629\\
13.4314	38.684\\
13.4407	38.739\\
13.4722	38.794\\
13.4948	38.848\\
13.4977	38.902\\
13.5349	38.956\\
13.5603	39.01\\
13.5977	39.064\\
13.6142	39.118\\
13.6169	39.172\\
13.6328	39.226\\
13.6424	39.28\\
13.6646	39.334\\
13.6796	39.388\\
13.7024	39.442\\
13.7051	39.496\\
13.7361	39.549\\
13.7528	39.602\\
13.7618	39.655\\
13.7969	39.708\\
13.8057	39.761\\
13.8281	39.814\\
13.8408	39.867\\
13.8718	39.92\\
13.881	39.973\\
13.9034	40.026\\
13.9184	40.079\\
13.9286	40.132\\
13.9383	40.185\\
13.9625	40.238\\
13.9851	40.29\\
13.9978	40.342\\
14.0195	40.394\\
14.0507	40.446\\
14.0795	40.498\\
14.0883	40.55\\
14.1262	40.602\\
14.1488	40.654\\
14.1611	40.706\\
14.1857	40.758\\
14.2115	40.81\\
14.2142	40.862\\
14.2303	40.914\\
14.243	40.966\\
14.2681	41.018\\
14.2708	41.069\\
14.3026	41.12\\
14.3246	41.171\\
14.3396	41.222\\
14.3561	41.273\\
14.3589	41.324\\
14.3814	41.375\\
14.3839	41.426\\
14.4252	41.477\\
14.4442	41.528\\
14.4533	41.579\\
14.4818	41.63\\
14.4849	41.681\\
14.5132	41.732\\
14.5292	41.783\\
14.5573	41.834\\
14.5763	41.884\\
14.579	41.934\\
14.6162	41.984\\
14.6265	42.034\\
14.6358	42.084\\
14.6578	42.134\\
14.6728	42.184\\
14.6831	42.234\\
14.6927	42.284\\
14.7172	42.334\\
14.7521	42.384\\
14.7866	42.434\\
14.8154	42.484\\
14.8181	42.534\\
14.8521	42.584\\
14.8614	42.634\\
14.8842	42.683\\
14.894	42.732\\
14.9158	42.781\\
14.9409	42.83\\
14.9559	42.879\\
14.9751	42.928\\
14.9975	42.977\\
15.0101	43.026\\
15.0322	43.075\\
15.063	43.124\\
15.0853	43.173\\
15.0949	43.222\\
15.1479	21.899\\
15.173	21.996\\
15.188	22.092\\
15.2038	22.188\\
15.2391	22.284\\
15.2549	22.379\\
15.2675	22.474\\
15.2922	22.568\\
15.3015	22.662\\
15.3204	22.756\\
15.3429	22.849\\
15.3554	22.942\\
15.3807	23.034\\
15.393	18.979\\
15.427	19.091\\
15.4613	19.202\\
15.4644	19.313\\
15.4808	19.423\\
15.4898	19.532\\
15.5263	19.641\\
15.5692	19.749\\
15.5998	19.856\\
15.6371	19.963\\
15.6672	20.069\\
15.7043	20.175\\
15.7473	20.28\\
15.7904	20.385\\
15.8333	20.489\\
15.8703	20.593\\
15.9132	20.696\\
15.935	20.798\\
15.95	20.9\\
15.9662	21.001\\
15.9751	21.102\\
15.9856	21.203\\
16.0116	21.303\\
16.0206	21.403\\
16.0367	21.502\\
16.0516	21.601\\
16.062	21.699\\
16.077	21.797\\
16.0871	21.894\\
16.1021	21.991\\
16.1155	22.087\\
16.1379	22.183\\
16.1529	22.279\\
16.1905	22.374\\
16.2219	22.469\\
16.232	22.563\\
16.2447	22.657\\
16.2573	22.751\\
16.2723	22.844\\
16.2853	22.937\\
16.3076	23.029\\
16.3105	23.121\\
16.3518	23.213\\
16.3761	23.304\\
16.3854	23.395\\
16.4021	23.486\\
16.4145	23.576\\
16.4272	23.666\\
16.4397	23.756\\
16.4523	23.845\\
16.4673	23.934\\
16.4864	24.023\\
16.5179	24.111\\
16.5401	24.199\\
16.5551	24.287\\
16.5656	24.374\\
16.5812	24.461\\
16.5968	24.548\\
16.6095	24.634\\
16.6346	24.72\\
16.6373	24.806\\
16.6659	24.891\\
16.6785	24.976\\
16.694	25.061\\
16.7224	25.146\\
16.7259	25.23\\
16.7479	25.314\\
16.7603	25.398\\
16.7631	25.481\\
16.7916	25.564\\
16.8042	25.647\\
16.8388	25.73\\
16.8731	25.812\\
16.8761	25.894\\
16.9011	25.976\\
16.9386	26.058\\
16.9672	26.139\\
16.97	26.22\\
17.0044	26.301\\
17.0138	26.382\\
17.0452	26.462\\
17.0612	26.542\\
17.0703	26.622\\
17.0872	26.702\\
17.0899	26.781\\
17.1182	26.86\\
17.1464	26.939\\
17.1783	27.018\\
17.2064	27.096\\
17.2216	27.174\\
17.2526	27.252\\
17.2628	27.33\\
17.2754	27.407\\
17.2942	27.484\\
17.3032	27.561\\
17.3193	27.638\\
17.3346	27.715\\
17.3667	27.791\\
17.3982	27.867\\
17.4263	27.943\\
17.4416	28.019\\
17.4575	28.095\\
17.4725	28.17\\
17.4957	28.245\\
17.4984	28.32\\
17.5359	28.395\\
17.5612	28.47\\
17.5923	28.544\\
17.6211	28.618\\
17.63	28.692\\
17.6521	28.766\\
17.6671	28.84\\
17.6835	28.913\\
17.6962	28.986\\
17.7299	29.059\\
17.7403	29.132\\
17.7553	29.205\\
17.778	29.277\\
17.7805	29.349\\
17.8093	29.421\\
17.8311	29.493\\
17.8527	29.565\\
17.8622	29.637\\
17.8725	29.708\\
17.8875	29.779\\
17.9094	29.85\\
17.919	29.921\\
17.9419	29.992\\
17.9755	30.063\\
};

\end{axis}
\end{tikzpicture}%

%% file: figures/vegas-cwnd-2.tex
%
%
\definecolor{mycolor1}{rgb}{0.00000,0.44700,0.74100}%
\definecolor{mycolor2}{rgb}{0.85000,0.32500,0.09800}%
\begin{tikzpicture}
\pgfplotsset{every tick label/.append style={font=\scriptsize}}

\begin{axis}[%
width=0.951\fwidth,
height=\fheight,
at={(0\fwidth,0\fheight)},
scale only axis,
xmajorgrids,
ymajorgrids,
xmin=0,
xmax=18,
xlabel style={font=\scriptsize\color{white!15!black}},
xlabel={Time (s)},
ymin=0,
ymax=90,
ylabel style={font=\scriptsize\color{white!15!black}},
ylabel={CWND (kB)},
axis background/.style={fill=white},
title style={font=\bfseries},
legend style={legend cell align=left, at={(0.5, 1.1)}, anchor=south, font=\footnotesize,  align=left, draw=white!15!black},
legend columns=2
]
\addplot [color=mycolor1,dashed,line width=1.2pt]
  table[row sep=crcr]{%
0.25657	4.38\\
0.27159	5.84\\
0.27159	7.3\\
0.30782	8.76\\
0.40544	10.22\\
0.41802	11.68\\
0.42123	13.14\\
0.44252	14.6\\
0.45509	16.059\\
0.45797	17.519\\
0.57329	18.979\\
0.57329	20.439\\
0.599	21.899\\
0.599	23.359\\
0.599	24.819\\
0.6357	26.279\\
0.6357	27.739\\
0.6357	29.199\\
0.67348	30.659\\
0.67348	32.119\\
0.67348	33.579\\
0.84773	35.039\\
0.84773	36.499\\
0.84773	37.959\\
0.88431	39.419\\
0.88431	40.879\\
0.91002	42.339\\
0.91002	43.799\\
0.91002	45.259\\
0.94704	46.719\\
0.94704	48.179\\
0.94704	49.639\\
0.98419	51.099\\
0.98419	52.558\\
0.98419	54.018\\
1.02093	55.478\\
1.02093	56.938\\
1.02093	58.398\\
1.0592	59.858\\
1.0592	61.318\\
1.0592	62.778\\
1.09537	64.238\\
1.09537	65.69\\
1.09537	67.15\\
1.13212	68.61\\
1.13212	70.07\\
1.44216	71.53\\
1.44216	72.99\\
1.44216	74.45\\
1.47875	75.91\\
1.47875	77.37\\
1.50418	78.83\\
1.50418	80.29\\
1.50418	81.75\\
1.54078	83.21\\
1.54078	84.67\\
1.5661	86.13\\
1.5661	87.59\\
1.60328	43.795\\
2.31211	43.843\\
2.31211	43.891\\
2.43227	21.945\\
3.34434	23.013\\
3.34434	23.105\\
3.34434	23.197\\
3.45268	23.84\\
3.79063	23.929\\
3.79063	24.018\\
3.79063	24.106\\
3.82726	24.194\\
3.82726	24.282\\
3.85534	24.369\\
3.85534	24.456\\
3.85534	24.543\\
3.89202	24.629\\
3.89202	24.715\\
3.91931	24.801\\
3.91931	24.886\\
3.91931	24.971\\
3.95572	25.056\\
3.95572	25.141\\
3.95572	25.225\\
4.31951	25.309\\
4.31951	25.393\\
4.31951	25.476\\
4.35614	25.559\\
4.35614	25.642\\
4.38266	25.725\\
4.38266	25.807\\
4.38266	25.889\\
4.41895	25.971\\
4.41895	26.053\\
4.44547	26.134\\
4.44547	26.215\\
4.44547	26.296\\
4.48176	26.377\\
4.48176	26.457\\
4.50759	26.537\\
4.50759	26.617\\
4.50759	26.697\\
4.54547	26.776\\
4.89685	26.855\\
4.89685	26.934\\
4.92216	27.013\\
4.92216	27.091\\
4.94794	27.169\\
4.94794	27.247\\
4.94794	27.325\\
4.98463	27.403\\
4.98463	27.48\\
5.01081	27.557\\
5.01081	27.634\\
5.01081	27.711\\
5.04804	27.787\\
5.04804	27.863\\
5.04804	27.939\\
5.08508	28.015\\
5.08508	28.091\\
5.11131	28.166\\
5.11131	28.241\\
5.49953	14.12\\
5.95844	14.27\\
5.95844	14.419\\
5.95844	14.566\\
5.99521	14.712\\
5.99521	14.856\\
5.99521	14.999\\
6.03259	15.141\\
6.03259	15.281\\
6.03259	15.42\\
6.0697	15.558\\
6.256	15.695\\
6.256	15.83\\
6.256	15.964\\
6.29277	16.097\\
6.29277	16.229\\
6.29277	16.36\\
6.33027	16.49\\
6.33027	16.619\\
6.33027	16.747\\
6.36658	16.874\\
6.36658	17\\
6.58178	17.125\\
6.58178	17.249\\
6.58178	17.372\\
6.61808	17.494\\
6.61808	17.615\\
6.6434	17.736\\
6.6434	17.856\\
6.66956	17.975\\
6.66956	18.093\\
6.66956	18.21\\
6.70587	18.327\\
6.70587	18.443\\
6.73244	18.558\\
6.94552	18.672\\
6.94552	18.786\\
6.97135	18.899\\
6.97135	19.011\\
6.97135	19.123\\
7.0084	19.234\\
7.0084	19.344\\
7.03417	19.454\\
7.03417	19.563\\
7.03417	19.671\\
7.07174	19.779\\
7.07174	19.886\\
7.07174	19.993\\
7.34776	20.099\\
7.34776	20.205\\
7.38499	20.31\\
7.38499	20.414\\
7.38499	20.518\\
7.42255	20.621\\
7.42255	20.724\\
7.42255	20.826\\
7.4596	20.928\\
7.4596	21.029\\
7.48577	21.13\\
7.48577	21.23\\
7.48577	21.33\\
7.52208	21.429\\
7.52208	21.528\\
7.76264	21.627\\
7.76264	21.725\\
7.8246	21.823\\
7.8246	21.92\\
7.85037	22.017\\
7.85037	22.113\\
7.85037	22.209\\
7.88668	22.304\\
7.88668	22.399\\
7.91325	22.494\\
7.91325	22.588\\
7.91325	22.682\\
7.95036	22.775\\
7.95036	22.868\\
8.2024	22.961\\
8.2024	23.053\\
8.2024	23.145\\
8.239	23.237\\
8.2779	23.328\\
8.2779	23.419\\
8.315	23.51\\
8.315	23.6\\
8.3403	23.69\\
8.3403	23.779\\
8.3661	23.868\\
8.3661	23.957\\
8.3661	24.045\\
8.4028	24.133\\
8.4028	24.221\\
8.4281	24.309\\
8.4281	24.396\\
8.68	24.483\\
8.7053	24.57\\
8.7053	24.656\\
8.7306	24.742\\
8.7306	24.828\\
8.7931	24.913\\
8.7931	24.998\\
8.8184	25.083\\
8.8184	25.167\\
8.8438	25.251\\
8.8438	25.335\\
8.8438	25.419\\
8.8805	25.502\\
8.8805	25.585\\
8.9062	25.668\\
8.9062	25.751\\
9.1464	25.833\\
9.1464	25.915\\
9.183	25.997\\
9.183	26.078\\
9.2084	26.159\\
9.2084	26.24\\
9.2084	26.321\\
9.246	26.401\\
9.3206	26.481\\
9.3206	26.561\\
9.3464	26.641\\
9.3464	26.72\\
9.3464	26.799\\
9.3839	26.878\\
9.3839	26.957\\
9.4096	27.036\\
9.4096	27.114\\
9.6604	27.192\\
9.6604	27.27\\
9.6604	27.348\\
9.6979	27.425\\
9.6979	27.502\\
9.7237	27.579\\
9.7237	27.656\\
9.7237	27.733\\
9.7624	27.809\\
9.7624	27.885\\
9.8115	27.961\\
9.8738	28.037\\
9.8738	28.113\\
9.9	28.188\\
9.9	28.263\\
9.9	28.338\\
9.9376	28.413\\
9.9376	28.488\\
9.9376	28.562\\
9.9747	28.636\\
10.2271	28.71\\
10.2271	28.784\\
10.2271	28.858\\
10.2647	28.931\\
10.2647	29.004\\
10.2647	29.077\\
10.3013	29.15\\
10.3013	29.223\\
10.3266	29.295\\
10.3266	29.367\\
10.3529	29.439\\
10.3904	29.511\\
10.4524	29.583\\
10.4524	29.655\\
10.4777	29.726\\
10.4777	29.797\\
10.5035	29.868\\
10.5035	29.939\\
10.5035	30.01\\
10.5402	30.081\\
10.793	30.151\\
10.793	30.221\\
10.793	30.291\\
10.8297	30.361\\
10.8297	30.431\\
10.8558	30.501\\
10.8558	30.57\\
10.8558	30.639\\
10.8925	30.708\\
10.8925	30.777\\
10.9191	30.846\\
10.9191	30.915\\
10.9554	30.983\\
11.0313	31.051\\
11.0313	31.119\\
11.0572	31.187\\
11.0572	31.255\\
11.0572	31.323\\
11.0942	31.391\\
11.0942	31.458\\
11.3475	31.525\\
11.3475	31.592\\
11.3475	31.659\\
11.3841	31.726\\
11.3841	31.793\\
11.4107	31.86\\
11.4107	31.926\\
11.4107	31.992\\
11.4479	32.058\\
11.4479	32.124\\
11.4479	32.19\\
11.4857	32.256\\
11.4857	32.322\\
11.5233	32.387\\
11.5233	32.452\\
11.6233	16.226\\
12.15	16.357\\
12.2971	16.487\\
12.2971	16.616\\
12.2971	16.744\\
12.3342	16.871\\
12.3342	16.997\\
12.3342	17.122\\
12.3706	17.246\\
12.3706	17.369\\
12.3706	17.491\\
12.4073	17.612\\
12.4707	17.733\\
12.4707	17.853\\
12.5077	17.972\\
12.6461	18.09\\
12.6723	18.207\\
12.6723	18.324\\
12.6723	18.44\\
12.7086	18.555\\
12.7086	18.669\\
12.7339	18.783\\
12.7339	18.896\\
12.7601	19.008\\
12.7601	19.12\\
12.7601	19.231\\
12.8361	19.341\\
12.8361	19.451\\
12.8859	19.56\\
13.012	19.668\\
13.012	19.776\\
13.0745	19.883\\
13.0745	19.99\\
13.0745	20.096\\
13.1112	20.202\\
13.1112	20.307\\
13.137	20.411\\
13.137	20.515\\
13.137	20.618\\
13.2129	20.721\\
13.2129	20.823\\
13.2501	20.925\\
13.2501	21.026\\
13.4138	21.127\\
13.4138	21.227\\
13.4138	21.327\\
13.477	21.426\\
13.5023	21.525\\
13.5023	21.624\\
13.5277	21.722\\
13.5277	21.82\\
13.5277	21.917\\
13.5645	22.014\\
13.5645	22.11\\
13.5645	22.206\\
13.6293	22.301\\
13.6293	22.396\\
13.6293	22.491\\
13.6665	22.585\\
13.842	22.679\\
13.8681	22.772\\
13.9062	22.865\\
13.9062	22.958\\
13.9438	23.05\\
13.9438	23.142\\
13.9438	23.234\\
13.9813	23.325\\
13.9813	23.416\\
13.9813	23.506\\
14.018	23.596\\
14.0569	23.686\\
14.0569	23.775\\
14.0569	23.864\\
14.0935	23.953\\
14.0935	24.041\\
14.2846	24.129\\
14.3462	24.217\\
14.3462	24.304\\
14.3728	24.391\\
14.3728	24.478\\
14.3728	24.565\\
14.4091	24.651\\
14.4091	24.737\\
14.4348	24.823\\
14.4348	24.908\\
14.4602	24.993\\
14.5105	25.078\\
14.5105	25.162\\
14.5105	25.246\\
14.5471	25.33\\
14.5471	25.414\\
14.5724	25.497\\
14.5724	25.58\\
14.7501	25.663\\
14.7991	25.746\\
14.8611	25.828\\
14.8611	25.91\\
14.8868	25.992\\
14.8868	26.074\\
14.9122	26.155\\
14.9122	26.236\\
14.9122	26.317\\
14.9502	26.397\\
14.9885	26.477\\
14.9885	26.557\\
14.9885	26.637\\
15.0248	26.717\\
15.0248	26.796\\
15.0501	26.875\\
15.0501	26.954\\
15.2404	27.033\\
15.2404	27.111\\
15.2767	27.189\\
15.2767	27.267\\
15.3275	27.345\\
15.3653	27.422\\
15.3653	27.499\\
15.3911	27.576\\
15.3911	27.653\\
15.3911	27.73\\
15.4294	27.806\\
15.4674	27.882\\
15.4674	27.958\\
15.4674	28.034\\
15.5037	28.11\\
15.5037	28.185\\
15.5295	28.26\\
15.5295	28.335\\
15.5295	28.41\\
15.5679	28.485\\
15.8063	28.559\\
15.8063	28.633\\
15.8063	28.707\\
15.8426	28.781\\
15.8807	28.855\\
15.8807	28.928\\
15.9304	29.001\\
15.9304	29.074\\
15.957	29.147\\
15.957	29.22\\
16.0069	29.292\\
16.0069	29.364\\
16.0069	29.436\\
16.0435	29.508\\
16.0435	29.58\\
16.0697	29.652\\
16.0697	29.723\\
16.0697	29.794\\
16.1084	29.865\\
16.1084	29.936\\
16.1084	30.007\\
16.3468	30.078\\
16.3468	30.148\\
16.3979	30.218\\
16.3979	30.288\\
16.473	30.358\\
16.473	30.428\\
16.4988	30.498\\
16.4988	30.567\\
16.4988	30.636\\
16.5372	30.705\\
16.5372	30.774\\
16.5372	30.843\\
16.5738	30.912\\
16.5738	30.98\\
16.5992	31.048\\
16.5992	31.116\\
16.6245	31.184\\
16.6245	31.252\\
16.6499	31.32\\
16.6499	31.388\\
16.6499	31.455\\
16.6874	31.522\\
16.9378	31.589\\
16.9378	31.656\\
16.9884	31.723\\
16.9884	31.79\\
17.0644	31.857\\
17.0644	31.923\\
17.0897	31.989\\
17.0897	32.055\\
17.1151	32.121\\
17.1151	32.187\\
17.1151	32.253\\
17.1518	32.319\\
17.1778	16.159\\
17.5534	16.29\\
17.5896	16.42\\
17.5896	16.549\\
17.5896	16.677\\
17.6269	16.804\\
17.6269	16.93\\
17.6269	17.055\\
17.6636	17.179\\
17.8232	17.303\\
17.8232	17.426\\
17.8232	17.548\\
17.86	17.669\\
17.86	17.789\\
17.86	17.908\\
17.8968	18.026\\
17.8968	18.144\\
17.8968	18.261\\
17.9347	18.377\\
17.9347	18.492\\
17.9347	18.607\\
};
\addlegendentry{Flow 1}

\addplot [color=mycolor2]
  table[row sep=crcr]{%
0.33764	4.38\\
0.35345	5.84\\
0.459	7.3\\
0.47482	8.76\\
0.5213	10.22\\
0.57985	11.68\\
0.6086	13.14\\
0.66457	14.6\\
0.70152	16.06\\
0.72691	17.52\\
0.74884	18.98\\
0.75199	20.44\\
0.78605	21.9\\
0.81181	23.36\\
0.82428	24.82\\
0.84594	26.28\\
0.87686	23.36\\
1.08907	21.9\\
1.27767	20.44\\
1.43908	18.98\\
18	18.98\\
};
\addlegendentry{Flow 2}

\end{axis}
\end{tikzpicture}%

%% file: figures/quic-cwnd-2.tex
%
%
\definecolor{mycolor1}{rgb}{0.00000,0.44700,0.74100}%
\definecolor{mycolor2}{rgb}{0.85000,0.32500,0.09800}%
\begin{tikzpicture}
\pgfplotsset{every tick label/.append style={font=\scriptsize}}

\begin{axis}[%
width=0.951\fwidth,
height=\fheight,
at={(0\fwidth,0\fheight)},
scale only axis,
xmajorgrids,
ymajorgrids,
xmin=0,
xmax=18,
xlabel style={font=\scriptsize\color{white!15!black}},
xlabel={Time (s)},
ymin=0,
ymax=90,
ylabel style={font=\scriptsize\color{white!15!black}},
ylabel={CWND (kB)},
axis background/.style={fill=white},
title style={font=\bfseries},
legend style={legend cell align=left, font=\scriptsize,  align=left, draw=white!15!black}
]
\addplot [color=mycolor1,dashed,line width=1.2pt]
  table[row sep=crcr]{%
0.25657	4.38\\
0.27159	5.84\\
0.27159	7.3\\
0.30782	8.76\\
0.40544	10.22\\
0.41802	11.68\\
0.42123	13.14\\
0.44252	14.6\\
0.45509	16.059\\
0.45797	17.519\\
0.57329	18.979\\
0.57329	20.439\\
0.599	21.899\\
0.599	23.359\\
0.599	24.819\\
0.6357	26.279\\
0.6357	27.739\\
0.6357	29.199\\
0.67348	30.659\\
0.67348	32.119\\
0.67348	33.579\\
0.84773	35.039\\
0.84773	36.499\\
0.84773	37.959\\
0.88431	39.419\\
0.88431	40.879\\
0.91002	42.339\\
0.91002	43.799\\
0.91002	45.259\\
0.94704	46.719\\
0.94704	48.179\\
0.94704	49.639\\
0.98419	51.099\\
0.98419	52.558\\
0.98419	54.018\\
1.02093	55.478\\
1.02093	56.938\\
1.02093	58.398\\
1.0592	59.858\\
1.0592	61.318\\
1.0592	62.778\\
1.09537	64.238\\
1.09537	65.69\\
1.09537	67.15\\
1.13212	68.61\\
1.13212	70.07\\
1.44216	71.53\\
1.44216	72.99\\
1.44216	74.45\\
1.47875	75.91\\
1.47875	77.37\\
1.50418	78.83\\
1.50418	80.29\\
1.50418	81.75\\
1.54078	83.21\\
1.54078	84.67\\
1.5661	86.13\\
1.5661	87.59\\
1.60328	43.795\\
2.31211	43.843\\
2.31211	43.891\\
2.43227	21.945\\
3.34434	23.013\\
3.34434	23.105\\
3.34434	23.197\\
3.45268	23.84\\
3.79063	23.929\\
3.79063	24.018\\
3.79063	24.106\\
3.82726	24.194\\
3.82726	24.282\\
3.85534	24.369\\
3.85534	24.456\\
3.85534	24.543\\
3.89202	24.629\\
3.89202	24.715\\
3.91931	24.801\\
3.91931	24.886\\
3.91931	24.971\\
3.95572	25.056\\
3.95572	25.141\\
3.95572	25.225\\
4.31951	25.309\\
4.31951	25.393\\
4.31951	25.476\\
4.35614	25.559\\
4.35614	25.642\\
4.38266	25.725\\
4.38266	25.807\\
4.38266	25.889\\
4.41895	25.971\\
4.41895	26.053\\
4.44547	26.134\\
4.44547	26.215\\
4.44547	26.296\\
4.48176	26.377\\
4.48176	26.457\\
4.50759	26.537\\
4.50759	26.617\\
4.50759	26.697\\
4.54547	26.776\\
4.89685	26.855\\
4.89685	26.934\\
4.92216	27.013\\
4.92216	27.091\\
4.94794	27.169\\
4.94794	27.247\\
4.94794	27.325\\
4.98463	27.403\\
4.98463	27.48\\
5.01081	27.557\\
5.01081	27.634\\
5.01081	27.711\\
5.04804	27.787\\
5.04804	27.863\\
5.04804	27.939\\
5.08508	28.015\\
5.08508	28.091\\
5.11131	28.166\\
5.11131	28.241\\
5.49953	14.12\\
5.95844	14.27\\
5.95844	14.419\\
5.95844	14.566\\
5.99521	14.712\\
5.99521	14.856\\
5.99521	14.999\\
6.03259	15.141\\
6.03259	15.281\\
6.03259	15.42\\
6.0697	15.558\\
6.256	15.695\\
6.256	15.83\\
6.256	15.964\\
6.29277	16.097\\
6.29277	16.229\\
6.29277	16.36\\
6.33027	16.49\\
6.33027	16.619\\
6.33027	16.747\\
6.36658	16.874\\
6.36658	17\\
6.58178	17.125\\
6.58178	17.249\\
6.58178	17.372\\
6.61808	17.494\\
6.61808	17.615\\
6.6434	17.736\\
6.6434	17.856\\
6.66956	17.975\\
6.66956	18.093\\
6.66956	18.21\\
6.70587	18.327\\
6.70587	18.443\\
6.73244	18.558\\
6.94552	18.672\\
6.94552	18.786\\
6.97135	18.899\\
6.97135	19.011\\
6.97135	19.123\\
7.0084	19.234\\
7.0084	19.344\\
7.03417	19.454\\
7.03417	19.563\\
7.03417	19.671\\
7.07174	19.779\\
7.07174	19.886\\
7.07174	19.993\\
7.34776	20.099\\
7.34776	20.205\\
7.38499	20.31\\
7.38499	20.414\\
7.38499	20.518\\
7.42255	20.621\\
7.42255	20.724\\
7.42255	20.826\\
7.4596	20.928\\
7.4596	21.029\\
7.48577	21.13\\
7.48577	21.23\\
7.48577	21.33\\
7.52208	21.429\\
7.52208	21.528\\
7.76264	21.627\\
7.76264	21.725\\
7.8246	21.823\\
7.8246	21.92\\
7.85037	22.017\\
7.85037	22.113\\
7.85037	22.209\\
7.88668	22.304\\
7.88668	22.399\\
7.91325	22.494\\
7.91325	22.588\\
7.91325	22.682\\
7.95036	22.775\\
7.95036	22.868\\
8.2024	22.961\\
8.2024	23.053\\
8.2024	23.145\\
8.239	23.237\\
8.2779	23.328\\
8.2779	23.419\\
8.315	23.51\\
8.315	23.6\\
8.3403	23.69\\
8.3403	23.779\\
8.3661	23.868\\
8.3661	23.957\\
8.3661	24.045\\
8.4028	24.133\\
8.4028	24.221\\
8.4281	24.309\\
8.4281	24.396\\
8.68	24.483\\
8.7053	24.57\\
8.7053	24.656\\
8.7306	24.742\\
8.7306	24.828\\
8.7931	24.913\\
8.7931	24.998\\
8.8184	25.083\\
8.8184	25.167\\
8.8438	25.251\\
8.8438	25.335\\
8.8438	25.419\\
8.8805	25.502\\
8.8805	25.585\\
8.9062	25.668\\
8.9062	25.751\\
9.1464	25.833\\
9.1464	25.915\\
9.183	25.997\\
9.183	26.078\\
9.2084	26.159\\
9.2084	26.24\\
9.2084	26.321\\
9.246	26.401\\
9.3206	26.481\\
9.3206	26.561\\
9.3464	26.641\\
9.3464	26.72\\
9.3464	26.799\\
9.3839	26.878\\
9.3839	26.957\\
9.4096	27.036\\
9.4096	27.114\\
9.6604	27.192\\
9.6604	27.27\\
9.6604	27.348\\
9.6979	27.425\\
9.6979	27.502\\
9.7237	27.579\\
9.7237	27.656\\
9.7237	27.733\\
9.7624	27.809\\
9.7624	27.885\\
9.8115	27.961\\
9.8738	28.037\\
9.8738	28.113\\
9.9	28.188\\
9.9	28.263\\
9.9	28.338\\
9.9376	28.413\\
9.9376	28.488\\
9.9376	28.562\\
9.9747	28.636\\
10.2271	28.71\\
10.2271	28.784\\
10.2271	28.858\\
10.2647	28.931\\
10.2647	29.004\\
10.2647	29.077\\
10.3013	29.15\\
10.3013	29.223\\
10.3266	29.295\\
10.3266	29.367\\
10.3529	29.439\\
10.3904	29.511\\
10.4524	29.583\\
10.4524	29.655\\
10.4777	29.726\\
10.4777	29.797\\
10.5035	29.868\\
10.5035	29.939\\
10.5035	30.01\\
10.5402	30.081\\
10.793	30.151\\
10.793	30.221\\
10.793	30.291\\
10.8297	30.361\\
10.8297	30.431\\
10.8558	30.501\\
10.8558	30.57\\
10.8558	30.639\\
10.8925	30.708\\
10.8925	30.777\\
10.9191	30.846\\
10.9191	30.915\\
10.9554	30.983\\
11.0313	31.051\\
11.0313	31.119\\
11.0572	31.187\\
11.0572	31.255\\
11.0572	31.323\\
11.0942	31.391\\
11.0942	31.458\\
11.3475	31.525\\
11.3475	31.592\\
11.3475	31.659\\
11.3841	31.726\\
11.3841	31.793\\
11.4107	31.86\\
11.4107	31.926\\
11.4107	31.992\\
11.4479	32.058\\
11.4479	32.124\\
11.4479	32.19\\
11.4857	32.256\\
11.4857	32.322\\
11.5233	32.387\\
11.5233	32.452\\
11.6233	16.226\\
12.15	16.357\\
12.2971	16.487\\
12.2971	16.616\\
12.2971	16.744\\
12.3342	16.871\\
12.3342	16.997\\
12.3342	17.122\\
12.3706	17.246\\
12.3706	17.369\\
12.3706	17.491\\
12.4073	17.612\\
12.4707	17.733\\
12.4707	17.853\\
12.5077	17.972\\
12.6461	18.09\\
12.6723	18.207\\
12.6723	18.324\\
12.6723	18.44\\
12.7086	18.555\\
12.7086	18.669\\
12.7339	18.783\\
12.7339	18.896\\
12.7601	19.008\\
12.7601	19.12\\
12.7601	19.231\\
12.8361	19.341\\
12.8361	19.451\\
12.8859	19.56\\
13.012	19.668\\
13.012	19.776\\
13.0745	19.883\\
13.0745	19.99\\
13.0745	20.096\\
13.1112	20.202\\
13.1112	20.307\\
13.137	20.411\\
13.137	20.515\\
13.137	20.618\\
13.2129	20.721\\
13.2129	20.823\\
13.2501	20.925\\
13.2501	21.026\\
13.4138	21.127\\
13.4138	21.227\\
13.4138	21.327\\
13.477	21.426\\
13.5023	21.525\\
13.5023	21.624\\
13.5277	21.722\\
13.5277	21.82\\
13.5277	21.917\\
13.5645	22.014\\
13.5645	22.11\\
13.5645	22.206\\
13.6293	22.301\\
13.6293	22.396\\
13.6293	22.491\\
13.6665	22.585\\
13.842	22.679\\
13.8681	22.772\\
13.9062	22.865\\
13.9062	22.958\\
13.9438	23.05\\
13.9438	23.142\\
13.9438	23.234\\
13.9813	23.325\\
13.9813	23.416\\
13.9813	23.506\\
14.018	23.596\\
14.0569	23.686\\
14.0569	23.775\\
14.0569	23.864\\
14.0935	23.953\\
14.0935	24.041\\
14.2846	24.129\\
14.3462	24.217\\
14.3462	24.304\\
14.3728	24.391\\
14.3728	24.478\\
14.3728	24.565\\
14.4091	24.651\\
14.4091	24.737\\
14.4348	24.823\\
14.4348	24.908\\
14.4602	24.993\\
14.5105	25.078\\
14.5105	25.162\\
14.5105	25.246\\
14.5471	25.33\\
14.5471	25.414\\
14.5724	25.497\\
14.5724	25.58\\
14.7501	25.663\\
14.7991	25.746\\
14.8611	25.828\\
14.8611	25.91\\
14.8868	25.992\\
14.8868	26.074\\
14.9122	26.155\\
14.9122	26.236\\
14.9122	26.317\\
14.9502	26.397\\
14.9885	26.477\\
14.9885	26.557\\
14.9885	26.637\\
15.0248	26.717\\
15.0248	26.796\\
15.0501	26.875\\
15.0501	26.954\\
15.2404	27.033\\
15.2404	27.111\\
15.2767	27.189\\
15.2767	27.267\\
15.3275	27.345\\
15.3653	27.422\\
15.3653	27.499\\
15.3911	27.576\\
15.3911	27.653\\
15.3911	27.73\\
15.4294	27.806\\
15.4674	27.882\\
15.4674	27.958\\
15.4674	28.034\\
15.5037	28.11\\
15.5037	28.185\\
15.5295	28.26\\
15.5295	28.335\\
15.5295	28.41\\
15.5679	28.485\\
15.8063	28.559\\
15.8063	28.633\\
15.8063	28.707\\
15.8426	28.781\\
15.8807	28.855\\
15.8807	28.928\\
15.9304	29.001\\
15.9304	29.074\\
15.957	29.147\\
15.957	29.22\\
16.0069	29.292\\
16.0069	29.364\\
16.0069	29.436\\
16.0435	29.508\\
16.0435	29.58\\
16.0697	29.652\\
16.0697	29.723\\
16.0697	29.794\\
16.1084	29.865\\
16.1084	29.936\\
16.1084	30.007\\
16.3468	30.078\\
16.3468	30.148\\
16.3979	30.218\\
16.3979	30.288\\
16.473	30.358\\
16.473	30.428\\
16.4988	30.498\\
16.4988	30.567\\
16.4988	30.636\\
16.5372	30.705\\
16.5372	30.774\\
16.5372	30.843\\
16.5738	30.912\\
16.5738	30.98\\
16.5992	31.048\\
16.5992	31.116\\
16.6245	31.184\\
16.6245	31.252\\
16.6499	31.32\\
16.6499	31.388\\
16.6499	31.455\\
16.6874	31.522\\
16.9378	31.589\\
16.9378	31.656\\
16.9884	31.723\\
16.9884	31.79\\
17.0644	31.857\\
17.0644	31.923\\
17.0897	31.989\\
17.0897	32.055\\
17.1151	32.121\\
17.1151	32.187\\
17.1151	32.253\\
17.1518	32.319\\
17.1778	16.159\\
17.5534	16.29\\
17.5896	16.42\\
17.5896	16.549\\
17.5896	16.677\\
17.6269	16.804\\
17.6269	16.93\\
17.6269	17.055\\
17.6636	17.179\\
17.8232	17.303\\
17.8232	17.426\\
17.8232	17.548\\
17.86	17.669\\
17.86	17.789\\
17.86	17.908\\
17.8968	18.026\\
17.8968	18.144\\
17.8968	18.261\\
17.9347	18.377\\
17.9347	18.492\\
17.9347	18.607\\
};

\addplot [color=mycolor2]
  table[row sep=crcr]{%
0.33764	4.124\\
0.33764	5.33\\
0.33764	6.536\\
0.33764	7.996\\
0.33764	9.456\\
0.33764	10.916\\
0.35345	12.376\\
0.45865	13.836\\
0.45865	15.296\\
0.45865	16.756\\
0.477	18.216\\
0.477	19.675\\
0.477	21.135\\
0.49928	22.595\\
0.49928	24.055\\
0.50833	25.515\\
0.55173	26.975\\
0.63897	26.976\\
0.63897	28.436\\
0.63897	29.896\\
0.65769	31.356\\
0.65769	32.816\\
0.65769	34.276\\
0.67588	35.736\\
0.67588	37.196\\
0.67588	38.656\\
0.69458	40.116\\
0.69458	41.576\\
0.69458	43.036\\
0.71311	44.496\\
0.71311	45.956\\
0.71311	47.416\\
0.75004	48.876\\
0.75004	50.336\\
0.78081	51.796\\
0.78081	53.256\\
0.97973	54.716\\
0.97973	56.175\\
0.97973	57.635\\
1.01052	59.095\\
1.01052	60.555\\
1.02334	62.015\\
1.02334	63.475\\
1.02334	64.935\\
1.04209	66.395\\
1.04209	67.855\\
1.04209	69.307\\
1.06053	70.767\\
1.06053	72.227\\
1.06053	73.687\\
1.0911	36.843\\
1.57124	36.9\\
1.57124	36.957\\
1.57124	37.014\\
1.66747	37.417\\
1.66747	37.473\\
1.66747	37.529\\
1.91758	37.642\\
1.91758	37.698\\
1.91758	37.754\\
1.96646	37.979\\
1.96646	38.035\\
1.98358	38.259\\
1.98358	38.314\\
1.98358	38.369\\
2.02558	38.646\\
2.02558	38.701\\
2.02558	38.756\\
2.04473	38.811\\
2.04473	38.865\\
2.04473	38.919\\
2.07586	38.973\\
2.07586	39.027\\
2.16209	39.081\\
2.16209	39.135\\
2.29062	39.189\\
2.29062	39.243\\
2.29062	39.297\\
2.30841	39.351\\
2.30841	39.405\\
2.30841	39.459\\
2.33989	19.729\\
2.74843	20.269\\
2.74843	20.374\\
2.74843	20.478\\
2.82148	20.582\\
2.82148	20.685\\
2.90845	20.788\\
2.90845	20.89\\
2.90845	20.992\\
2.9278	21.093\\
2.9278	21.194\\
2.9278	21.294\\
2.94635	21.394\\
2.94635	21.493\\
2.94635	21.592\\
2.96456	21.69\\
2.96456	21.788\\
2.96456	21.885\\
2.99542	21.982\\
2.99542	22.078\\
3.12828	22.174\\
3.12828	22.27\\
3.12828	22.365\\
3.14724	22.46\\
3.14724	22.554\\
3.14724	22.648\\
3.16605	22.742\\
3.16605	22.835\\
3.16605	22.928\\
3.18483	23.02\\
3.18483	23.112\\
3.18483	23.204\\
3.21031	23.295\\
3.21031	23.386\\
3.21031	23.477\\
3.37286	23.567\\
3.37286	23.657\\
3.37286	23.747\\
3.39184	23.836\\
3.39184	23.925\\
3.39184	24.014\\
3.4108	24.102\\
3.4108	24.19\\
3.4108	24.278\\
3.42958	24.365\\
3.42958	24.452\\
3.42958	24.539\\
3.4486	24.625\\
3.4486	24.711\\
3.4486	24.797\\
3.4798	24.882\\
3.4798	24.967\\
3.4798	25.052\\
3.64297	25.137\\
3.64297	25.221\\
3.65598	25.305\\
3.65598	25.389\\
3.65598	25.472\\
3.675	25.555\\
3.675	25.638\\
3.675	25.721\\
3.69358	25.803\\
3.69358	25.885\\
3.69358	25.967\\
3.71313	26.049\\
3.71313	26.13\\
3.71313	26.211\\
3.73134	26.292\\
3.73134	26.373\\
3.73134	26.453\\
3.77465	26.533\\
3.77465	26.613\\
3.92044	26.693\\
3.92044	26.772\\
3.92044	26.851\\
3.94506	26.93\\
3.94506	27.009\\
3.94506	27.087\\
3.96997	27.165\\
3.96997	27.243\\
3.96997	27.321\\
3.98878	27.399\\
3.98878	27.476\\
3.98878	27.553\\
4.00773	27.63\\
4.00773	27.707\\
4.00773	27.783\\
4.02652	27.859\\
4.02652	27.935\\
4.02652	28.011\\
4.07726	28.087\\
4.07726	28.162\\
4.20867	28.237\\
4.22168	28.312\\
4.22168	28.387\\
4.22168	28.462\\
4.24679	28.536\\
4.24679	28.61\\
4.24679	28.684\\
4.2656	28.758\\
4.2656	28.832\\
4.2656	28.905\\
4.28459	28.978\\
4.28459	29.051\\
4.28459	29.124\\
4.30354	29.197\\
4.30354	29.27\\
4.30354	29.342\\
4.32828	29.414\\
4.32828	29.486\\
4.32828	29.558\\
4.4924	29.63\\
4.4924	29.701\\
4.4924	29.772\\
4.51748	29.843\\
4.51748	29.914\\
4.51748	29.985\\
4.54262	30.056\\
4.54262	30.126\\
4.54262	30.196\\
4.5614	30.266\\
4.5614	30.336\\
4.5614	30.406\\
4.58055	30.476\\
4.58055	30.545\\
4.58055	30.614\\
4.59897	30.683\\
4.59897	30.752\\
4.59897	30.821\\
4.63106	30.89\\
4.63106	30.959\\
4.7874	31.027\\
4.7874	31.095\\
4.7874	31.163\\
4.80676	31.231\\
4.80676	31.299\\
4.80676	31.367\\
4.83207	31.434\\
4.83207	31.501\\
4.83207	31.568\\
4.85681	31.635\\
4.85681	31.702\\
4.85681	31.769\\
4.88822	31.836\\
4.88822	31.902\\
4.88822	31.968\\
4.907	32.034\\
4.907	32.1\\
4.907	32.166\\
4.93768	32.232\\
4.93768	32.298\\
4.96318	32.363\\
4.96318	32.428\\
4.96318	32.493\\
5.12057	32.558\\
5.12057	32.623\\
5.12057	32.688\\
5.14545	32.753\\
5.14545	32.818\\
5.14545	32.882\\
5.1706	32.946\\
5.1706	33.01\\
5.1706	33.074\\
5.19571	33.138\\
5.19571	33.202\\
5.19571	33.266\\
5.21526	33.33\\
5.21526	33.393\\
5.21526	33.456\\
5.23367	33.519\\
5.23367	33.582\\
5.23367	33.645\\
5.25938	33.708\\
5.25938	33.771\\
5.25938	33.834\\
5.29023	33.897\\
5.32107	33.959\\
5.32107	34.021\\
5.45441	34.083\\
5.45441	34.145\\
5.45441	34.207\\
5.47925	34.269\\
5.47925	34.331\\
5.47925	34.393\\
5.51066	34.454\\
5.51066	34.515\\
5.51066	34.576\\
5.52968	34.637\\
5.52968	34.698\\
5.52968	34.759\\
5.54846	34.82\\
5.54846	34.881\\
5.54846	34.942\\
5.56721	35.003\\
5.56721	35.063\\
5.56721	35.123\\
5.5983	35.183\\
5.5983	35.243\\
5.64254	35.303\\
5.64254	35.363\\
5.64254	35.423\\
5.69201	35.483\\
5.69201	35.543\\
5.81202	35.602\\
5.81202	35.661\\
5.81202	35.72\\
5.83101	35.779\\
5.83101	35.838\\
5.83101	35.897\\
5.85669	35.956\\
5.85669	36.015\\
5.85669	36.074\\
5.8753	36.133\\
5.8753	36.191\\
5.8753	36.249\\
5.89428	36.307\\
5.89428	36.365\\
5.89428	36.423\\
5.91304	36.481\\
5.91304	36.539\\
5.91304	36.597\\
5.94372	36.655\\
5.96903	36.713\\
5.96903	36.771\\
6.00035	36.828\\
6.03916	36.885\\
6.03916	36.942\\
6.06955	36.999\\
6.16416	18.499\\
6.43187	18.614\\
6.43187	18.728\\
6.43187	18.841\\
6.45025	18.954\\
6.45025	19.066\\
6.45025	19.177\\
6.49342	19.288\\
6.49342	19.398\\
6.49342	19.507\\
6.5304	19.616\\
6.5304	19.724\\
6.59819	19.832\\
6.59819	19.939\\
6.59819	20.045\\
6.61694	20.151\\
6.61694	20.256\\
6.61694	20.361\\
6.64814	20.465\\
6.64814	20.569\\
6.67821	20.672\\
6.67821	20.775\\
6.67821	20.877\\
6.70925	20.979\\
6.70925	21.08\\
6.77256	21.181\\
6.77256	21.281\\
6.77256	21.381\\
6.80993	21.48\\
6.80993	21.579\\
6.80993	21.677\\
6.84206	21.775\\
6.84206	21.872\\
6.84206	21.969\\
6.86055	22.066\\
6.86055	22.162\\
6.86055	22.258\\
6.91672	22.353\\
6.91672	22.448\\
6.96712	22.542\\
6.96712	22.636\\
6.96712	22.73\\
7.00431	22.823\\
7.00431	22.916\\
7.03632	23.008\\
7.03632	23.1\\
7.03632	23.192\\
7.0614	23.283\\
7.0614	23.374\\
7.0614	23.465\\
7.09229	23.555\\
7.09229	23.645\\
7.13118	23.735\\
7.13118	23.824\\
7.1613	23.913\\
7.1613	24.002\\
7.19927	24.09\\
7.22513	24.178\\
7.22513	24.266\\
7.24466	24.353\\
7.24466	24.44\\
7.24466	24.527\\
7.26906	24.613\\
7.26906	24.699\\
7.26906	24.785\\
7.29991	24.871\\
7.29991	24.956\\
7.32557	25.041\\
7.35684	25.126\\
7.35684	25.21\\
7.35684	25.294\\
7.38825	25.378\\
7.43832	25.461\\
7.43832	25.544\\
7.46343	25.627\\
7.46343	25.71\\
7.47664	25.792\\
7.47664	25.874\\
7.47664	25.956\\
7.50214	26.038\\
7.50214	26.119\\
7.50214	26.2\\
7.53302	26.281\\
7.53302	26.362\\
7.53302	26.442\\
7.57058	26.522\\
7.57058	26.602\\
7.6139	26.682\\
7.6407	26.761\\
7.69001	26.84\\
7.69001	26.919\\
7.69001	26.998\\
7.70956	27.076\\
7.70956	27.154\\
7.70956	27.232\\
7.7404	27.31\\
7.7404	27.388\\
7.7404	27.465\\
7.76574	27.542\\
7.76574	27.619\\
7.76574	27.696\\
7.79062	27.772\\
7.79062	27.848\\
7.79062	27.924\\
7.82262	28\\
7.866	28.076\\
7.90395	28.151\\
7.94224	28.226\\
7.94224	28.301\\
7.94224	28.376\\
7.96663	28.451\\
7.96663	28.525\\
7.96663	28.599\\
7.99191	28.673\\
7.99191	28.747\\
7.99191	28.821\\
8.0169	28.894\\
8.0169	28.967\\
8.0169	29.04\\
8.0485	29.113\\
8.0485	29.186\\
8.0485	29.259\\
8.0736	29.331\\
8.0736	29.403\\
8.0736	29.475\\
8.1232	29.547\\
8.1232	29.619\\
8.1671	29.69\\
8.1671	29.761\\
8.2369	29.832\\
8.2369	29.903\\
8.2369	29.974\\
8.262	30.045\\
8.262	30.115\\
8.262	30.185\\
8.2875	30.255\\
8.2875	30.325\\
8.2875	30.395\\
8.3188	30.465\\
8.3188	30.534\\
8.3188	30.603\\
8.3376	30.672\\
8.3376	30.741\\
8.3376	30.81\\
8.3686	30.879\\
8.4007	30.948\\
8.4007	31.016\\
8.4393	31.084\\
8.4393	31.152\\
8.4393	31.22\\
8.4884	31.288\\
8.5388	31.356\\
8.5388	31.423\\
8.5388	31.49\\
8.5578	31.557\\
8.5578	31.624\\
8.5578	31.691\\
8.5831	31.758\\
8.5831	31.825\\
8.5831	31.891\\
8.6148	31.957\\
8.6148	32.023\\
8.6148	32.089\\
8.6454	32.155\\
8.6454	32.221\\
8.6708	32.287\\
8.6708	32.353\\
8.696	32.418\\
8.696	32.483\\
8.7273	32.548\\
8.7273	32.613\\
8.7976	32.678\\
8.7976	32.743\\
8.8277	32.808\\
8.8414	32.872\\
8.8414	32.936\\
8.8414	33\\
8.8598	33.064\\
8.8598	33.128\\
8.8598	33.192\\
8.8908	33.256\\
8.904	33.32\\
8.904	33.383\\
8.904	33.446\\
8.9358	33.509\\
8.9358	33.572\\
8.9358	33.635\\
8.9725	33.698\\
8.9725	33.761\\
9.0041	33.824\\
9.0041	33.887\\
9.0041	33.949\\
9.0413	34.011\\
9.1059	34.073\\
9.1059	34.135\\
9.1358	34.197\\
9.1358	34.259\\
9.1496	34.321\\
9.1496	34.383\\
9.1496	34.444\\
9.1678	34.505\\
9.1678	34.566\\
9.1678	34.627\\
9.1989	34.688\\
9.2253	34.749\\
9.2253	34.81\\
9.2253	34.871\\
9.2432	34.932\\
9.2432	34.993\\
9.2432	35.053\\
9.2803	35.113\\
9.2803	35.173\\
9.3062	35.233\\
9.3062	35.293\\
9.3062	35.353\\
9.3689	35.413\\
9.4263	35.473\\
9.4263	35.533\\
9.4263	35.592\\
9.4575	35.651\\
9.4575	35.71\\
9.4575	35.769\\
9.4764	35.828\\
9.4764	35.887\\
9.4764	35.946\\
9.5071	36.005\\
9.5324	36.064\\
9.5324	36.123\\
9.5514	36.182\\
9.5514	36.24\\
9.5514	36.298\\
9.5763	36.356\\
9.5763	36.414\\
9.5763	36.472\\
9.6089	36.53\\
9.6089	36.588\\
9.6268	36.646\\
9.6268	36.704\\
9.6268	36.762\\
9.6963	36.819\\
9.6963	36.876\\
9.7395	36.933\\
9.7782	36.99\\
9.7782	37.047\\
9.7782	37.104\\
9.797	37.161\\
9.797	37.218\\
9.797	37.275\\
9.8279	37.332\\
9.853	37.389\\
9.853	37.446\\
9.8727	37.502\\
9.8727	37.558\\
9.8727	37.614\\
9.8974	37.67\\
9.8974	37.726\\
9.8974	37.782\\
9.9164	37.838\\
9.9164	37.894\\
9.9164	37.95\\
9.954	38.006\\
9.954	38.062\\
9.9787	38.118\\
9.9787	38.173\\
9.9787	38.228\\
10.1427	19.114\\
10.4816	19.225\\
10.5125	19.335\\
10.5125	19.445\\
10.5125	19.554\\
10.5434	19.663\\
10.5434	19.771\\
10.5933	19.878\\
10.5933	19.985\\
10.5933	20.091\\
10.6123	20.197\\
10.6123	20.302\\
10.6123	20.406\\
10.7436	20.51\\
10.7436	20.613\\
10.7877	20.716\\
10.7877	20.818\\
10.7877	20.92\\
10.8184	21.021\\
10.8184	21.122\\
10.8563	21.222\\
10.8563	21.322\\
10.8563	21.421\\
10.8883	21.52\\
10.8883	21.619\\
10.8883	21.717\\
11.0259	21.815\\
11.0259	21.912\\
11.0259	22.009\\
11.0762	22.105\\
11.0762	22.201\\
11.0762	22.297\\
11.1073	22.392\\
11.1073	22.487\\
11.1398	22.581\\
11.1398	22.675\\
11.1701	22.769\\
11.1701	22.862\\
11.1833	22.955\\
11.1833	23.047\\
11.1833	23.139\\
11.2267	23.231\\
11.3341	23.322\\
11.3341	23.413\\
11.3785	23.504\\
11.3785	23.594\\
11.3785	23.684\\
11.4094	23.774\\
11.4094	23.863\\
11.4406	23.952\\
11.4406	24.04\\
11.4735	24.128\\
11.4735	24.216\\
11.4735	24.304\\
11.4976	24.391\\
11.4976	24.478\\
11.4976	24.565\\
11.5349	24.651\\
11.5668	24.737\\
11.5668	24.823\\
11.6988	24.908\\
11.6988	24.993\\
11.6988	25.078\\
11.7299	25.162\\
11.7299	25.246\\
11.7679	25.33\\
11.7679	25.414\\
11.7872	25.497\\
11.7872	25.58\\
11.7872	25.663\\
11.8062	25.746\\
11.8062	25.828\\
11.8062	25.91\\
11.8689	25.992\\
11.8817	26.074\\
11.8817	26.155\\
11.8817	26.236\\
11.9126	26.317\\
11.9377	26.397\\
12.0391	26.477\\
12.0391	26.557\\
12.0391	26.637\\
12.1014	26.717\\
12.1014	26.796\\
12.1148	26.875\\
12.1148	26.954\\
12.1148	27.033\\
12.1394	27.111\\
12.1394	27.189\\
12.1394	27.267\\
12.2026	27.345\\
12.2026	27.422\\
12.2026	27.499\\
12.2276	27.576\\
12.2276	27.653\\
12.2276	27.73\\
12.2724	27.806\\
12.2724	27.882\\
12.3907	27.958\\
12.3907	28.034\\
12.3907	28.11\\
12.4342	28.185\\
12.4342	28.26\\
12.4609	28.335\\
12.4609	28.41\\
12.4609	28.484\\
12.4794	28.558\\
12.4794	28.632\\
12.4794	28.706\\
12.5228	28.78\\
12.5493	28.854\\
12.5493	28.927\\
12.5493	29\\
12.5805	29.073\\
12.5805	29.146\\
12.6111	29.219\\
12.6362	29.291\\
12.7431	29.363\\
12.7431	29.435\\
12.8061	14.717\\
13.0351	15.006\\
13.0351	15.148\\
13.0351	15.288\\
13.0909	15.427\\
13.0909	15.565\\
13.0909	15.701\\
13.1275	15.836\\
13.1275	15.97\\
13.1275	16.103\\
13.1828	16.235\\
13.1828	16.366\\
13.2447	16.496\\
13.2447	16.625\\
13.2706	16.753\\
13.2706	16.88\\
13.2706	17.006\\
13.3078	17.131\\
13.3078	17.255\\
13.3334	17.378\\
13.3334	17.5\\
13.3772	17.621\\
13.3772	17.741\\
13.3772	17.861\\
13.427	17.98\\
13.4715	18.098\\
13.4715	18.215\\
13.4715	18.332\\
13.5098	18.448\\
13.5098	18.563\\
13.5098	18.677\\
13.5468	18.791\\
13.5468	18.904\\
13.5721	19.016\\
13.6093	19.128\\
13.6093	19.239\\
13.6538	19.349\\
13.6676	19.459\\
13.6676	19.568\\
13.6676	19.676\\
13.7104	19.784\\
13.7104	19.891\\
13.7355	19.998\\
13.7611	20.104\\
13.7611	20.21\\
13.7611	20.315\\
13.7984	20.419\\
13.8296	20.523\\
13.8296	20.626\\
13.8614	20.729\\
13.8614	20.831\\
13.9051	20.933\\
13.9051	21.034\\
13.937	21.135\\
13.937	21.235\\
13.963	21.335\\
13.9875	21.434\\
13.9875	21.533\\
13.9875	21.631\\
14.0182	21.729\\
14.0182	21.827\\
14.0563	21.924\\
14.0563	22.021\\
14.0936	22.117\\
14.0936	22.213\\
14.1446	22.308\\
14.1446	22.403\\
14.1758	22.498\\
14.1758	22.592\\
14.2076	22.686\\
14.2264	22.779\\
14.2264	22.872\\
14.2264	22.965\\
14.2575	23.057\\
14.2575	23.149\\
14.2895	23.241\\
14.2895	23.332\\
14.2895	23.423\\
14.3391	23.514\\
14.3391	23.604\\
14.3706	23.694\\
14.3706	23.783\\
14.4212	23.872\\
14.4212	23.961\\
14.4536	24.049\\
14.4536	24.137\\
14.4904	24.225\\
14.4904	24.312\\
14.5109	24.399\\
14.5109	24.486\\
14.5109	24.573\\
14.5475	24.659\\
14.5475	24.745\\
14.5475	24.831\\
14.5787	24.916\\
14.5787	25.001\\
14.63	25.086\\
14.63	25.17\\
14.6804	25.254\\
14.6804	25.338\\
14.7297	25.422\\
14.7297	25.505\\
14.7552	25.588\\
14.7552	25.671\\
14.768	25.754\\
14.768	25.836\\
14.768	25.918\\
14.7994	26\\
14.7994	26.081\\
14.8364	26.162\\
14.8364	26.243\\
14.8617	26.324\\
14.8617	26.404\\
14.9062	26.484\\
14.9062	26.564\\
14.9062	26.644\\
14.9495	26.724\\
14.9815	26.803\\
14.9815	26.882\\
15.0253	26.961\\
15.0253	27.04\\
15.0381	27.118\\
15.0381	27.196\\
15.0381	27.274\\
15.0822	27.352\\
15.0822	27.429\\
15.12	27.506\\
15.12	27.583\\
15.1328	27.66\\
15.1328	27.737\\
15.1328	27.813\\
15.1891	27.889\\
15.1891	27.965\\
15.1891	28.041\\
15.22	28.117\\
15.22	28.192\\
15.271	28.267\\
15.271	28.342\\
15.3155	28.417\\
15.3155	28.492\\
15.3155	28.566\\
15.3404	28.64\\
15.3404	28.714\\
15.3404	28.788\\
15.4025	28.861\\
15.4025	28.934\\
15.4221	29.007\\
15.4221	29.08\\
15.4221	29.153\\
15.4529	29.226\\
15.497	29.298\\
15.497	29.37\\
15.497	29.442\\
15.5281	29.514\\
15.5281	29.586\\
15.5731	29.658\\
15.5731	29.729\\
15.5731	29.8\\
15.6105	29.871\\
15.6364	29.942\\
15.6364	30.013\\
15.6364	30.084\\
15.6611	30.154\\
15.6611	30.224\\
15.6611	30.294\\
15.7184	30.364\\
15.7184	30.434\\
15.7184	30.504\\
15.7688	30.573\\
15.7688	30.642\\
15.7988	30.711\\
15.7988	30.78\\
15.8187	30.849\\
15.8187	30.918\\
15.8187	30.986\\
15.8749	31.054\\
15.8749	31.122\\
15.8749	31.19\\
15.9063	31.258\\
15.9063	31.326\\
15.9498	31.394\\
15.9498	31.461\\
15.9753	31.528\\
15.9753	31.595\\
15.9753	31.662\\
16.0064	31.729\\
16.0064	31.796\\
16.0516	31.863\\
16.089	31.929\\
16.089	31.995\\
16.089	32.061\\
16.1206	32.127\\
16.1206	32.193\\
16.145	32.259\\
16.145	32.325\\
16.145	32.39\\
16.2078	32.455\\
16.2078	32.52\\
16.2078	32.585\\
16.2389	32.65\\
16.2389	32.715\\
16.2717	32.78\\
16.2717	32.845\\
16.3024	32.909\\
16.3024	32.973\\
16.3024	33.037\\
16.3336	33.101\\
16.3336	33.165\\
16.3599	33.229\\
16.3599	33.293\\
16.4155	33.357\\
16.4155	33.42\\
16.4534	33.483\\
16.4534	33.546\\
16.4534	33.609\\
16.4726	33.672\\
16.4726	33.735\\
16.4726	33.798\\
16.5033	33.861\\
16.5033	33.923\\
16.5548	33.985\\
16.5548	34.047\\
16.5548	34.109\\
16.6114	34.171\\
16.6114	34.233\\
16.6114	34.295\\
16.6426	34.357\\
16.6426	34.419\\
16.6426	34.48\\
16.6737	34.541\\
16.7	34.602\\
16.7	34.663\\
16.7301	34.724\\
16.7868	17.362\\
17.1451	17.484\\
17.1451	17.605\\
17.1769	17.726\\
17.1769	17.846\\
17.1769	17.965\\
17.1951	18.083\\
17.1951	18.2\\
17.1951	18.317\\
17.2257	18.433\\
17.2257	18.548\\
17.2257	18.662\\
17.3058	18.776\\
17.3058	18.889\\
17.3058	19.001\\
17.3549	19.113\\
17.3549	19.224\\
17.3683	19.334\\
17.3683	19.444\\
17.3683	19.553\\
17.3871	19.662\\
17.3871	19.77\\
17.3871	19.877\\
17.4181	19.984\\
17.4875	20.09\\
17.4875	20.196\\
17.4875	20.301\\
17.5184	20.405\\
17.5184	20.509\\
17.5435	20.612\\
17.5573	20.715\\
17.5573	20.817\\
17.5573	20.919\\
17.5877	21.02\\
17.5877	21.121\\
17.5877	21.221\\
17.6133	21.321\\
17.6133	21.42\\
17.6133	21.519\\
17.669	21.618\\
17.7137	21.716\\
17.7137	21.814\\
17.7137	21.911\\
17.7392	22.008\\
17.7392	22.104\\
17.7392	22.2\\
17.7771	22.296\\
17.7771	22.391\\
17.7771	22.486\\
17.808	22.58\\
17.808	22.674\\
17.808	22.768\\
17.8452	22.861\\
17.8902	22.954\\
17.8902	23.046\\
17.9213	23.138\\
17.947	23.23\\
17.947	23.321\\
17.947	23.412\\
17.9713	23.502\\
17.9713	23.592\\
17.9713	23.682\\
};

\end{axis}

\end{tikzpicture}%

%% file: figures/newreno-rtt-2.tex
%
%
\definecolor{mycolor1}{rgb}{0.00000,0.44700,0.74100}%
\definecolor{mycolor2}{rgb}{0.85000,0.32500,0.09800}%
\begin{tikzpicture}
\pgfplotsset{every tick label/.append style={font=\scriptsize}}

\begin{axis}[%
width=0.951\fwidth,
height=\fheight,
at={(0\fwidth,0\fheight)},
scale only axis,
xmajorgrids,
ymajorgrids,
xmin=0,
xmax=18,
xlabel style={font=\scriptsize\color{white!15!black}},
xlabel={Time (s)},
ymin=0.1,
ymax=0.6,
ylabel style={font=\scriptsize\color{white!15!black}},
ylabel={Measured RTT (s)},
axis background/.style={fill=white},
title style={font=\bfseries},
legend style={legend cell align=left, font=\scriptsize,  align=left, draw=white!15!black}
]
\addplot [color=mycolor1,dashed,line width=1.2pt]
  table[row sep=crcr]{%
0.23764	0.12045\\
0.25326	0.126069\\
0.36265	0.115013\\
0.38443	0.120988\\
0.38687	0.108614\\
0.47651	0.113859\\
0.49987	0.127219\\
0.50275	0.120097\\
0.52112	0.134249\\
0.54322	0.126669\\
0.60978	0.118289\\
0.62066	0.120791\\
0.64279	0.140039\\
0.64567	0.132917\\
0.66439	0.143269\\
0.69264	0.151513\\
0.6952	0.151983\\
0.7262	0.154282\\
0.76404	0.143377\\
0.78618	0.155521\\
0.79521	0.152421\\
0.81361	0.157947\\
0.83245	0.168055\\
0.86088	0.176487\\
0.86381	0.171175\\
0.8856	0.190402\\
0.88834	0.183137\\
0.91065	0.195453\\
0.92621	0.200014\\
0.95741	0.211219\\
0.98551	0.221469\\
1.00743	0.21222\\
1.03193	0.226716\\
1.06037	0.246752\\
1.06298	0.249368\\
1.08158	0.249132\\
1.10054	0.248092\\
1.12266	0.258848\\
1.12539	0.261584\\
1.15028	0.264679\\
1.17837	0.290029\\
1.18114	0.282805\\
1.20348	0.295146\\
1.21845	0.297609\\
1.24967	0.313461\\
1.2655	0.319295\\
1.28086	0.323445\\
1.30899	0.341575\\
1.31822	0.332705\\
1.3589	0.351465\\
1.36164	0.354213\\
1.38712	0.355189\\
1.40274	0.370813\\
1.41794	0.376011\\
1.52143	0.376036\\
1.55887	0.380499\\
1.5957	0.382219\\
1.6766	0.401091\\
1.69535	0.386356\\
1.7421	0.380455\\
1.77344	0.386324\\
1.80457	0.391643\\
1.86018	0.397937\\
1.88412	0.397591\\
1.91403	0.379936\\
1.95128	0.382409\\
2.01074	0.416909\\
2.07368	0.464737\\
2.10087	0.466\\
2.132	0.476038\\
2.16234	0.472909\\
2.21664	0.498274\\
2.28284	0.545631\\
2.31305	0.550953\\
2.3433	0.543745\\
2.40136	0.558141\\
2.46265	0.528625\\
2.47765	0.516377\\
2.5411	0.430231\\
2.56601	0.414007\\
2.60326	0.420915\\
2.62099	0.384347\\
2.62411	0.331267\\
2.66125	0.328199\\
2.69235	0.303163\\
2.73468	0.310488\\
2.80156	0.341157\\
2.83132	0.333667\\
2.8689	0.337451\\
2.90534	0.319336\\
2.93598	0.322724\\
2.97253	0.301283\\
2.99741	0.285064\\
3.02556	0.28088\\
3.04056	0.28588\\
3.05107	0.23951\\
3.06605	0.224731\\
3.08807	0.236749\\
3.10322	0.224316\\
3.12493	0.236029\\
3.13408	0.218732\\
3.14418	0.21884\\
3.15918	0.2132\\
3.1845	0.211961\\
3.1982	0.200783\\
3.22648	0.204041\\
3.2391	0.203544\\
3.25159	0.201033\\
3.25428	0.192655\\
3.2767	0.20065\\
3.29167	0.205617\\
3.30218	0.200096\\
3.31718	0.200523\\
3.35243	0.198245\\
3.37158	0.192396\\
3.37411	0.189611\\
3.39289	0.194693\\
3.4214	0.203207\\
3.43409	0.195505\\
3.44348	0.189199\\
3.47798	0.203936\\
3.4906	0.203903\\
3.50309	0.201426\\
3.51572	0.20298\\
3.54699	0.207621\\
3.56196	0.212588\\
3.57536	0.201255\\
3.60709	0.214203\\
3.63749	0.224597\\
3.65665	0.213173\\
3.68849	0.200513\\
3.71088	0.207932\\
3.72334	0.210243\\
3.76703	0.228972\\
3.78924	0.213875\\
3.81132	0.215841\\
3.82035	0.213253\\
3.8424	0.225304\\
3.85211	0.204625\\
3.88395	0.207009\\
3.91536	0.2071\\
3.93707	0.216196\\
3.95243	0.21582\\
3.97464	0.228037\\
4.00901	0.219769\\
4.04041	0.220068\\
4.07189	0.219773\\
4.09427	0.222491\\
4.10333	0.219373\\
4.13438	0.230431\\
4.15062	0.235263\\
4.15371	0.219257\\
4.18771	0.24008\\
4.19705	0.225001\\
4.23482	0.23518\\
4.24513	0.236125\\
4.25485	0.225846\\
4.27657	0.236157\\
4.28926	0.229251\\
4.3116	0.219948\\
4.33948	0.236155\\
4.34217	0.228848\\
4.396	0.242286\\
4.42044	0.236609\\
4.46474	0.24314\\
4.48672	0.231866\\
4.50281	0.237957\\
4.51777	0.242923\\
4.52812	0.240989\\
4.54309	0.241516\\
4.56225	0.230657\\
4.59096	0.238785\\
4.59389	0.231712\\
4.62223	0.239001\\
4.62529	0.232063\\
4.64731	0.238768\\
4.66231	0.243768\\
4.69375	0.250096\\
4.72643	0.239711\\
4.74778	0.251067\\
4.76044	0.247632\\
4.7731	0.245321\\
4.78806	0.249943\\
4.79837	0.245288\\
4.8074	0.245144\\
4.83916	0.245279\\
4.85476	0.250875\\
4.86973	0.255841\\
4.8862	0.250903\\
4.88909	0.243796\\
4.91127	0.253404\\
4.92627	0.253964\\
4.93658	0.254272\\
4.98282	0.259436\\
5.00599	0.248206\\
5.00871	0.235959\\
5.0246	0.251504\\
5.03706	0.25396\\
5.05923	0.251831\\
5.08112	0.2536\\
5.09067	0.251504\\
5.10626	0.2571\\
5.12123	0.256471\\
5.13444	0.245348\\
5.17567	0.253837\\
5.18809	0.25182\\
5.20328	0.256135\\
5.23162	0.269691\\
5.24659	0.250452\\
5.26976	0.261049\\
5.27245	0.253744\\
5.2949	0.260304\\
5.32631	0.267079\\
5.329	0.259773\\
5.35758	0.266908\\
5.36027	0.259601\\
5.39208	0.257639\\
5.41389	0.279453\\
5.41679	0.272348\\
5.43904	0.279377\\
5.45186	0.263876\\
5.4761	0.278004\\
5.48637	0.273088\\
5.50188	0.278601\\
5.51688	0.275255\\
5.54597	0.273521\\
5.54887	0.266416\\
5.56499	0.272539\\
5.57765	0.272744\\
5.60507	0.276069\\
5.61558	0.266983\\
5.6246	0.264332\\
5.64702	0.26702\\
5.65587	0.263797\\
5.68748	0.270691\\
5.70972	0.272939\\
5.72847	0.279439\\
5.74347	0.279472\\
5.75905	0.285044\\
5.76877	0.282671\\
5.78483	0.288461\\
5.79748	0.285044\\
5.83126	0.294384\\
5.84161	0.282739\\
5.85443	0.279439\\
5.86692	0.276961\\
5.87938	0.279417\\
5.90155	0.276947\\
5.93299	0.277115\\
5.95508	0.279443\\
5.9646	0.277119\\
5.9981	0.290626\\
6.02412	0.285081\\
6.03674	0.283264\\
6.05209	0.27332\\
6.07451	0.27912\\
6.08969	0.269863\\
6.13108	0.279476\\
6.13398	0.267403\\
6.15636	0.279443\\
6.17206	0.280171\\
6.19939	0.297843\\
6.21574	0.282748\\
6.24714	0.282548\\
6.27539	0.290968\\
6.30054	0.289417\\
6.30323	0.282107\\
6.32564	0.289217\\
6.34061	0.293872\\
6.35072	0.298633\\
6.36662	0.294933\\
6.38829	0.298601\\
6.39102	0.291328\\
6.43473	0.300753\\
6.45136	0.297623\\
6.46636	0.3\\
6.47651	0.294451\\
6.49147	0.299417\\
6.50451	0.288777\\
6.53625	0.289104\\
6.56467	0.297523\\
6.57736	0.289821\\
6.58654	0.283316\\
6.62121	0.297667\\
6.63621	0.300571\\
6.65249	0.301877\\
6.66745	0.306172\\
6.6776	0.300983\\
6.68699	0.295972\\
6.72166	0.310645\\
6.75926	0.307897\\
6.76232	0.300959\\
6.77844	0.302081\\
6.79359	0.292119\\
6.81578	0.311265\\
6.81867	0.304159\\
6.84725	0.311\\
6.8506	0.304356\\
6.87256	0.316312\\
6.88753	0.310516\\
6.90377	0.317225\\
6.90649	0.309948\\
6.92907	0.317312\\
6.94407	0.31286\\
6.96071	0.310156\\
6.9732	0.310716\\
6.9759	0.298443\\
7.01328	0.306172\\
7.02379	0.306679\\
7.03894	0.304931\\
7.07041	0.298088\\
7.10182	0.298224\\
7.13036	0.311689\\
7.13309	0.304413\\
7.17027	0.319667\\
7.18082	0.310912\\
7.19578	0.313223\\
7.20589	0.308368\\
7.21475	0.30826\\
7.23061	0.324117\\
7.23986	0.313368\\
7.26224	0.322608\\
7.27724	0.323168\\
7.29368	0.322408\\
7.30617	0.320275\\
7.32465	0.328757\\
7.33379	0.310512\\
7.35656	0.322209\\
7.37156	0.322621\\
7.39667	0.326261\\
7.40698	0.316811\\
7.41653	0.314715\\
7.43842	0.31701\\
7.44797	0.314887\\
7.46989	0.317044\\
7.48251	0.319667\\
7.50997	0.319155\\
7.5295	0.314755\\
7.56074	0.320884\\
7.58909	0.329229\\
7.59181	0.319011\\
7.62056	0.3297\\
7.63302	0.329333\\
7.64564	0.326956\\
7.6863	0.314739\\
7.70851	0.326956\\
7.7212	0.329645\\
7.73386	0.327189\\
7.74268	0.326145\\
7.7649	0.338364\\
7.77461	0.32664\\
7.7967	0.329463\\
7.81185	0.331957\\
7.82777	0.332913\\
7.84043	0.335569\\
7.8626	0.333095\\
7.8942	0.333463\\
7.91593	0.335064\\
7.92528	0.333463\\
7.95373	0.335377\\
7.95652	0.325956\\
7.97881	0.33328\\
7.99399	0.337787\\
8.0314	0.335073\\
8.0544	0.335297\\
8.067	0.333513\\
8.082	0.339317\\
8.1073	0.332673\\
8.136	0.341387\\
8.1546	0.347357\\
8.1673	0.342152\\
8.1926	0.339841\\
8.2179	0.335501\\
8.2366	0.342439\\
8.2392	0.334965\\
8.2555	0.341255\\
8.2643	0.339001\\
8.2959	0.339369\\
8.3183	0.342119\\
8.3333	0.344496\\
8.3557	0.351719\\
8.3684	0.354408\\
8.3812	0.339812\\
8.3937	0.342301\\
8.4063	0.339645\\
8.4213	0.344267\\
8.447	0.339667\\
8.4817	0.345657\\
8.4852	0.330741\\
8.5071	0.341959\\
8.5194	0.339842\\
8.5223	0.332393\\
8.5447	0.339845\\
8.5597	0.344812\\
8.57	0.340007\\
8.5787	0.339492\\
8.5948	0.345583\\
8.6097	0.345441\\
8.6198	0.355553\\
8.6349	0.344985\\
8.6512	0.35532\\
8.6885	0.359609\\
8.6976	0.345265\\
8.7202	0.354509\\
8.7329	0.352199\\
8.7521	0.345871\\
8.7547	0.338344\\
8.792	0.345087\\
8.8274	0.355181\\
8.846	0.360804\\
8.8485	0.34396\\
8.8836	0.361292\\
8.8864	0.344352\\
8.9091	0.354385\\
8.924	0.345339\\
8.9465	0.357795\\
8.9778	0.358078\\
9.0297	0.353501\\
9.0612	0.349405\\
9.0986	0.353389\\
9.1424	0.357679\\
9.1728	0.333317\\
9.2096	0.313149\\
9.2463	0.312345\\
9.277	0.301711\\
9.3079	0.305381\\
9.345	0.305345\\
9.3607	0.31104\\
9.3755	0.294289\\
9.4186	0.30004\\
9.449	0.266213\\
9.486	0.266437\\
9.5113	0.254957\\
9.5362	0.239197\\
9.5736	0.245719\\
9.6029	0.232217\\
9.606	0.220496\\
9.6343	0.238779\\
9.643	0.2144\\
9.6594	0.220772\\
9.6687	0.209695\\
9.6791	0.210036\\
9.6919	0.195868\\
9.7107	0.204685\\
9.7201	0.188784\\
9.7423	0.196127\\
9.7548	0.198616\\
9.8048	0.208101\\
9.8177	0.201688\\
9.8326	0.206655\\
9.8618	0.193049\\
9.8744	0.195701\\
9.8894	0.200327\\
9.9086	0.188464\\
9.9312	0.191488\\
9.9839	0.20094\\
10.0067	0.189373\\
10.0193	0.191655\\
10.0444	0.197891\\
10.0533	0.191516\\
10.0755	0.201429\\
10.0882	0.203744\\
10.1008	0.201433\\
10.1098	0.201256\\
10.1323	0.203944\\
10.1447	0.20351\\
10.1696	0.21256\\
10.1846	0.217527\\
10.2139	0.210569\\
10.2265	0.209835\\
10.2545	0.201252\\
10.277	0.20394\\
10.2897	0.20414\\
10.3113	0.201484\\
10.3397	0.209868\\
10.3429	0.200632\\
10.3772	0.219097\\
10.3863	0.206636\\
10.434	0.2234\\
10.4367	0.212732\\
10.4589	0.219988\\
10.4742	0.225333\\
10.4968	0.222703\\
10.5091	0.222135\\
10.512	0.210053\\
10.5471	0.215608\\
10.5658	0.222947\\
10.5973	0.224425\\
10.6	0.213744\\
10.628	0.221464\\
10.6374	0.220807\\
10.654	0.217375\\
10.6665	0.219864\\
10.6851	0.228415\\
10.6978	0.228289\\
10.7231	0.228848\\
10.726	0.218521\\
10.748	0.225964\\
10.7629	0.230931\\
10.7798	0.22038\\
10.795	0.210449\\
10.8355	0.235463\\
10.8678	0.230377\\
10.8801	0.232666\\
10.8927	0.228064\\
10.9176	0.23768\\
10.9326	0.236936\\
10.9525	0.226528\\
10.9742	0.238219\\
10.9837	0.225744\\
11.0058	0.232832\\
11.037	0.242055\\
11.0562	0.234896\\
11.0714	0.240073\\
11.1024	0.24396\\
11.1191	0.239198\\
11.1315	0.241479\\
11.1632	0.235584\\
11.1757	0.233073\\
11.191	0.238496\\
11.2229	0.239228\\
11.2383	0.24462\\
11.251	0.247243\\
11.2845	0.253772\\
11.3012	0.245333\\
11.3162	0.249396\\
11.3326	0.251243\\
11.3476	0.256243\\
11.3702	0.254465\\
11.3852	0.256075\\
11.3955	0.25142\\
11.408	0.253909\\
11.4206	0.245092\\
11.4356	0.249947\\
11.4457	0.254746\\
11.455	0.243992\\
11.4835	0.250572\\
11.4866	0.238236\\
11.5215	0.248148\\
11.5466	0.248533\\
11.559	0.247799\\
11.5842	0.254665\\
11.5968	0.254168\\
11.6156	0.254632\\
11.6306	0.250347\\
11.647	0.251853\\
11.6597	0.253605\\
11.6721	0.251619\\
11.6871	0.255915\\
11.7036	0.257977\\
11.7124	0.257436\\
11.7381	0.251508\\
11.7789	0.262155\\
11.8042	0.257636\\
11.8071	0.250492\\
11.842	0.26116\\
11.857	0.262803\\
11.8671	0.257948\\
11.8824	0.256265\\
11.9049	0.260492\\
11.93	0.257948\\
11.9424	0.259701\\
11.9706	0.258208\\
12.0015	0.2634\\
12.0304	0.262189\\
12.0555	0.264276\\
12.0646	0.257472\\
12.1085	0.271801\\
12.1246	0.271993\\
12.1372	0.270204\\
12.171	0.268931\\
12.1874	0.272552\\
12.2001	0.270237\\
12.2127	0.272748\\
12.2285	0.27258\\
12.2532	0.262976\\
12.2817	0.280223\\
12.2844	0.272944\\
12.3005	0.279007\\
12.3348	0.270288\\
12.3574	0.272552\\
12.3724	0.277552\\
12.395	0.273144\\
12.41	0.275455\\
12.4203	0.270801\\
12.436	0.276522\\
12.4579	0.272389\\
12.4706	0.273147\\
12.4957	0.272948\\
12.5208	0.278684\\
12.5298	0.276592\\
12.5522	0.278688\\
12.5618	0.277356\\
12.5773	0.282886\\
12.5923	0.281231\\
12.6212	0.286355\\
12.6245	0.279608\\
12.64	0.285172\\
12.6526	0.284612\\
12.6778	0.286105\\
12.6928	0.287715\\
12.7155	0.284608\\
12.7407	0.286301\\
12.7533	0.285367\\
12.7658	0.282889\\
12.7809	0.288033\\
12.797	0.289021\\
12.812	0.293988\\
12.8223	0.292515\\
12.8348	0.295004\\
12.8538	0.291984\\
12.8566	0.28484\\
12.8788	0.290941\\
12.8938	0.291533\\
12.9042	0.291841\\
12.9132	0.278744\\
12.9417	0.291699\\
12.9567	0.291732\\
12.9671	0.292077\\
12.9824	0.294588\\
13.0046	0.30184\\
13.0169	0.304096\\
13.0198	0.291905\\
13.0422	0.304329\\
13.0548	0.304184\\
13.0673	0.30164\\
13.0823	0.306495\\
13.0989	0.303792\\
13.1141	0.307072\\
13.1397	0.292437\\
13.1618	0.305167\\
13.1745	0.298256\\
13.1871	0.297697\\
13.2021	0.298256\\
13.2183	0.305143\\
13.2211	0.297864\\
13.2435	0.304893\\
13.2585	0.306143\\
13.2686	0.301844\\
13.2842	0.307132\\
13.3066	0.304584\\
13.3217	0.291932\\
13.3441	0.304356\\
13.3568	0.304589\\
13.3694	0.302244\\
13.3846	0.306717\\
13.4218	0.312856\\
13.4384	0.300153\\
13.4417	0.301928\\
13.4731	0.300717\\
13.5011	0.314329\\
13.5161	0.318951\\
13.5262	0.314096\\
13.5359	0.304884\\
13.5576	0.316575\\
13.5726	0.319151\\
13.5889	0.320425\\
13.6014	0.322213\\
13.6205	0.317308\\
13.6355	0.318955\\
13.6583	0.316844\\
13.671	0.316252\\
13.6836	0.3145\\
13.6986	0.319189\\
13.7087	0.309923\\
13.7237	0.314923\\
13.7467	0.308227\\
13.7492	0.307528\\
13.7808	0.307696\\
13.803	0.310683\\
13.818	0.306899\\
13.8442	0.308256\\
13.8657	0.319747\\
13.8684	0.312468\\
13.897	0.328737\\
13.9096	0.323168\\
13.9349	0.321024\\
13.9598	0.328708\\
13.9748	0.319856\\
13.9915	0.323172\\
14.0041	0.320861\\
14.0168	0.323201\\
14.0319	0.323313\\
14.0635	0.319842\\
14.0858	0.326579\\
14.1009	0.320148\\
14.1423	0.328737\\
14.155	0.323135\\
14.1638	0.319612\\
14.1895	0.32108\\
14.2178	0.329104\\
14.2328	0.325879\\
14.2493	0.327385\\
14.2617	0.329808\\
14.3058	0.335432\\
14.3183	0.329492\\
14.3436	0.327148\\
14.3742	0.347379\\
14.3875	0.336025\\
14.3965	0.332972\\
14.4285	0.327575\\
14.4504	0.339464\\
14.4657	0.338437\\
14.4972	0.333372\\
14.5286	0.3391\\
14.5658	0.33738\\
14.5825	0.335795\\
14.5949	0.335627\\
14.6233	0.323913\\
14.6515	0.335065\\
14.6642	0.333313\\
14.6894	0.333484\\
14.723	0.357068\\
14.7481	0.351607\\
14.7614	0.332837\\
14.7837	0.335236\\
14.7987	0.338339\\
14.8306	0.333368\\
14.8625	0.3339\\
14.8931	0.344463\\
14.9254	0.332949\\
14.9568	0.333532\\
14.991	0.347521\\
15.0038	0.342352\\
15.0163	0.339841\\
15.0289	0.34246\\
15.079	0.345965\\
15.0916	0.348621\\
15.1167	0.355359\\
15.1194	0.348076\\
15.1541	0.359721\\
15.1914	0.362969\\
15.2235	0.351005\\
15.2361	0.353628\\
15.2705	0.345089\\
15.2992	0.353825\\
15.3021	0.345289\\
15.3366	0.359963\\
15.3618	0.360823\\
15.3744	0.358391\\
15.4306	0.356989\\
15.4434	0.35446\\
15.4738	0.359865\\
15.4872	0.347969\\
15.5018	0.337788\\
15.539	0.331895\\
15.5699	0.336351\\
15.6005	0.33792\\
15.6244	0.292744\\
15.6612	0.289353\\
15.6984	0.299232\\
15.7288	0.304679\\
15.7595	0.293879\\
15.7966	0.274751\\
15.8178	0.285916\\
15.8273	0.268236\\
15.864	0.253551\\
15.8887	0.244315\\
15.932	0.223652\\
15.9635	0.214656\\
15.9957	0.189091\\
16.0297	0.213109\\
16.0431	0.196049\\
16.0557	0.198672\\
16.0808	0.196795\\
16.0998	0.191069\\
16.1317	0.179413\\
16.1407	0.177176\\
16.1694	0.173698\\
16.1719	0.176209\\
16.1878	0.192115\\
16.2006	0.184705\\
16.2258	0.182849\\
16.2384	0.18536\\
16.2511	0.183049\\
16.2764	0.183216\\
16.323	0.182304\\
16.3456	0.185126\\
16.355	0.183073\\
16.3866	0.173804\\
16.4084	0.185668\\
16.4209	0.182649\\
16.4236	0.174335\\
16.446	0.182649\\
16.455	0.178683\\
16.5047	0.205996\\
16.5213	0.198373\\
16.5338	0.200863\\
16.5586	0.203664\\
16.5779	0.193744\\
16.5806	0.194023\\
16.6032	0.197416\\
16.613	0.189369\\
16.6498	0.19476\\
16.6851	0.199616\\
16.71	0.195272\\
16.725	0.19298\\
16.7416	0.195276\\
16.7566	0.200276\\
16.7667	0.195276\\
16.7756	0.19496\\
16.8076	0.194592\\
16.8229	0.209896\\
16.8354	0.212385\\
16.8606	0.210837\\
16.8795	0.209253\\
16.8921	0.211872\\
16.9262	0.205595\\
16.9602	0.22524\\
16.9707	0.218427\\
16.9925	0.216961\\
16.9952	0.209672\\
17.0115	0.215979\\
17.024	0.216425\\
17.0268	0.209271\\
17.0585	0.200635\\
17.0934	0.214251\\
17.1056	0.215503\\
17.1338	0.207616\\
17.1689	0.212272\\
17.2001	0.2166\\
17.203	0.210494\\
17.2314	0.226141\\
17.2342	0.212709\\
17.2558	0.228973\\
17.2652	0.218703\\
17.2879	0.229413\\
17.2908	0.211923\\
17.3405	0.237125\\
17.3572	0.228463\\
17.3822	0.227996\\
17.3947	0.230452\\
17.4229	0.219901\\
17.439	0.225974\\
17.4542	0.219908\\
17.4702	0.225981\\
17.4852	0.230981\\
17.5021	0.222899\\
17.5109	0.220043\\
17.5268	0.225945\\
17.5519	0.234933\\
17.5833	0.226171\\
17.6148	0.232565\\
17.6178	0.224955\\
17.655	0.232124\\
17.6922	0.242596\\
17.7089	0.244725\\
17.7116	0.230707\\
17.7432	0.232357\\
17.7654	0.244558\\
17.7812	0.243771\\
17.7967	0.244892\\
17.8117	0.249747\\
17.822	0.245092\\
17.8598	0.246806\\
17.8887	0.233665\\
17.9252	0.250129\\
17.9357	0.23344\\
17.9453	0.233697\\
17.9765	0.233297\\
17.9975	0.254253\\
};

\addplot [color=mycolor2]
  table[row sep=crcr]{%
0.33764	0.12045\\
0.35345	0.126257\\
0.459	0.111362\\
0.47482	0.12137\\
0.48982	0.12619\\
0.5213	0.142697\\
0.57985	0.110846\\
0.6086	0.108778\\
0.66457	0.154754\\
0.70169	0.160382\\
0.7267	0.136855\\
0.74545	0.136856\\
0.79539	0.157103\\
0.85455	0.179983\\
0.86702	0.182453\\
0.90688	0.195195\\
0.917	0.195311\\
0.92602	0.199317\\
0.95449	0.209036\\
0.95723	0.211773\\
0.98787	0.222728\\
0.99799	0.222845\\
1.02323	0.227841\\
1.03823	0.232841\\
1.08177	0.266377\\
1.12878	0.264228\\
1.14399	0.27697\\
1.16611	0.278689\\
1.18133	0.283906\\
1.19716	0.290279\\
1.20638	0.289382\\
1.22853	0.30251\\
1.23739	0.311366\\
1.25934	0.313779\\
1.26823	0.313747\\
1.28423	0.327005\\
1.29329	0.326061\\
1.32088	0.343657\\
1.33081	0.33294\\
1.34645	0.348459\\
1.36183	0.353843\\
1.38397	0.361016\\
1.39935	0.366116\\
1.41498	0.376752\\
1.42998	0.381752\\
1.48618	0.394409\\
1.54572	0.379604\\
1.57142	0.374257\\
1.63358	0.376195\\
1.68276	0.389471\\
1.76455	0.392723\\
1.91449	0.402484\\
1.9446	0.375709\\
1.97573	0.404309\\
2.02958	0.429317\\
2.09549	0.470355\\
2.14343	0.475269\\
2.18068	0.473984\\
2.28148	0.531869\\
2.31257	0.533024\\
2.3614	0.555743\\
2.42875	0.56474\\
2.50213	0.516405\\
2.52509	0.475511\\
2.55722	0.441727\\
2.58211	0.418684\\
2.59711	0.406432\\
2.62242	0.330936\\
2.65286	0.330291\\
2.65511	0.322539\\
2.69593	0.314535\\
2.70455	0.306513\\
2.74078	0.302028\\
2.80758	0.342699\\
2.83743	0.325295\\
2.86834	0.323603\\
2.89258	0.312842\\
2.92349	0.316074\\
2.94817	0.305754\\
2.99111	0.310557\\
3.01941	0.294855\\
3.02194	0.271154\\
3.04395	0.28317\\
3.05892	0.241337\\
3.06963	0.242049\\
3.08462	0.227189\\
3.10047	0.222131\\
3.11547	0.227131\\
3.14056	0.207072\\
3.16228	0.218787\\
3.17182	0.213647\\
3.18779	0.219615\\
3.21993	0.209969\\
3.23278	0.200848\\
3.24524	0.203304\\
3.25786	0.203911\\
3.27035	0.201433\\
3.28532	0.205144\\
3.29566	0.201044\\
3.30832	0.197301\\
3.32403	0.183471\\
3.34605	0.195483\\
3.36105	0.200483\\
3.39343	0.211613\\
3.42774	0.197816\\
3.4404	0.195505\\
3.45537	0.200127\\
3.4718	0.20145\\
3.47453	0.194176\\
3.49674	0.201425\\
3.51171	0.206051\\
3.52475	0.200717\\
3.54081	0.206771\\
3.5533	0.209261\\
3.57158	0.214988\\
3.59103	0.214912\\
3.60603	0.219912\\
3.64112	0.227685\\
3.65391	0.213852\\
3.66909	0.218692\\
3.69121	0.22584\\
3.70035	0.205515\\
3.72604	0.201292\\
3.75112	0.216372\\
3.77337	0.222021\\
3.78599	0.220187\\
3.80501	0.214089\\
3.81743	0.21586\\
3.85121	0.225188\\
3.86805	0.21446\\
3.88305	0.218604\\
3.89929	0.220199\\
3.91429	0.225199\\
3.93076	0.216199\\
3.93978	0.21374\\
3.95584	0.219795\\
3.97084	0.224795\\
3.98405	0.212924\\
4.00626	0.222887\\
4.02126	0.22292\\
4.0375	0.222491\\
4.0525	0.221611\\
4.06894	0.228051\\
4.08394	0.219208\\
4.10058	0.22198\\
4.12569	0.229023\\
4.13814	0.228855\\
4.16615	0.226364\\
4.19793	0.217095\\
4.21367	0.229627\\
4.22596	0.241916\\
4.25378	0.236983\\
4.28291	0.234869\\
4.29553	0.229015\\
4.31053	0.231048\\
4.32697	0.233032\\
4.34271	0.231593\\
4.37653	0.240836\\
4.40465	0.243779\\
4.4247	0.226768\\
4.45186	0.243929\\
4.47171	0.23304\\
4.47441	0.225737\\
4.50911	0.242217\\
4.52157	0.244673\\
4.54685	0.240776\\
4.57829	0.241319\\
4.59713	0.244417\\
4.60972	0.247007\\
4.62854	0.239201\\
4.64351	0.244168\\
4.66336	0.238656\\
4.68489	0.250191\\
4.69463	0.229755\\
4.71669	0.242279\\
4.75409	0.248333\\
4.76675	0.247632\\
4.7794	0.245321\\
4.80468	0.244935\\
4.82327	0.253519\\
4.83827	0.247632\\
4.87603	0.253968\\
4.8925	0.25342\\
4.90747	0.253964\\
4.92937	0.266009\\
4.94562	0.250991\\
4.96151	0.256879\\
4.97647	0.261845\\
4.99897	0.253612\\
5.01213	0.245525\\
5.02744	0.250696\\
5.04336	0.251649\\
5.05599	0.254272\\
5.08743	0.253616\\
5.11274	0.255028\\
5.12792	0.260213\\
5.15956	0.242093\\
5.191	0.245384\\
5.21588	0.254376\\
5.24443	0.257952\\
5.24764	0.251163\\
5.28835	0.266229\\
5.30121	0.260337\\
5.31383	0.260467\\
5.34507	0.266739\\
5.37634	0.266569\\
5.38896	0.26568\\
5.4352	0.283596\\
5.44554	0.265709\\
5.45488	0.263877\\
5.4858	0.26992\\
5.51792	0.270288\\
5.53963	0.272511\\
5.55463	0.277511\\
5.5713	0.270433\\
5.58627	0.275055\\
5.60903	0.272852\\
5.62169	0.264272\\
5.63669	0.269272\\
5.65313	0.264276\\
5.66848	0.268973\\
5.70267	0.288531\\
5.72522	0.270343\\
5.74657	0.291689\\
5.75591	0.281583\\
5.77235	0.286547\\
5.79133	0.28526\\
5.8007	0.28278\\
5.83835	0.270413\\
5.86057	0.278625\\
5.87319	0.276928\\
5.88568	0.279417\\
5.89831	0.276928\\
5.92255	0.29032\\
5.93209	0.281587\\
5.94853	0.28486\\
5.96122	0.28274\\
5.97997	0.291491\\
5.99497	0.279277\\
6.01773	0.282511\\
6.04305	0.277136\\
6.04574	0.273393\\
6.07721	0.276511\\
6.10882	0.270467\\
6.13739	0.279276\\
6.15005	0.279476\\
6.1686	0.282915\\
6.17149	0.275812\\
6.21183	0.291176\\
6.23674	0.294648\\
6.24608	0.287004\\
6.28173	0.291761\\
6.29419	0.289217\\
6.30681	0.291836\\
6.31929	0.289017\\
6.32869	0.282944\\
6.36319	0.297448\\
6.36605	0.28884\\
6.40365	0.294833\\
6.42587	0.307045\\
6.43578	0.288385\\
6.47267	0.301175\\
6.49544	0.29384\\
6.53234	0.2975\\
6.55832	0.297713\\
6.57084	0.300235\\
6.58363	0.289787\\
6.59863	0.294442\\
6.61779	0.289108\\
6.64631	0.297516\\
6.67376	0.307707\\
6.68407	0.297916\\
6.69926	0.303101\\
6.72488	0.289109\\
6.74678	0.301007\\
6.77139	0.315618\\
6.78053	0.297864\\
6.79697	0.304304\\
6.81233	0.304345\\
6.83454	0.316556\\
6.84954	0.304188\\
6.86638	0.311028\\
6.87887	0.310005\\
6.88156	0.298245\\
6.92273	0.314097\\
6.94495	0.307157\\
6.96685	0.308205\\
6.97948	0.310828\\
7.00462	0.310005\\
7.01728	0.308019\\
7.02629	0.301411\\
7.06047	0.315581\\
7.08665	0.306112\\
7.10503	0.310312\\
7.11772	0.310201\\
7.1365	0.314172\\
7.15147	0.319139\\
7.17427	0.314729\\
7.19322	0.311655\\
7.2122	0.310368\\
7.2272	0.315368\\
7.25589	0.320821\\
7.26492	0.319967\\
7.28737	0.323168\\
7.29982	0.322968\\
7.31248	0.320657\\
7.3219	0.304969\\
7.35334	0.307284\\
7.38098	0.324927\\
7.39032	0.316844\\
7.41329	0.3145\\
7.44473	0.314696\\
7.46371	0.317211\\
7.4762	0.314733\\
7.47889	0.307429\\
7.50128	0.3147\\
7.51377	0.320549\\
7.52659	0.3145\\
7.54159	0.318844\\
7.55783	0.320624\\
7.57283	0.325624\\
7.59539	0.330475\\
7.60445	0.319532\\
7.62683	0.329467\\
7.64183	0.3295\\
7.65484	0.322939\\
7.67093	0.329023\\
7.68321	0.329876\\
7.71485	0.324533\\
7.72751	0.327189\\
7.74013	0.329812\\
7.75871	0.33308\\
7.77117	0.335536\\
7.79019	0.333113\\
7.79921	0.320312\\
7.82142	0.332528\\
7.83408	0.335184\\
7.84673	0.334913\\
7.85935	0.333076\\
7.87817	0.341581\\
7.89059	0.339004\\
7.90938	0.341008\\
7.92207	0.339241\\
7.94668	0.353852\\
7.95562	0.331407\\
7.98171	0.326873\\
8.01	0.345179\\
8.0128	0.331281\\
8.0352	0.338328\\
8.0478	0.340984\\
8.0607	0.333513\\
8.0756	0.338135\\
8.0921	0.333372\\
8.095	0.325725\\
8.1383	0.339115\\
8.1611	0.341608\\
8.1736	0.342152\\
8.1862	0.339841\\
8.2113	0.339641\\
8.2397	0.336573\\
8.2617	0.341808\\
8.2767	0.342352\\
8.2923	0.347924\\
8.312	0.341985\\
8.321	0.339264\\
8.3526	0.339868\\
8.375	0.342523\\
8.3873	0.342156\\
8.3902	0.330085\\
8.4223	0.330255\\
8.4379	0.342927\\
8.4529	0.337825\\
8.4841	0.335824\\
8.5008	0.342497\\
8.5132	0.339808\\
8.5161	0.332564\\
8.5384	0.339845\\
8.551	0.342501\\
8.5637	0.339808\\
8.5761	0.34226\\
8.6074	0.345624\\
8.6166	0.344305\\
8.6387	0.351944\\
8.6537	0.341832\\
8.6795	0.338968\\
8.6955	0.344888\\
8.7043	0.341664\\
8.7265	0.35388\\
8.7415	0.355888\\
8.7583	0.348359\\
8.7674	0.345019\\
8.7832	0.350873\\
8.7984	0.349945\\
8.8208	0.357872\\
8.8611	0.344967\\
8.8901	0.354385\\
8.9154	0.352041\\
8.9277	0.353989\\
8.9616	0.345069\\
8.9839	0.357355\\
8.9988	0.349595\\
9.0213	0.357595\\
9.0363	0.362595\\
9.055	0.350681\\
9.0774	0.353411\\
9.0921	0.337151\\
9.1109	0.343557\\
9.1334	0.356013\\
9.1418	0.333357\\
9.1642	0.345813\\
9.1765	0.345172\\
9.1789	0.333117\\
9.2102	0.329048\\
9.2405	0.315069\\
9.2776	0.327164\\
9.3085	0.314527\\
9.3324	0.301108\\
9.3695	0.294548\\
9.4063	0.285364\\
9.4306	0.268835\\
9.4649	0.275948\\
9.4676	0.268639\\
9.4898	0.269655\\
9.499	0.248532\\
9.5265	0.266012\\
9.5368	0.249197\\
9.5519	0.254364\\
9.5647	0.246285\\
9.5803	0.227912\\
9.6092	0.229725\\
9.6242	0.234692\\
9.6467	0.220424\\
9.6748	0.224204\\
9.6949	0.195943\\
9.7226	0.204043\\
9.732	0.1955\\
9.7484	0.198273\\
9.7582	0.181363\\
9.7763	0.195964\\
9.8113	0.205445\\
9.8337	0.189515\\
9.8554	0.195672\\
9.8681	0.198361\\
9.8807	0.195901\\
9.8957	0.200868\\
9.9058	0.190912\\
9.9208	0.188832\\
9.9375	0.191488\\
9.9465	0.188249\\
9.9716	0.195356\\
10.0004	0.193797\\
10.0032	0.179808\\
10.0285	0.184809\\
10.056	0.202285\\
10.0659	0.199928\\
10.0818	0.201429\\
10.0945	0.203744\\
10.1071	0.2014\\
10.1221	0.205712\\
10.1476	0.201083\\
10.1784	0.211972\\
10.1916	0.200036\\
10.2077	0.204488\\
10.2227	0.209488\\
10.2392	0.210647\\
10.2419	0.193771\\
10.2706	0.204753\\
10.2833	0.203973\\
10.296	0.20414\\
10.3084	0.201596\\
10.3326	0.214952\\
10.342	0.206251\\
10.3613	0.203777\\
10.3985	0.210076\\
10.4148	0.223145\\
10.4177	0.216036\\
10.4525	0.219788\\
10.4552	0.216073\\
10.4807	0.218524\\
10.5029	0.222303\\
10.5154	0.219821\\
10.5407	0.218847\\
10.5532	0.221333\\
10.5682	0.212183\\
10.5903	0.228936\\
10.6036	0.221952\\
10.6159	0.224275\\
10.6189	0.210416\\
10.6351	0.220268\\
10.6475	0.229833\\
10.6602	0.22272\\
10.6694	0.214104\\
10.6914	0.226157\\
10.701	0.220247\\
10.7294	0.229048\\
10.7418	0.228848\\
10.7445	0.216572\\
10.7765	0.213331\\
10.7985	0.220384\\
10.8074	0.217136\\
10.8394	0.220459\\
10.8613	0.232321\\
10.8739	0.228819\\
10.8961	0.226751\\
10.9241	0.234781\\
10.9366	0.235577\\
10.949	0.238067\\
10.9678	0.241809\\
10.9828	0.243419\\
10.9994	0.235143\\
11.0086	0.23212\\
11.0246	0.238045\\
11.0395	0.243012\\
11.0625	0.235343\\
11.0811	0.241701\\
11.084	0.234593\\
11.1128	0.241512\\
11.1157	0.229436\\
11.1442	0.238103\\
11.1469	0.230793\\
11.1695	0.23304\\
11.1785	0.229463\\
11.2069	0.245333\\
11.2195	0.241682\\
11.2544	0.245739\\
11.2855	0.235919\\
11.3173	0.236175\\
11.3487	0.235933\\
11.364	0.248384\\
11.379	0.253384\\
11.4042	0.257256\\
11.4235	0.24498\\
11.4577	0.259044\\
11.467	0.249567\\
11.49	0.248704\\
11.4988	0.244461\\
11.5308	0.24538\\
11.5529	0.247256\\
11.5619	0.244625\\
11.5778	0.250551\\
11.5928	0.255517\\
11.6121	0.237528\\
11.628	0.248997\\
11.6407	0.251653\\
11.6533	0.254309\\
11.6683	0.254164\\
11.688	0.244406\\
11.7215	0.267828\\
11.7505	0.251676\\
11.7821	0.251308\\
11.8194	0.257468\\
11.8355	0.263561\\
11.8504	0.262603\\
11.8607	0.257948\\
11.8698	0.257608\\
11.8985	0.260495\\
11.9112	0.258181\\
11.9262	0.262803\\
11.9453	0.257254\\
11.9677	0.25972\\
11.9829	0.2649\\
12.0199	0.275008\\
12.0401	0.258004\\
12.0617	0.269507\\
12.0766	0.269436\\
12.087	0.267605\\
12.096	0.260531\\
12.1182	0.272752\\
12.1496	0.279889\\
12.1523	0.27258\\
12.1686	0.27884\\
12.1811	0.270437\\
12.1937	0.272551\\
12.2251	0.279856\\
12.2279	0.262983\\
12.2564	0.278672\\
12.2691	0.276181\\
12.2878	0.284965\\
12.3028	0.269432\\
12.3224	0.262636\\
12.3509	0.279272\\
12.3636	0.276961\\
12.3791	0.26338\\
12.414	0.2708\\
12.4326	0.280257\\
12.4354	0.273116\\
12.4516	0.272405\\
12.4642	0.270637\\
12.4893	0.270601\\
12.492	0.266888\\
12.5173	0.269168\\
12.5585	0.277136\\
12.5835	0.285637\\
12.5934	0.271018\\
12.6085	0.286152\\
12.6235	0.280844\\
12.6561	0.276971\\
12.6714	0.292345\\
12.6841	0.285371\\
12.6866	0.277895\\
12.7092	0.282859\\
12.7188	0.283331\\
12.7341	0.288669\\
12.747	0.284824\\
12.7594	0.282889\\
12.7688	0.276788\\
12.7907	0.288641\\
12.8001	0.282716\\
12.8286	0.290956\\
12.8411	0.293445\\
12.875	0.294188\\
12.8852	0.291795\\
12.9002	0.296762\\
12.9105	0.29136\\
12.9255	0.287376\\
12.9481	0.291995\\
12.9509	0.284853\\
12.9761	0.289476\\
12.9983	0.301697\\
13.0134	0.294649\\
13.0356	0.306871\\
13.0485	0.304384\\
13.0701	0.301328\\
13.1016	0.30156\\
13.1295	0.309268\\
13.1388	0.299624\\
13.1681	0.304529\\
13.1808	0.295745\\
13.1934	0.298223\\
13.2059	0.295712\\
13.2216	0.3006\\
13.2371	0.301677\\
13.2465	0.295604\\
13.2812	0.305136\\
13.2836	0.30754\\
13.3002	0.303845\\
13.3092	0.295804\\
13.3378	0.304588\\
13.3504	0.304788\\
13.3723	0.302132\\
13.4033	0.301728\\
13.4314	0.309468\\
13.4407	0.301928\\
13.4722	0.313368\\
13.4948	0.313984\\
13.4977	0.306875\\
13.5349	0.318408\\
13.5603	0.31382\\
13.5977	0.321068\\
13.6142	0.320597\\
13.6169	0.30766\\
13.6328	0.32352\\
13.6424	0.307948\\
13.6646	0.316844\\
13.7024	0.309943\\
13.7051	0.30266\\
13.7361	0.312484\\
13.7528	0.312044\\
13.7618	0.301096\\
13.7969	0.311584\\
13.8057	0.30806\\
13.8281	0.310745\\
13.8408	0.313368\\
13.8718	0.323869\\
13.881	0.32068\\
13.9034	0.322792\\
13.9184	0.327792\\
13.9286	0.320824\\
13.9383	0.321435\\
13.9625	0.320148\\
13.9851	0.323172\\
14.0105	0.320861\\
14.0195	0.317109\\
14.0507	0.325596\\
14.0795	0.329801\\
14.0883	0.326444\\
14.1262	0.320512\\
14.1488	0.322792\\
14.1611	0.322935\\
14.1857	0.332612\\
14.2115	0.330476\\
14.2142	0.3232\\
14.2303	0.329293\\
14.243	0.328953\\
14.2681	0.329759\\
14.2708	0.322449\\
14.3026	0.320298\\
14.3246	0.329492\\
14.3561	0.336605\\
14.3589	0.329464\\
14.3814	0.33068\\
14.3839	0.323204\\
14.4252	0.336972\\
14.4442	0.335648\\
14.4533	0.327044\\
14.4818	0.335593\\
14.4849	0.325484\\
14.5132	0.33884\\
14.5292	0.34478\\
14.5573	0.333104\\
14.5763	0.335995\\
14.579	0.32372\\
14.6162	0.345444\\
14.6265	0.335991\\
14.6358	0.333169\\
14.6578	0.335082\\
14.6728	0.338139\\
14.6831	0.333484\\
14.6927	0.333755\\
14.7172	0.333368\\
14.7521	0.348245\\
14.7866	0.33334\\
14.8154	0.333571\\
14.8181	0.3332\\
14.8521	0.347612\\
14.8614	0.337592\\
14.8842	0.344996\\
14.894	0.324499\\
14.9158	0.336804\\
14.9409	0.342315\\
14.9559	0.347315\\
14.9751	0.339328\\
14.9975	0.341609\\
15.0101	0.342319\\
15.0322	0.339496\\
15.063	0.345792\\
15.0853	0.347737\\
15.0949	0.342737\\
15.1229	0.358425\\
15.1567	0.350077\\
15.173	0.354941\\
15.188	0.350209\\
15.2038	0.355981\\
15.2391	0.345089\\
15.2549	0.350952\\
15.2675	0.353571\\
15.2922	0.357705\\
15.3015	0.350033\\
15.3204	0.345341\\
15.3429	0.357831\\
15.3554	0.360367\\
15.3807	0.358188\\
15.3952	0.353032\\
15.427	0.344417\\
15.4613	0.366436\\
15.4644	0.349803\\
15.4808	0.347923\\
15.4898	0.34188\\
15.5263	0.328273\\
15.5692	0.320131\\
15.5998	0.309937\\
15.6371	0.296661\\
15.6672	0.289131\\
15.7043	0.299051\\
15.7473	0.300241\\
15.7904	0.270887\\
15.8333	0.254155\\
15.8703	0.250454\\
15.9132	0.235939\\
15.935	0.247797\\
15.95	0.235733\\
15.9662	0.241905\\
15.9751	0.217786\\
15.9856	0.218298\\
16.0116	0.200969\\
16.0206	0.187232\\
16.0367	0.193331\\
16.0516	0.198297\\
16.062	0.191674\\
16.077	0.196674\\
16.0871	0.196819\\
16.1021	0.188893\\
16.1155	0.170465\\
16.1379	0.177892\\
16.1529	0.16778\\
16.1905	0.179832\\
16.2069	0.18635\\
16.2219	0.191317\\
16.232	0.18536\\
16.2447	0.183049\\
16.2573	0.184785\\
16.2723	0.18536\\
16.2853	0.173279\\
16.3076	0.192071\\
16.3105	0.175185\\
16.3518	0.184348\\
16.3761	0.198592\\
16.3854	0.184956\\
16.4021	0.183713\\
16.4145	0.182646\\
16.4272	0.18408\\
16.4397	0.182649\\
16.4523	0.184348\\
16.4673	0.184956\\
16.4864	0.191084\\
16.5179	0.187067\\
16.5401	0.199293\\
16.5551	0.191084\\
16.5656	0.191596\\
16.5812	0.195791\\
16.5968	0.197616\\
16.6095	0.197416\\
16.6346	0.197412\\
16.6373	0.185131\\
16.6659	0.203577\\
16.6785	0.201233\\
16.694	0.187239\\
16.7224	0.204535\\
16.7259	0.1877\\
16.7479	0.19782\\
16.7603	0.195276\\
16.7631	0.186875\\
16.7916	0.200403\\
16.8042	0.197383\\
16.8388	0.201517\\
16.8731	0.207187\\
16.8761	0.200236\\
16.9011	0.207048\\
16.9296	0.215391\\
16.9386	0.212715\\
16.9672	0.221269\\
16.97	0.212121\\
17.0044	0.221048\\
17.0138	0.212183\\
17.0452	0.218656\\
17.0612	0.222384\\
17.0703	0.211804\\
17.0872	0.211057\\
17.0899	0.20374\\
17.1182	0.217149\\
17.1464	0.207784\\
17.1783	0.208273\\
17.2064	0.216767\\
17.2216	0.221937\\
17.2376	0.223851\\
17.2526	0.224867\\
17.2628	0.225009\\
17.2754	0.220176\\
17.2942	0.223885\\
17.3032	0.216059\\
17.3193	0.229441\\
17.3346	0.224783\\
17.3667	0.220293\\
17.3982	0.219868\\
17.4263	0.228244\\
17.4416	0.224603\\
17.4575	0.225937\\
17.4725	0.224265\\
17.4957	0.222899\\
17.4984	0.210651\\
17.5359	0.222696\\
17.5612	0.216601\\
17.5923	0.225611\\
17.6211	0.234613\\
17.63	0.231874\\
17.6521	0.243929\\
17.6671	0.243924\\
17.6835	0.247269\\
17.6962	0.244549\\
17.7299	0.243419\\
17.7403	0.241861\\
17.7553	0.236516\\
17.778	0.242093\\
17.7805	0.234612\\
17.8093	0.248021\\
17.8311	0.23882\\
17.8527	0.250377\\
17.8622	0.24452\\
17.8725	0.24249\\
17.8875	0.237313\\
17.9094	0.247335\\
17.919	0.241908\\
17.9419	0.23344\\
17.9662	0.247676\\
17.9755	0.235583\\
};

\end{axis}

\end{tikzpicture}%

%% file: figures/vegas-rtt-2.tex
%
%
\definecolor{mycolor1}{rgb}{0.00000,0.44700,0.74100}%
\definecolor{mycolor2}{rgb}{0.85000,0.32500,0.09800}%
\begin{tikzpicture}
\pgfplotsset{every tick label/.append style={font=\scriptsize}}

\begin{axis}[%
width=0.951\fwidth,
height=\fheight,
at={(0\fwidth,0\fheight)},
scale only axis,
xmajorgrids,
ymajorgrids,
xmin=0,
xmax=18,
xlabel style={font=\scriptsize\color{white!15!black}},
xlabel={Time (s)},
ymin=0.1,
ymax=0.2,
ylabel style={font=\scriptsize\color{white!15!black}},
ylabel={Measured RTT (s)},
axis background/.style={fill=white},
title style={font=\bfseries},
legend style={legend cell align=left, font=\scriptsize,  align=left, draw=white!15!black}
]
\addplot [color=mycolor1,dashed,line width=1.2pt]
  table[row sep=crcr]{%
0.23764	0.12045\\
0.25326	0.126069\\
0.36265	0.115013\\
0.38443	0.120988\\
0.38687	0.108614\\
0.47651	0.113859\\
0.49987	0.127219\\
0.50275	0.120097\\
0.52112	0.134249\\
0.54322	0.126669\\
0.60978	0.118289\\
0.62066	0.120791\\
0.64279	0.140039\\
0.64567	0.132917\\
0.66439	0.143269\\
0.6954	0.133483\\
0.73231	0.160398\\
0.75522	0.135259\\
0.76782	0.137152\\
0.78374	0.119347\\
0.80845	0.124063\\
0.8367	0.141301\\
0.86176	0.146507\\
0.87705	0.151799\\
0.91437	0.162055\\
0.95193	0.158188\\
0.98044	0.171988\\
0.98301	0.17456\\
1.01777	0.181076\\
1.05514	0.196673\\
1.06775	0.1958\\
1.08671	0.189656\\
1.10567	0.181299\\
1.12067	0.181339\\
1.1499	0.178263\\
1.16847	0.185452\\
1.18743	0.184703\\
1.19967	0.186951\\
1.23147	0.185015\\
1.26902	0.188917\\
1.29447	0.178618\\
1.32211	0.177457\\
1.33804	0.175969\\
1.3472	0.160049\\
1.36399	0.166377\\
1.37921	0.166569\\
1.395	0.172365\\
1.42644	0.172606\\
1.44556	0.164209\\
1.45804	0.166686\\
1.47068	0.163844\\
1.48331	0.166481\\
1.50137	0.169256\\
1.51072	0.149339\\
1.52752	0.153524\\
1.54271	0.1535\\
1.57768	0.15768\\
1.60319	0.14744\\
1.60613	0.135568\\
1.62844	0.14776\\
1.64088	0.145237\\
1.6658	0.155077\\
1.67821	0.157488\\
1.6965	0.158788\\
1.72245	0.150052\\
1.74438	0.138256\\
1.77285	0.14702\\
1.81037	0.147921\\
1.82301	0.147212\\
1.83562	0.144852\\
1.8607	0.144656\\
1.8757	0.153252\\
1.88873	0.144348\\
1.91019	0.155808\\
1.92016	0.150656\\
1.94227	0.159423\\
1.95727	0.159456\\
1.96738	0.159569\\
1.98002	0.159652\\
1.99263	0.157292\\
2.00785	0.16204\\
2.03	0.169377\\
2.03942	0.15372\\
2.06156	0.162832\\
2.07404	0.165308\\
2.09336	0.1544\\
2.10836	0.155892\\
2.13094	0.153552\\
2.13405	0.1417\\
2.17481	0.156961\\
2.19307	0.16522\\
2.22222	0.13566\\
2.25683	0.153281\\
2.26627	0.135331\\
2.28822	0.154169\\
2.30322	0.149477\\
2.31336	0.149618\\
2.32833	0.140997\\
2.37284	0.150613\\
2.41019	0.158109\\
2.4205	0.154237\\
2.43296	0.147159\\
2.46674	0.15352\\
2.47685	0.153487\\
2.49185	0.149144\\
2.51445	0.161745\\
2.52942	0.15658\\
2.54609	0.153561\\
2.56106	0.158528\\
2.5714	0.151205\\
2.58382	0.153124\\
2.58661	0.140504\\
2.62393	0.147192\\
2.63424	0.147204\\
2.64968	0.147832\\
2.66568	0.153836\\
2.68092	0.156471\\
2.70309	0.163669\\
2.71574	0.159652\\
2.72836	0.157308\\
2.74336	0.161967\\
2.75983	0.153524\\
2.77229	0.15598\\
2.78511	0.15118\\
2.80011	0.155676\\
2.829	0.153128\\
2.84822	0.148289\\
2.86084	0.147755\\
2.8636	0.135544\\
2.89141	0.153048\\
2.9109	0.154221\\
2.92355	0.15372\\
2.9425	0.147388\\
2.96674	0.15352\\
2.97628	0.15306\\
3.00204	0.138444\\
3.03049	0.147363\\
3.0558	0.145052\\
3.0708	0.149907\\
3.08092	0.145052\\
3.09341	0.150905\\
3.12501	0.148724\\
3.14683	0.144792\\
3.1689	0.147\\
3.2065	0.151009\\
3.21916	0.153161\\
3.23178	0.150972\\
3.28836	0.151205\\
3.30731	0.150476\\
3.31026	0.143432\\
3.35335	0.159448\\
3.37022	0.151209\\
3.38288	0.15372\\
3.39533	0.151209\\
3.40796	0.153832\\
3.42041	0.151176\\
3.43582	0.156588\\
3.4647	0.15444\\
3.48329	0.153332\\
3.49829	0.158332\\
3.52761	0.147392\\
3.53057	0.13538\\
3.55273	0.147392\\
3.56535	0.145048\\
3.58393	0.153332\\
3.59662	0.150801\\
3.6149	0.159084\\
3.63439	0.151213\\
3.64936	0.155876\\
3.66599	0.154188\\
3.67828	0.156477\\
3.6804	0.149835\\
3.70359	0.153357\\
3.71856	0.155835\\
3.735	0.156872\\
3.74769	0.153565\\
3.76011	0.15118\\
3.77511	0.141068\\
3.79174	0.147355\\
3.80674	0.143876\\
3.82971	0.149309\\
3.83275	0.142349\\
3.85446	0.154021\\
3.86942	0.155831\\
3.89875	0.153749\\
3.91137	0.151372\\
3.92662	0.156312\\
3.94261	0.153283\\
3.95761	0.155676\\
3.97429	0.153524\\
3.98362	0.140868\\
4.00552	0.152769\\
4.02049	0.156031\\
4.04962	0.156877\\
4.05274	0.141676\\
4.08648	0.165108\\
4.09582	0.155672\\
4.11792	0.165116\\
4.12726	0.159648\\
4.1439	0.160276\\
4.15659	0.15355\\
4.16901	0.153487\\
4.18401	0.150204\\
4.20697	0.154232\\
4.23774	0.155504\\
4.26918	0.153364\\
4.2943	0.157036\\
4.30364	0.149548\\
4.32028	0.15138\\
4.32939	0.135384\\
4.35172	0.147708\\
4.36672	0.1474\\
4.37703	0.147712\\
4.39244	0.141484\\
4.4144	0.15344\\
4.4294	0.1472\\
4.44517	0.152972\\
4.48031	0.15092\\
4.50804	0.158947\\
4.51799	0.141276\\
4.55284	0.150397\\
4.55559	0.143148\\
4.5967	0.1573\\
4.62181	0.156859\\
4.63429	0.159345\\
4.65344	0.153132\\
4.67856	0.160564\\
4.68767	0.139988\\
4.70387	0.14828\\
4.73534	0.149893\\
4.77307	0.154032\\
4.79818	0.151377\\
4.81717	0.145252\\
4.83258	0.144903\\
4.85453	0.156856\\
4.86953	0.155468\\
4.88597	0.1571\\
4.90097	0.146988\\
4.91714	0.15316\\
4.92668	0.141264\\
4.94312	0.147704\\
4.95872	0.148196\\
4.99247	0.159897\\
5.0208	0.156076\\
5.03111	0.15158\\
5.04611	0.149948\\
5.06188	0.150912\\
5.07122	0.14454\\
5.10644	0.153324\\
5.13155	0.149456\\
5.1567	0.15101\\
5.16635	0.135552\\
5.1885	0.147204\\
5.20411	0.147999\\
5.21974	0.14852\\
5.23474	0.144076\\
5.25118	0.150516\\
5.26618	0.147392\\
5.27629	0.147504\\
5.31072	0.144372\\
5.3392	0.152853\\
5.34216	0.14346\\
5.38306	0.153132\\
5.4075	0.162764\\
5.41684	0.155472\\
5.43348	0.157304\\
5.4489	0.147603\\
5.4841	0.144903\\
5.48586	0.143703\\
5.5215	0.149677\\
5.54678	0.150569\\
5.55927	0.153058\\
5.58455	0.153687\\
5.59392	0.145021\\
5.63079	0.161893\\
5.6411	0.155241\\
5.6561	0.150549\\
5.66621	0.150661\\
5.69698	0.15684\\
5.70632	0.149351\\
5.7384	0.144488\\
5.77338	0.14986\\
5.77634	0.135548\\
5.79853	0.147233\\
5.81115	0.145047\\
5.84238	0.151172\\
5.87402	0.153524\\
5.8833	0.1449\\
5.91478	0.13844\\
5.94346	0.147592\\
5.95608	0.147559\\
5.981	0.147504\\
6.01267	0.148497\\
6.02788	0.144576\\
6.05044	0.147441\\
6.05952	0.14474\\
6.07551	0.150736\\
6.09051	0.150408\\
6.12175	0.15308\\
6.13819	0.15952\\
6.15343	0.153964\\
6.18443	0.161758\\
6.21305	0.165172\\
6.22618	0.15666\\
6.2387	0.152995\\
6.25802	0.147504\\
6.30486	0.156664\\
6.33549	0.16206\\
6.35846	0.15092\\
6.37728	0.151413\\
6.39224	0.155876\\
6.41454	0.156528\\
6.44012	0.153644\\
6.45512	0.158644\\
6.47159	0.156737\\
6.48656	0.161704\\
6.503	0.154192\\
6.51549	0.157029\\
6.53782	0.135576\\
6.57536	0.1478\\
6.59735	0.159789\\
6.61231	0.162\\
6.62879	0.160349\\
6.64375	0.162159\\
6.66043	0.160549\\
6.6695	0.141504\\
6.69136	0.15354\\
6.7007	0.14302\\
6.73257	0.147216\\
6.74841	0.153053\\
6.76338	0.156031\\
6.78586	0.160429\\
6.80082	0.162039\\
6.82394	0.154437\\
6.82686	0.14736\\
6.86188	0.147192\\
6.87437	0.149681\\
6.88933	0.14322\\
6.90577	0.147363\\
6.92077	0.14408\\
6.95623	0.145405\\
6.95915	0.138328\\
6.98064	0.153776\\
6.98998	0.143424\\
7.00642	0.149864\\
7.01908	0.145055\\
7.03769	0.153324\\
7.05051	0.147163\\
7.0693	0.153332\\
7.0944	0.153988\\
7.10352	0.144368\\
7.12564	0.156492\\
7.14064	0.161492\\
7.15732	0.153319\\
7.16994	0.153516\\
7.18855	0.156533\\
7.20352	0.155831\\
7.21383	0.151172\\
7.22883	0.14934\\
7.25458	0.141064\\
7.28936	0.153528\\
7.30818	0.154221\\
7.3208	0.153487\\
7.33962	0.157338\\
7.35244	0.153891\\
7.37736	0.153335\\
7.41113	0.162304\\
7.42168	0.1472\\
7.4341	0.149623\\
7.45292	0.150573\\
7.46574	0.147559\\
7.48998	0.156832\\
7.49993	0.150311\\
7.52527	0.135572\\
7.55356	0.153857\\
7.56638	0.145248\\
7.59149	0.144604\\
7.62906	0.151176\\
7.64758	0.159692\\
7.65672	0.143412\\
7.67928	0.154012\\
7.70393	0.158796\\
7.71327	0.149711\\
7.72971	0.151184\\
7.74216	0.15364\\
7.77363	0.150353\\
7.78646	0.145052\\
7.80146	0.14474\\
7.8179	0.147004\\
7.83018	0.150901\\
7.84301	0.145052\\
7.85801	0.14474\\
7.87448	0.147191\\
7.88968	0.149976\\
7.91787	0.15316\\
7.93119	0.145048\\
7.94644	0.14998\\
7.96226	0.150809\\
7.97512	0.147033\\
7.98774	0.144848\\
8.0027	0.149736\\
8.0192	0.147\\
8.0342	0.149743\\
8.0631	0.153421\\
8.0662	0.135548\\
8.0945	0.15332\\
8.1095	0.149903\\
8.1197	0.147392\\
8.1323	0.145048\\
8.1511	0.153157\\
8.1639	0.147355\\
8.1975	0.14408\\
8.2141	0.147912\\
8.2519	0.1472\\
8.2644	0.144881\\
8.277	0.147359\\
8.3013	0.156636\\
8.3108	0.149707\\
8.3272	0.15118\\
8.3422	0.15618\\
8.3526	0.145052\\
8.368	0.153844\\
8.3988	0.155976\\
8.4153	0.151006\\
8.4279	0.153487\\
8.4523	0.152964\\
8.4718	0.151009\\
8.4847	0.147233\\
8.4973	0.145081\\
8.5099	0.1472\\
8.5284	0.150373\\
8.541	0.152996\\
8.5602	0.148293\\
8.5754	0.135184\\
8.6131	0.147808\\
8.6355	0.153724\\
8.648	0.15118\\
8.6507	0.143423\\
8.6792	0.156509\\
8.6942	0.155839\\
8.7045	0.151213\\
8.714	0.138579\\
8.7425	0.1472\\
8.7575	0.1522\\
8.7895	0.138772\\
8.8231	0.162316\\
8.8269	0.150812\\
8.8491	0.159844\\
8.8743	0.160361\\
8.8898	0.149864\\
8.9119	0.159256\\
8.9436	0.156677\\
8.9467	0.137504\\
8.9809	0.161648\\
8.9898	0.143379\\
9.0067	0.147596\\
9.0193	0.145252\\
9.0381	0.148316\\
9.047	0.1472\\
9.063	0.153228\\
9.0784	0.156308\\
9.1001	0.163164\\
9.1197	0.153328\\
9.1347	0.158328\\
9.1514	0.147359\\
9.1664	0.149703\\
9.1765	0.144844\\
9.1858	0.138836\\
9.2208	0.147592\\
9.2453	0.156868\\
9.2603	0.161868\\
9.2708	0.151409\\
9.2833	0.153328\\
9.3021	0.157337\\
9.3149	0.153361\\
9.3276	0.151209\\
9.3365	0.140668\\
9.3642	0.158448\\
9.3991	0.156177\\
9.4092	0.153732\\
9.4217	0.151413\\
9.447	0.150936\\
9.4658	0.15372\\
9.5037	0.147559\\
9.5288	0.145048\\
9.5413	0.147537\\
9.554	0.144885\\
9.5666	0.147237\\
9.5793	0.145081\\
9.5917	0.147504\\
9.5947	0.135348\\
9.6232	0.147396\\
9.6483	0.147508\\
9.6637	0.147808\\
9.6797	0.153869\\
9.6924	0.153331\\
9.6949	0.141108\\
9.7173	0.15332\\
9.73	0.151009\\
9.7426	0.153324\\
9.758	0.143628\\
9.7804	0.147192\\
9.793	0.144848\\
9.8172	0.158788\\
9.8266	0.152908\\
9.8489	0.153936\\
9.8622	0.147592\\
9.8747	0.147392\\
9.8873	0.145048\\
9.9026	0.149947\\
9.931	0.153523\\
9.946	0.155634\\
9.9563	0.150732\\
9.9713	0.155732\\
9.9995	0.152964\\
10.0194	0.147196\\
10.0224	0.13518\\
10.0508	0.153519\\
10.0759	0.154384\\
10.0909	0.149899\\
10.1011	0.145078\\
10.1137	0.1472\\
10.1483	0.135544\\
10.1705	0.148076\\
10.1954	0.153128\\
10.2205	0.15949\\
10.233	0.157138\\
10.2456	0.159493\\
10.2583	0.157308\\
10.2707	0.159619\\
10.2857	0.159652\\
10.3087	0.150405\\
10.3213	0.153028\\
10.3363	0.146764\\
10.3716	0.151217\\
10.4094	0.153723\\
10.422	0.151409\\
10.4346	0.153728\\
10.45	0.154332\\
10.4657	0.159993\\
10.4807	0.159848\\
10.491	0.159648\\
10.506	0.155672\\
10.5222	0.161941\\
10.5351	0.151205\\
10.5478	0.153857\\
10.5604	0.151401\\
10.5728	0.153483\\
10.5853	0.150973\\
10.6041	0.159481\\
10.6191	0.155831\\
10.6296	0.15392\\
10.642	0.151376\\
10.6608	0.159656\\
10.6758	0.155676\\
10.6862	0.153924\\
10.6986	0.151409\\
10.7113	0.153524\\
10.7239	0.15118\\
10.7575	0.159408\\
10.7678	0.15372\\
10.7805	0.151409\\
10.7931	0.153487\\
10.8056	0.150728\\
10.8244	0.153357\\
10.837	0.151205\\
10.8497	0.153516\\
10.8623	0.151205\\
10.875	0.153686\\
10.8872	0.150972\\
10.906	0.159789\\
10.9189	0.151405\\
10.9315	0.15372\\
10.9442	0.151376\\
10.9468	0.143652\\
10.9881	0.153724\\
11.0008	0.151413\\
11.0132	0.153524\\
11.0258	0.15118\\
11.0411	0.156113\\
11.0698	0.153728\\
11.0824	0.151417\\
11.095	0.153491\\
11.1075	0.150736\\
11.1227	0.150652\\
11.1516	0.15138\\
11.1641	0.153324\\
11.167	0.14128\\
11.2042	0.153076\\
11.2145	0.153388\\
11.2272	0.145081\\
11.2396	0.147188\\
11.2522	0.144844\\
11.2837	0.150932\\
11.3025	0.154188\\
11.3149	0.153353\\
11.3276	0.150597\\
11.3426	0.155564\\
11.3592	0.145085\\
11.3719	0.147196\\
11.3843	0.144877\\
11.3993	0.149703\\
11.4096	0.145048\\
11.4246	0.150048\\
11.4472	0.150764\\
11.4599	0.147233\\
11.4723	0.144877\\
11.4873	0.149699\\
11.5166	0.147355\\
11.5194	0.135145\\
11.5415	0.147196\\
11.5542	0.144852\\
11.5729	0.153132\\
11.6046	0.147392\\
11.6074	0.13518\\
11.6354	0.153023\\
11.6449	0.144324\\
11.6675	0.148112\\
11.6863	0.147396\\
11.6989	0.147363\\
11.7175	0.150979\\
11.7325	0.149347\\
11.7672	0.159864\\
11.7806	0.153487\\
11.7956	0.15062\\
11.8059	0.150932\\
11.8209	0.155932\\
11.8466	0.135132\\
11.8812	0.153525\\
11.9063	0.154192\\
11.9188	0.156681\\
11.9441	0.153365\\
11.9633	0.147204\\
11.985	0.138416\\
12.0319	0.155388\\
12.0563	0.16492\\
12.0657	0.159452\\
12.076	0.159493\\
12.0887	0.157337\\
12.1139	0.157504\\
12.139	0.154\\
12.1705	0.155613\\
12.1958	0.15074\\
12.2108	0.15574\\
12.2211	0.145477\\
12.2336	0.147589\\
12.2462	0.145278\\
12.2589	0.147901\\
12.284	0.147704\\
12.3154	0.147708\\
12.3462	0.15704\\
12.3555	0.149548\\
12.372	0.15118\\
12.3845	0.153328\\
12.3872	0.141116\\
12.4245	0.153324\\
12.4348	0.153635\\
12.4498	0.147395\\
12.4661	0.153632\\
12.4811	0.147192\\
12.4971	0.153164\\
12.5167	0.147388\\
12.5262	0.138972\\
12.5547	0.147755\\
12.567	0.150043\\
12.5699	0.135376\\
12.5984	0.153358\\
12.6134	0.149344\\
12.63	0.153728\\
12.6423	0.151176\\
12.6706	0.144372\\
12.6927	0.156525\\
12.7054	0.154021\\
12.7242	0.159481\\
12.7391	0.149348\\
12.7684	0.145052\\
12.8085	0.153108\\
12.8281	0.1377\\
12.8503	0.147563\\
12.8627	0.145052\\
12.8877	0.153524\\
12.8972	0.13804\\
12.9533	0.153067\\
12.9726	0.144456\\
13.0003	0.152623\\
13.0098	0.14954\\
13.0413	0.15598\\
13.0573	0.160135\\
13.0723	0.155275\\
13.0948	0.167768\\
13.1039	0.140661\\
13.1301	0.137891\\
13.1645	0.154677\\
13.1769	0.1571\\
13.1896	0.159823\\
13.2024	0.151128\\
13.2212	0.153725\\
13.2362	0.153917\\
13.2525	0.160223\\
13.262	0.13804\\
13.2841	0.154007\\
13.3155	0.155584\\
13.3185	0.143991\\
13.3407	0.151213\\
13.3533	0.153653\\
13.3849	0.153528\\
13.4223	0.160331\\
13.4413	0.159312\\
13.4697	0.14126\\
13.4917	0.153249\\
13.5044	0.153683\\
13.5356	0.159956\\
13.5673	0.160193\\
13.601	0.141268\\
13.624	0.154244\\
13.6269	0.137508\\
13.6548	0.153027\\
13.6641	0.147156\\
13.6805	0.153596\\
13.693	0.145056\\
13.7582	0.147196\\
13.781	0.154069\\
13.8065	0.142392\\
13.8251	0.150976\\
13.8401	0.149344\\
13.8627	0.156529\\
13.8776	0.161496\\
13.903	0.156172\\
13.9133	0.145044\\
13.9257	0.147533\\
13.9384	0.144878\\
13.951	0.147501\\
13.9825	0.1472\\
14.0076	0.147749\\
14.0226	0.149903\\
14.0329	0.145248\\
14.0454	0.147737\\
14.058	0.145048\\
14.0705	0.147037\\
14.0831	0.144885\\
14.0981	0.149707\\
14.1082	0.144852\\
14.1232	0.149852\\
14.1584	0.153628\\
14.1614	0.143787\\
14.1836	0.151009\\
14.1962	0.153161\\
14.2089	0.151009\\
14.2213	0.15332\\
14.2241	0.141108\\
14.2465	0.153319\\
14.2717	0.153324\\
14.2872	0.149957\\
14.3152	0.167973\\
14.3241	0.143179\\
14.341	0.147396\\
14.3536	0.145085\\
14.3785	0.154441\\
14.3815	0.137695\\
14.4039	0.147393\\
14.4165	0.145049\\
14.4407	0.158792\\
14.4503	0.143579\\
14.4857	0.147196\\
14.4981	0.147159\\
14.5131	0.147192\\
14.5357	0.154237\\
14.5486	0.147392\\
14.561	0.147163\\
14.5916	0.152768\\
14.6176	0.147897\\
14.6301	0.150349\\
14.6429	0.144848\\
14.6581	0.149759\\
14.6741	0.151009\\
14.6834	0.13518\\
14.7058	0.147159\\
14.721	0.146832\\
14.737	0.152827\\
14.752	0.147391\\
14.7686	0.154028\\
14.7811	0.153513\\
14.7937	0.151169\\
14.8087	0.15586\\
14.825	0.156922\\
14.8377	0.154237\\
14.8505	0.147359\\
14.8655	0.14978\\
14.9007	0.153529\\
14.9132	0.15121\\
14.9259	0.147037\\
14.9385	0.144852\\
14.9537	0.149796\\
14.9826	0.147312\\
14.9854	0.135384\\
15.0195	0.158792\\
15.0291	0.144404\\
15.0513	0.156561\\
15.0611	0.137859\\
15.083	0.144848\\
15.098	0.149344\\
15.124	0.13866\\
15.1583	0.153124\\
15.1733	0.158124\\
15.2054	0.144368\\
15.2278	0.147159\\
15.243	0.149788\\
15.2683	0.144288\\
15.3093	0.155621\\
15.3124	0.143836\\
15.3472	0.154025\\
15.3622	0.158992\\
15.3727	0.147404\\
15.3879	0.149992\\
15.4133	0.144992\\
15.4567	0.158528\\
15.4668	0.154432\\
15.4822	0.150152\\
15.4981	0.156013\\
15.5109	0.15372\\
15.5234	0.151176\\
15.5386	0.155716\\
15.5609	0.162988\\
15.5759	0.167988\\
15.6046	0.16164\\
15.6179	0.151176\\
15.6304	0.153365\\
15.6557	0.147596\\
15.6683	0.145252\\
15.6833	0.149748\\
15.7272	0.156264\\
15.7493	0.149848\\
15.7688	0.15121\\
15.7814	0.153524\\
15.7939	0.151013\\
15.8063	0.153469\\
15.8316	0.153124\\
15.8631	0.149964\\
15.882	0.144848\\
15.897	0.149848\\
15.9135	0.151009\\
15.9262	0.147392\\
15.9388	0.145081\\
15.9515	0.147591\\
15.9641	0.145244\\
15.9764	0.1475\\
16.0078	0.153829\\
16.0331	0.151209\\
16.0481	0.155872\\
16.0582	0.151209\\
16.0711	0.147559\\
16.0836	0.145081\\
16.096	0.147188\\
16.1087	0.144877\\
16.1213	0.147192\\
16.1401	0.150573\\
16.1525	0.153029\\
16.178	0.148264\\
16.1905	0.147396\\
16.2032	0.145085\\
16.2156	0.147396\\
16.2283	0.145052\\
16.2408	0.147196\\
16.2534	0.144885\\
16.2684	0.149711\\
16.2785	0.144856\\
16.2935	0.143224\\
16.3285	0.153057\\
16.3315	0.143264\\
16.3539	0.150813\\
16.3665	0.153324\\
16.3914	0.153153\\
16.3945	0.141276\\
16.4167	0.15332\\
16.4294	0.150975\\
16.4479	0.159252\\
16.4854	0.167764\\
16.4986	0.147388\\
16.5112	0.147355\\
16.53	0.151172\\
16.5425	0.153661\\
16.5452	0.150696\\
16.5676	0.15352\\
16.5826	0.155831\\
16.5929	0.150744\\
16.6079	0.149748\\
16.6243	0.15138\\
16.6495	0.151413\\
16.6644	0.155835\\
16.6838	0.138608\\
16.7125	0.147592\\
16.7503	0.1472\\
16.7688	0.150972\\
16.8005	0.151171\\
16.8034	0.143903\\
16.8406	0.156748\\
16.8511	0.1474\\
16.8664	0.143819\\
16.8824	0.144881\\
16.8973	0.149848\\
16.9138	0.151176\\
16.929	0.149992\\
16.9547	0.141352\\
16.9823	0.158968\\
16.9917	0.141072\\
17.0145	0.153169\\
17.0271	0.147367\\
17.0459	0.153491\\
17.0609	0.153524\\
17.0773	0.153365\\
17.0898	0.150736\\
17.0927	0.143692\\
17.1086	0.153873\\
17.1214	0.147004\\
17.1613	0.155472\\
17.178	0.153483\\
17.1904	0.150972\\
17.2219	0.151009\\
17.241	0.148308\\
17.2438	0.141064\\
17.2659	0.147359\\
17.3063	0.153063\\
17.3166	0.141932\\
17.332	0.14402\\
17.3478	0.144436\\
17.363	0.14968\\
17.3854	0.151176\\
17.3945	0.150696\\
17.4105	0.156692\\
17.4229	0.159144\\
17.4506	0.159644\\
17.4672	0.150973\\
17.4828	0.15608\\
17.5174	0.159459\\
17.5451	0.16206\\
17.5616	0.157093\\
17.5711	0.135632\\
17.5932	0.147704\\
17.6308	0.153332\\
17.6624	0.150716\\
17.6652	0.13538\\
17.6937	0.153861\\
17.7086	0.149836\\
17.7253	0.154152\\
17.7282	0.137659\\
17.7694	0.151176\\
17.7881	0.144848\\
17.8031	0.149848\\
17.8135	0.148276\\
17.8386	0.153528\\
17.8542	0.150511\\
17.8698	0.15118\\
17.8789	0.1507\\
17.9076	0.14966\\
17.9477	0.149348\\
17.958	0.144852\\
17.9734	0.14976\\
};

\addplot [color=mycolor2]
  table[row sep=crcr]{%
0.33764	0.12045\\
0.35345	0.126257\\
0.459	0.111362\\
0.47482	0.12137\\
0.48982	0.12619\\
0.5213	0.142697\\
0.57985	0.110846\\
0.6086	0.108778\\
0.66457	0.154754\\
0.70152	0.133835\\
0.72691	0.127059\\
0.74884	0.140243\\
0.75199	0.123705\\
0.78605	0.14776\\
0.81181	0.137235\\
0.82428	0.139705\\
0.84594	0.134421\\
0.87686	0.149953\\
0.89289	0.155981\\
0.90176	0.15292\\
0.91775	0.165763\\
0.92681	0.164819\\
0.94896	0.176971\\
0.96396	0.164864\\
0.98657	0.177468\\
0.99562	0.171346\\
1.02697	0.181029\\
1.04882	0.192878\\
1.05175	0.185809\\
1.08907	0.187308\\
1.10859	0.181784\\
1.14339	0.184423\\
1.15602	0.178091\\
1.17102	0.1754\\
1.18111	0.185484\\
1.19611	0.190484\\
1.21235	0.185377\\
1.21544	0.178472\\
1.24035	0.178604\\
1.30082	0.17268\\
1.31323	0.175091\\
1.32823	0.172288\\
1.34506	0.164042\\
1.34817	0.156877\\
1.37902	0.163573\\
1.40446	0.164109\\
1.43908	0.169013\\
1.44218	0.154513\\
1.46436	0.166686\\
1.47699	0.163846\\
1.49889	0.15072\\
1.5212	0.153295\\
1.5301	0.15108\\
1.55554	0.151081\\
1.59616	0.162009\\
1.60951	0.147792\\
1.63459	0.145269\\
1.64363	0.144736\\
1.65948	0.150589\\
1.67445	0.155556\\
1.68476	0.154665\\
1.69382	0.143728\\
1.70981	0.154268\\
1.72481	0.149576\\
1.74074	0.1555\\
1.76653	0.14702\\
1.76964	0.13516\\
1.79435	0.140724\\
1.81669	0.147212\\
1.84452	0.140696\\
1.87966	0.144852\\
1.91722	0.150692\\
1.91996	0.150328\\
1.94541	0.15106\\
1.9737	0.159652\\
1.98867	0.161979\\
2.0019	0.157376\\
2.02368	0.169157\\
2.03628	0.171764\\
2.07024	0.162492\\
2.08687	0.156909\\
2.08981	0.149849\\
2.10581	0.160397\\
2.12081	0.155705\\
2.14654	0.144641\\
2.18113	0.159546\\
2.19693	0.138285\\
2.22557	0.145325\\
2.23464	0.144832\\
2.26547	0.149472\\
2.29101	0.144472\\
2.32917	0.135716\\
2.34772	0.140796\\
2.36988	0.152952\\
2.38488	0.150236\\
2.41398	0.159332\\
2.42667	0.147893\\
2.43591	0.144904\\
2.47287	0.156536\\
2.49266	0.144936\\
2.50829	0.160564\\
2.5172	0.14948\\
2.53956	0.159489\\
2.55453	0.155631\\
2.56487	0.155975\\
2.57984	0.153519\\
2.60515	0.149544\\
2.61529	0.149689\\
2.63026	0.147392\\
2.64673	0.154073\\
2.64949	0.146828\\
2.67814	0.160936\\
2.68073	0.153524\\
2.69692	0.15972\\
2.70941	0.159852\\
2.72207	0.157541\\
2.73453	0.159653\\
2.75367	0.153833\\
2.75646	0.141312\\
2.79378	0.153524\\
2.81318	0.143688\\
2.84169	0.150965\\
2.84484	0.137728\\
2.86697	0.145044\\
2.88197	0.149899\\
2.89507	0.138608\\
2.91723	0.150765\\
2.93219	0.155732\\
2.95516	0.14198\\
2.95774	0.144568\\
2.9864	0.15322\\
2.99889	0.154041\\
3.01385	0.149307\\
3.03328	0.138212\\
3.04911	0.154041\\
3.06197	0.147037\\
3.08724	0.145052\\
3.10898	0.141239\\
3.1366	0.158852\\
3.14614	0.147396\\
3.16241	0.15352\\
3.17206	0.138776\\
3.18772	0.154437\\
3.20014	0.15686\\
3.21283	0.15372\\
3.22548	0.151409\\
3.2529	0.153316\\
3.26934	0.160361\\
3.2725	0.143983\\
3.30031	0.154176\\
3.31046	0.150536\\
3.33222	0.160164\\
3.33498	0.15292\\
3.36389	0.151209\\
3.36685	0.144019\\
3.38917	0.151376\\
3.40417	0.156376\\
3.41428	0.151379\\
3.4232	0.150696\\
3.45751	0.165008\\
3.47699	0.156537\\
3.47992	0.14946\\
3.5088	0.144906\\
3.51038	0.143535\\
3.53394	0.147392\\
3.5464	0.147225\\
3.55905	0.144881\\
3.56794	0.14474\\
3.59046	0.147404\\
3.60546	0.152404\\
3.61856	0.138644\\
3.65036	0.139981\\
3.67758	0.157197\\
3.68712	0.156536\\
3.69743	0.153491\\
3.71243	0.153524\\
3.73183	0.143892\\
3.7631	0.14454\\
3.78562	0.147199\\
3.79453	0.144168\\
3.82272	0.15266\\
3.83206	0.144936\\
3.8483	0.151176\\
3.8633	0.155672\\
3.87994	0.148108\\
3.88269	0.150863\\
3.90508	0.153716\\
3.91413	0.151028\\
3.9454	0.150868\\
3.96792	0.153691\\
3.98292	0.150868\\
3.99919	0.15332\\
4.01416	0.155864\\
4.02746	0.144769\\
4.0652	0.151067\\
4.09351	0.159381\\
4.09652	0.151116\\
4.1279	0.149784\\
4.15026	0.154217\\
4.16522	0.156031\\
4.18166	0.1542\\
4.20064	0.153324\\
4.20996	0.144765\\
4.24165	0.145136\\
4.27272	0.144824\\
4.30133	0.153425\\
4.30428	0.144023\\
4.3266	0.15138\\
4.3416	0.141269\\
4.37003	0.160072\\
4.38614	0.144492\\
4.41755	0.144828\\
4.44882	0.14454\\
4.46465	0.150369\\
4.47727	0.152992\\
4.49609	0.159289\\
4.50537	0.143764\\
4.54581	0.159664\\
4.55913	0.153127\\
4.56804	0.150492\\
4.59985	0.151032\\
4.63048	0.155831\\
4.65579	0.15042\\
4.67538	0.150016\\
4.71298	0.14494\\
4.74451	0.144656\\
4.76658	0.156725\\
4.77923	0.159381\\
4.79186	0.151376\\
4.80134	0.135552\\
4.82962	0.154239\\
4.83873	0.143656\\
4.86718	0.1542\\
4.87006	0.147072\\
4.90448	0.159976\\
4.92417	0.147592\\
4.92721	0.135664\\
4.95558	0.154236\\
4.95853	0.147192\\
4.97456	0.15322\\
4.9898	0.141068\\
5.02478	0.154724\\
5.05224	0.16218\\
5.06891	0.14474\\
5.07175	0.14454\\
5.10266	0.155452\\
5.1191	0.150564\\
5.14067	0.140868\\
5.20392	0.135006\\
5.23223	0.150481\\
5.23544	0.143688\\
5.25751	0.144848\\
5.27251	0.153412\\
5.28541	0.14474\\
5.30793	0.147563\\
5.32314	0.152774\\
5.34553	0.153161\\
5.35461	0.150697\\
5.37674	0.153287\\
5.38602	0.150583\\
5.41453	0.159097\\
5.41766	0.149956\\
5.4514	0.165992\\
5.45505	0.149943\\
5.47711	0.159176\\
5.48657	0.1444\\
5.51447	0.15986\\
5.53078	0.144765\\
5.55311	0.147397\\
5.56248	0.14482\\
5.59088	0.15322\\
5.62212	0.157064\\
5.64746	0.150889\\
5.66937	0.138584\\
5.685	0.154212\\
5.69448	0.143692\\
5.72296	0.150484\\
5.72575	0.143272\\
5.76093	0.147708\\
5.77971	0.145051\\
5.7922	0.147537\\
5.81402	0.144652\\
5.84534	0.150864\\
5.87068	0.144932\\
5.89913	0.153524\\
5.91413	0.158524\\
5.93713	0.14723\\
5.94009	0.135377\\
5.9738	0.15978\\
5.97746	0.143823\\
5.99661	0.141268\\
6.02509	0.154412\\
6.03421	0.143668\\
6.05673	0.147407\\
6.07197	0.144724\\
6.08797	0.150835\\
6.10297	0.16288\\
6.11308	0.153291\\
6.1224	0.14494\\
6.14748	0.150867\\
6.17575	0.159611\\
6.18524	0.151028\\
6.22284	0.140864\\
6.25082	0.15266\\
6.26036	0.147392\\
6.27326	0.140864\\
6.30191	0.154437\\
6.30466	0.147189\\
6.33614	0.1409\\
6.37112	0.14828\\
6.3739	0.141068\\
6.39623	0.153392\\
6.41123	0.140868\\
6.4243	0.141035\\
6.45593	0.141268\\
6.4872	0.141068\\
6.52433	0.150424\\
6.54116	0.147558\\
6.55616	0.149935\\
6.56906	0.144765\\
6.59102	0.156722\\
6.60598	0.161689\\
6.62246	0.156525\\
6.62541	0.14948\\
6.6541	0.156893\\
6.65722	0.150016\\
6.70134	0.135181\\
6.72943	0.160365\\
6.73238	0.14346\\
6.76419	0.141731\\
6.79526	0.149848\\
6.81758	0.160363\\
6.83258	0.155863\\
6.84273	0.154221\\
6.85769	0.156348\\
6.8682	0.147\\
6.87732	0.14494\\
6.89964	0.147563\\
6.90856	0.144372\\
6.93424	0.13898\\
6.96252	0.147563\\
6.97752	0.149749\\
6.98767	0.145086\\
7.00264	0.149911\\
7.01278	0.145089\\
7.02183	0.144508\\
7.0533	0.14474\\
7.07858	0.144339\\
7.10599	0.161752\\
7.11553	0.156332\\
7.13197	0.159448\\
7.14146	0.143787\\
7.17281	0.140983\\
7.20985	0.156547\\
7.22668	0.148105\\
7.22936	0.150777\\
7.25726	0.158985\\
7.2666	0.151064\\
7.29215	0.150696\\
7.31451	0.15352\\
7.32384	0.151033\\
7.34615	0.153565\\
7.36136	0.141513\\
7.3898	0.160448\\
7.39259	0.153236\\
7.41498	0.165627\\
7.42797	0.147167\\
7.44991	0.1379\\
7.4782	0.154351\\
7.48098	0.147136\\
7.49701	0.153165\\
7.51198	0.155827\\
7.52861	0.147392\\
7.53753	0.144936\\
7.56009	0.147792\\
7.57505	0.150069\\
7.59436	0.144452\\
7.62528	0.155364\\
7.63542	0.154441\\
7.64507	0.138063\\
7.66683	0.144852\\
7.68183	0.1443\\
7.69194	0.154412\\
7.70142	0.143896\\
7.72358	0.153495\\
7.73882	0.14446\\
7.75482	0.150455\\
7.76982	0.155455\\
7.78924	0.144172\\
7.81157	0.146804\\
7.82657	0.149548\\
7.84596	0.144539\\
7.86815	0.147192\\
7.8774	0.138572\\
7.90501	0.156188\\
7.91435	0.149341\\
7.9249	0.145078\\
7.93378	0.144536\\
7.96879	0.147225\\
7.99053	0.144569\\
8.0221	0.144739\\
8.0561	0.15868\\
8.0787	0.144904\\
8.107	0.153221\\
8.1352	0.14462\\
8.1576	0.147229\\
8.1667	0.144569\\
8.2008	0.148989\\
8.2233	0.144656\\
8.2456	0.1472\\
8.2581	0.151064\\
8.2801	0.144904\\
8.3113	0.144624\\
8.3334	0.156731\\
8.3484	0.161664\\
8.3678	0.144452\\
8.3899	0.156609\\
8.3995	0.143716\\
8.4366	0.156496\\
8.456	0.144704\\
8.4783	0.147233\\
8.491	0.147559\\
8.5036	0.145281\\
8.5125	0.144736\\
8.544	0.144484\\
8.5665	0.147559\\
8.5815	0.152559\\
8.6005	0.144488\\
8.6292	0.153173\\
8.6323	0.143991\\
8.6634	0.140872\\
8.6858	0.153229\\
8.7007	0.156755\\
8.7111	0.147396\\
8.726	0.14954\\
8.7362	0.144681\\
8.7451	0.144536\\
8.7739	0.144736\\
8.7769	0.144535\\
8.8052	0.153349\\
8.8142	0.150828\\
8.8301	0.156689\\
8.8427	0.159345\\
8.8554	0.159645\\
8.8807	0.159611\\
8.8959	0.150864\\
8.9214	0.144569\\
8.9592	0.14494\\
8.9879	0.147763\\
8.9908	0.135752\\
9.013	0.147596\\
9.0255	0.145085\\
9.0592	0.163284\\
9.0756	0.154211\\
9.0782	0.156792\\
9.1037	0.144568\\
9.1355	0.144736\\
9.1575	0.147192\\
9.1725	0.149503\\
9.1883	0.150304\\
9.1978	0.149844\\
9.2144	0.14524\\
9.2236	0.145391\\
9.2542	0.155932\\
9.2643	0.160565\\
9.2771	0.153528\\
9.2991	0.143828\\
9.3213	0.15372\\
9.3339	0.151376\\
9.3525	0.154685\\
9.362	0.138357\\
9.3778	0.154186\\
9.3903	0.156642\\
9.4029	0.159265\\
9.4154	0.151213\\
9.443	0.153724\\
9.4594	0.160367\\
9.4624	0.143819\\
9.4995	0.152884\\
9.5126	0.140668\\
9.535	0.147196\\
9.5477	0.144885\\
9.5603	0.147429\\
9.573	0.145081\\
9.598	0.145048\\
9.6006	0.141197\\
9.6167	0.154241\\
9.6295	0.147396\\
9.6513	0.138612\\
9.6727	0.160068\\
9.682	0.149707\\
9.6985	0.153487\\
9.711	0.151009\\
9.7236	0.15332\\
9.7363	0.151009\\
9.7517	0.141064\\
9.774	0.147149\\
9.7867	0.145081\\
9.8053	0.154037\\
9.8082	0.14696\\
9.8301	0.158848\\
9.8559	0.147425\\
9.8589	0.13538\\
9.8905	0.13878\\
9.9152	0.143524\\
9.9523	0.15332\\
9.9627	0.154469\\
9.972	0.143752\\
10.0124	0.15354\\
10.0258	0.147229\\
10.0348	0.144368\\
10.0601	0.1449\\
10.0823	0.147229\\
10.0947	0.149652\\
10.1166	0.14462\\
10.1475	0.1504\\
10.1794	0.144536\\
10.2047	0.144572\\
10.2393	0.159185\\
10.252	0.159652\\
10.277	0.160452\\
10.2893	0.162708\\
10.2924	0.155864\\
10.3245	0.145112\\
10.3528	0.148084\\
10.3554	0.150672\\
10.3779	0.153336\\
10.3902	0.155625\\
10.403	0.151217\\
10.4157	0.15372\\
10.4283	0.151409\\
10.4374	0.144968\\
10.4745	0.162092\\
10.4847	0.160164\\
10.4941	0.149647\\
10.5288	0.153716\\
10.5414	0.153545\\
10.5541	0.151201\\
10.5667	0.153679\\
10.5821	0.144704\\
10.6106	0.153361\\
10.6231	0.155817\\
10.6381	0.15352\\
10.6577	0.143829\\
10.6796	0.155818\\
10.6925	0.151379\\
10.705	0.153524\\
10.7176	0.151213\\
10.7393	0.147196\\
10.7613	0.159181\\
10.7742	0.151409\\
10.8018	0.15372\\
10.818	0.16036\\
10.8212	0.143983\\
10.8434	0.153716\\
10.856	0.151405\\
10.8687	0.15372\\
10.8839	0.144564\\
10.9126	0.153557\\
10.9252	0.15392\\
10.9379	0.151609\\
10.9594	0.1414\\
10.9816	0.160401\\
10.9944	0.153724\\
11.0071	0.153691\\
11.0195	0.151213\\
11.0223	0.143623\\
11.0507	0.156868\\
11.0537	0.149824\\
11.0761	0.151417\\
11.0887	0.153527\\
11.1012	0.151013\\
11.1103	0.150836\\
11.1261	0.156697\\
11.1356	0.144023\\
11.1578	0.151213\\
11.1727	0.155668\\
11.1978	0.155535\\
11.2208	0.147592\\
11.2334	0.147355\\
11.2484	0.147388\\
11.2647	0.154469\\
11.2675	0.147224\\
11.296	0.160365\\
11.2989	0.143619\\
11.3213	0.15352\\
11.3339	0.151209\\
11.3529	0.145082\\
11.356	0.1382\\
11.378	0.144877\\
11.393	0.149536\\
11.4033	0.144848\\
11.4124	0.144904\\
11.4439	0.144936\\
11.4662	0.147558\\
11.4812	0.149902\\
11.4976	0.144732\\
11.5004	0.144369\\
11.5227	0.147159\\
11.5352	0.147196\\
11.557	0.144656\\
11.5824	0.13858\\
11.6107	0.147192\\
11.6257	0.149537\\
11.6388	0.138412\\
11.68	0.147229\\
11.6829	0.135184\\
11.7145	0.1377\\
11.7303	0.147908\\
11.7395	0.14702\\
11.7772	0.141508\\
11.7932	0.154404\\
11.8086	0.149956\\
11.8395	0.156548\\
11.85	0.147358\\
11.8591	0.144568\\
11.8844	0.144908\\
11.915	0.155536\\
11.9403	0.15308\\
11.9563	0.152864\\
11.9656	0.146984\\
11.9814	0.152756\\
12.0036	0.144544\\
12.0349	0.1505\\
12.069	0.164612\\
12.0823	0.157337\\
12.095	0.159993\\
12.1076	0.157304\\
12.117	0.14136\\
12.1485	0.1449\\
12.1798	0.144936\\
12.2023	0.1476\\
12.2146	0.149885\\
12.2274	0.145644\\
12.2399	0.147592\\
12.2617	0.144768\\
12.2931	0.144624\\
12.3247	0.144908\\
12.3532	0.153392\\
12.3561	0.143548\\
12.3781	0.150813\\
12.3908	0.153328\\
12.4159	0.154124\\
12.4378	0.144684\\
12.4692	0.144536\\
12.5005	0.144448\\
12.5231	0.147388\\
12.538	0.149899\\
12.5484	0.145277\\
12.5608	0.1477\\
12.5636	0.135373\\
12.5952	0.137532\\
12.6237	0.154436\\
12.6266	0.147392\\
12.6549	0.154441\\
12.6581	0.137895\\
12.6866	0.15349\\
12.699	0.151009\\
12.718	0.154408\\
12.7208	0.147196\\
12.737	0.153392\\
12.752	0.156752\\
12.7621	0.147163\\
12.7715	0.144824\\
12.7999	0.15322\\
12.8123	0.154204\\
12.8273	0.149536\\
12.8566	0.145049\\
12.8594	0.141368\\
12.8845	0.137336\\
12.9159	0.144488\\
12.9318	0.150317\\
12.9467	0.155284\\
12.9697	0.147204\\
12.9887	0.147563\\
12.9978	0.138444\\
13.02	0.150601\\
13.029	0.144508\\
13.0605	0.144532\\
13.0828	0.156856\\
13.0921	0.135376\\
13.1144	0.1477\\
13.1294	0.147392\\
13.1486	0.140831\\
13.1858	0.156776\\
13.2054	0.144908\\
13.2367	0.144624\\
13.2592	0.147363\\
13.2746	0.149959\\
13.2934	0.144769\\
13.3219	0.153687\\
13.3341	0.155977\\
13.3562	0.150784\\
13.3786	0.153528\\
13.3878	0.151032\\
13.4033	0.156609\\
13.4183	0.161576\\
13.4343	0.164912\\
13.4436	0.159033\\
13.4537	0.160328\\
13.4687	0.155468\\
13.4854	0.153357\\
13.4981	0.151205\\
13.5166	0.160444\\
13.5196	0.1534\\
13.5483	0.160513\\
13.551	0.153268\\
13.5827	0.139056\\
13.6115	0.147608\\
13.6303	0.147567\\
13.6453	0.149911\\
13.6646	0.14506\\
13.6869	0.147767\\
13.696	0.144912\\
13.7183	0.147567\\
13.7273	0.144628\\
13.7615	0.158852\\
13.7897	0.149144\\
13.8	0.144681\\
13.8092	0.144536\\
13.8406	0.144652\\
13.8565	0.15056\\
13.8715	0.15556\\
13.8875	0.160176\\
13.8966	0.149703\\
13.9069	0.150011\\
13.9194	0.145077\\
13.9321	0.147733\\
13.947	0.147392\\
13.9665	0.137532\\
13.9918	0.141147\\
14.0139	0.147233\\
14.0266	0.145048\\
14.039	0.142425\\
14.0517	0.145081\\
14.0642	0.147229\\
14.0768	0.144885\\
14.1044	0.147396\\
14.1239	0.137528\\
14.146	0.154233\\
14.1551	0.143783\\
14.1773	0.15332\\
14.1899	0.151009\\
14.2026	0.15352\\
14.2152	0.151176\\
14.2178	0.143619\\
14.2401	0.151009\\
14.2551	0.155976\\
14.2654	0.150972\\
14.2743	0.150468\\
14.2904	0.156497\\
14.3053	0.161464\\
14.3222	0.147596\\
14.3252	0.137895\\
14.3473	0.145085\\
14.3568	0.13898\\
14.3849	0.147396\\
14.3975	0.147395\\
14.4195	0.145192\\
14.4508	0.135468\\
14.4667	0.14454\\
14.4792	0.154041\\
14.5009	0.144168\\
14.5423	0.147392\\
14.564	0.144456\\
14.5954	0.144624\\
14.6112	0.150452\\
14.6262	0.149507\\
14.6457	0.14474\\
14.6676	0.156697\\
14.6826	0.156335\\
14.7085	0.144568\\
14.73	0.156024\\
14.7399	0.144532\\
14.7621	0.156721\\
14.7748	0.153357\\
14.7874	0.151201\\
14.7965	0.150828\\
14.8339	0.156227\\
14.8442	0.148093\\
14.8531	0.144536\\
14.8847	0.144772\\
14.9197	0.147563\\
14.9414	0.144944\\
14.97	0.153461\\
14.998	0.144908\\
15.0265	0.143724\\
15.0283	0.143539\\
15.0636	0.158848\\
15.0858	0.144368\\
15.1203	0.150359\\
15.1425	0.144535\\
15.1742	0.145901\\
15.1902	0.142069\\
15.2056	0.147505\\
15.2215	0.146083\\
15.2305	0.144736\\
15.259	0.153188\\
15.2873	0.144773\\
15.3157	0.153533\\
15.3248	0.150671\\
15.3407	0.156529\\
15.3557	0.155276\\
15.3662	0.150577\\
15.3755	0.144908\\
15.3976	0.147204\\
15.4126	0.153612\\
15.4356	0.148279\\
15.4382	0.150864\\
15.4696	0.144765\\
15.5046	0.15392\\
15.5261	0.150696\\
15.5578	0.149956\\
15.5864	0.153832\\
15.6145	0.14494\\
15.6369	0.147563\\
15.6491	0.149852\\
15.6711	0.14494\\
15.6966	0.138811\\
15.731	0.15322\\
15.7532	0.138664\\
15.7751	0.150569\\
15.7876	0.153025\\
15.8002	0.153132\\
15.8152	0.153324\\
15.8253	0.1542\\
15.8348	0.143684\\
15.8506	0.154033\\
15.8658	0.139585\\
15.8976	0.144368\\
15.9198	0.147033\\
15.9325	0.147392\\
15.9452	0.145081\\
15.9578	0.147621\\
15.9793	0.144539\\
16.0109	0.145104\\
16.0268	0.150965\\
16.0418	0.155932\\
16.0648	0.147433\\
16.0772	0.147392\\
16.0899	0.145047\\
16.1023	0.147188\\
16.1369	0.138059\\
16.1616	0.1407\\
16.1842	0.147396\\
16.1969	0.145085\\
16.2095	0.147708\\
16.222	0.145085\\
16.2247	0.137495\\
16.2471	0.144885\\
16.2597	0.1474\\
16.2813	0.144376\\
16.3068	0.145112\\
16.3349	0.153524\\
16.3475	0.153324\\
16.3602	0.151013\\
16.3694	0.144624\\
16.3977	0.15332\\
16.4231	0.151009\\
16.4319	0.15066\\
16.4574	0.150659\\
16.4924	0.147555\\
16.4954	0.135544\\
16.5237	0.154316\\
16.5265	0.147104\\
16.5614	0.153687\\
16.5889	0.156952\\
16.6057	0.148313\\
16.6084	0.150985\\
16.6338	0.138415\\
16.6684	0.153024\\
16.6809	0.154437\\
16.6959	0.149703\\
16.706	0.149848\\
16.7189	0.145081\\
16.7439	0.145048\\
16.7469	0.141163\\
16.7627	0.154316\\
16.772	0.1436\\
16.7878	0.154033\\
16.8028	0.149341\\
16.8191	0.15561\\
16.8341	0.153316\\
16.8446	0.15372\\
16.8537	0.137691\\
16.8761	0.145081\\
16.8853	0.138441\\
16.9167	0.14474\\
16.9537	0.155676\\
16.9704	0.153728\\
16.9798	0.135756\\
17.0081	0.154412\\
17.0111	0.137699\\
17.0395	0.154208\\
17.0423	0.146996\\
17.071	0.15424\\
17.0961	0.149819\\
17.1244	0.144539\\
17.1552	0.155364\\
17.1748	0.143684\\
17.2059	0.14352\\
17.2284	0.145081\\
17.244	0.150652\\
17.2596	0.150732\\
17.2687	0.144369\\
17.2848	0.150398\\
17.2997	0.155365\\
17.3101	0.144881\\
17.3191	0.144368\\
17.3415	0.147\\
17.3508	0.144904\\
17.382	0.143619\\
17.4293	0.160561\\
17.4417	0.162984\\
17.4512	0.141504\\
17.4797	0.160536\\
17.4826	0.153492\\
17.5137	0.162912\\
17.524	0.153528\\
17.5331	0.151032\\
17.5704	0.156932\\
17.596	0.14474\\
17.612	0.150769\\
17.627	0.155736\\
17.6434	0.150801\\
17.6464	0.143752\\
17.6778	0.144739\\
17.7002	0.147433\\
17.7126	0.149856\\
17.7314	0.147893\\
17.7407	0.144704\\
17.7627	0.156693\\
17.7721	0.13518\\
17.8063	0.159864\\
17.8226	0.144769\\
17.8509	0.153054\\
17.854	0.143791\\
17.8761	0.15098\\
17.8911	0.150456\\
17.9169	0.144708\\
17.9392	0.147204\\
17.9542	0.152204\\
17.9671	0.14454\\
17.9956	0.153132\\
};

\end{axis}

\end{tikzpicture}%

%% file: figures/quic-rtt-2.tex
%
%
\definecolor{mycolor1}{rgb}{0.00000,0.44700,0.74100}%
\definecolor{mycolor2}{rgb}{0.85000,0.32500,0.09800}%
\begin{tikzpicture}
\pgfplotsset{every tick label/.append style={font=\scriptsize}}

\begin{axis}[%
width=0.951\fwidth,
height=\fheight,
at={(0\fwidth,0\fheight)},
scale only axis,
xmajorgrids,
ymajorgrids,
xmin=0,
xmax=18,
xlabel style={font=\scriptsize\color{white!15!black}},
xlabel={Time (s)},
ymin=0.1,
ymax=0.55,
ylabel style={font=\scriptsize\color{white!15!black}},
ylabel={Measured RTT (s)},
axis background/.style={fill=white},
title style={font=\bfseries},
legend style={legend cell align=left, font=\scriptsize,  align=left, draw=white!15!black}
]
\addplot [color=mycolor1,dashed,line width=1.2pt]
  table[row sep=crcr]{%
0.23764	0.12045\\
0.25326	0.126069\\
0.36247	0.124825\\
0.38065	0.143005\\
0.39918	0.135743\\
0.41516	0.136901\\
0.51444	0.151977\\
0.53947	0.158825\\
0.55798	0.177337\\
0.57654	0.177361\\
0.59489	0.195709\\
0.61359	0.19843\\
0.64422	0.202635\\
0.74407	0.229624\\
0.77469	0.235223\\
0.7873	0.247833\\
0.80576	0.256291\\
0.82488	0.266895\\
0.84348	0.266935\\
0.86185	0.285307\\
0.88058	0.294039\\
0.89918	0.304287\\
0.91794	0.304351\\
0.93648	0.322887\\
0.95488	0.331287\\
0.98587	0.352283\\
1.14039	0.39632\\
1.63881	0.473611\\
1.76379	0.536243\\
1.79955	0.541156\\
1.84714	0.509425\\
1.8888	0.526249\\
1.91855	0.512793\\
2.13153	0.472717\\
2.16473	0.365175\\
2.1831	0.383547\\
2.20184	0.382291\\
2.22166	0.374519\\
2.44074	0.309207\\
2.45912	0.327591\\
2.4777	0.346171\\
2.50873	0.344007\\
2.53996	0.356859\\
2.55894	0.357095\\
2.57772	0.356063\\
2.63811	0.365855\\
2.68149	0.371653\\
2.78469	0.315569\\
2.81549	0.246553\\
2.83471	0.196594\\
2.85309	0.214978\\
2.87184	0.21373\\
2.89023	0.208741\\
2.99659	0.2119\\
3.01514	0.230451\\
3.03412	0.218632\\
3.05294	0.218232\\
3.07209	0.218999\\
3.09087	0.219028\\
3.10969	0.21946\\
3.22259	0.225996\\
3.24174	0.226597\\
3.26035	0.226232\\
3.27913	0.226196\\
3.29829	0.226197\\
3.3169	0.226032\\
3.33588	0.226196\\
3.36709	0.2374\\
3.48655	0.244808\\
3.50493	0.253192\\
3.52428	0.245148\\
3.5427	0.253564\\
3.56188	0.24498\\
3.58063	0.253732\\
3.59985	0.23276\\
3.62456	0.247472\\
3.7757	0.270772\\
3.79421	0.269924\\
3.81316	0.27046\\
3.83194	0.27006\\
3.85076	0.270124\\
3.86971	0.269863\\
3.88869	0.264132\\
3.9196	0.275043\\
3.95717	0.280475\\
4.04571	0.270012\\
4.07675	0.282548\\
4.09591	0.282748\\
4.11452	0.28258\\
4.13331	0.282548\\
4.15246	0.282749\\
4.17124	0.282552\\
4.19679	0.267185\\
4.34072	0.295008\\
4.36007	0.28332\\
4.37909	0.283184\\
4.39787	0.283348\\
4.41682	0.283516\\
4.43564	0.28318\\
4.47357	0.276788\\
4.55526	0.296357\\
4.63056	0.28984\\
4.64958	0.289504\\
4.66853	0.289439\\
4.68731	0.28944\\
4.70613	0.289304\\
4.72491	0.289272\\
4.74406	0.28964\\
4.76268	0.289104\\
4.86965	0.294396\\
4.96967	0.320087\\
4.98865	0.320117\\
5.00726	0.319952\\
5.02641	0.320287\\
5.04519	0.320284\\
5.06381	0.319748\\
5.08279	0.320116\\
5.10847	0.304913\\
5.28445	0.314781\\
5.32837	0.321112\\
5.34695	0.329692\\
5.36631	0.321112\\
5.38452	0.329328\\
5.40371	0.320912\\
5.42246	0.329664\\
5.49766	0.314584\\
5.59935	0.3149\\
5.62395	0.329503\\
5.65539	0.327012\\
5.6861	0.339143\\
5.69891	0.34196\\
5.71809	0.333572\\
5.73685	0.342324\\
5.75586	0.333404\\
5.77445	0.341988\\
5.80013	0.326557\\
5.93796	0.338616\\
6.03866	0.339744\\
6.10725	0.351385\\
6.12623	0.351785\\
6.14502	0.344889\\
6.22124	0.352435\\
6.28975	0.351789\\
6.38317	0.347479\\
6.40146	0.256449\\
6.43263	0.267616\\
6.46887	0.169117\\
6.52463	0.178913\\
6.54934	0.166169\\
6.56755	0.166089\\
6.60469	0.183225\\
6.64761	0.178741\\
6.67898	0.190109\\
6.71002	0.185392\\
6.72847	0.179132\\
6.74745	0.1799\\
6.79768	0.182988\\
6.87934	0.19036\\
6.92343	0.194964\\
6.94239	0.194932\\
6.99842	0.200739\\
7.01122	0.203544\\
7.09961	0.188985\\
7.13065	0.207212\\
7.16257	0.220181\\
7.1878	0.189381\\
7.31928	0.219672\\
7.33863	0.207988\\
7.38216	0.219591\\
7.40134	0.21354\\
7.42662	0.207933\\
7.53975	0.220468\\
7.56483	0.235548\\
7.5967	0.214544\\
7.62141	0.229256\\
7.64017	0.238828\\
7.65918	0.23256\\
7.80929	0.244463\\
7.82209	0.247264\\
7.84817	0.22676\\
7.8671	0.235689\\
7.8915	0.251336\\
7.91045	0.251276\\
8.055	0.266541\\
8.0859	0.263819\\
8.1048	0.262752\\
8.1304	0.263255\\
8.1616	0.274459\\
8.1806	0.270124\\
8.199	0.278508\\
8.2248	0.263469\\
8.2686	0.269633\\
8.3195	0.270157\\
8.3637	0.258808\\
8.3944	0.264015\\
8.4073	0.266988\\
8.4574	0.276824\\
8.4823	0.283348\\
8.5015	0.27662\\
8.5262	0.281364\\
8.602	0.282467\\
8.6589	0.264513\\
8.6901	0.282776\\
8.7286	0.271169\\
8.7467	0.279272\\
8.7657	0.283344\\
8.7971	0.2956\\
8.8222	0.296\\
8.8855	0.273501\\
8.9796	0.28948\\
9.0365	0.289848\\
9.0551	0.28948\\
9.0743	0.277256\\
9.1053	0.298264\\
9.1931	0.307624\\
9.2874	0.307828\\
9.3258	0.289304\\
9.3507	0.304183\\
9.3702	0.295885\\
9.3881	0.313784\\
9.4075	0.302127\\
9.4637	0.321308\\
9.5018	0.308656\\
9.5703	0.326024\\
9.6083	0.320912\\
9.6457	0.319916\\
9.6651	0.314421\\
9.6902	0.320003\\
9.7211	0.332976\\
9.7406	0.333122\\
9.7651	0.337685\\
9.8223	0.320552\\
9.8912	0.332356\\
9.9415	0.3332\\
10.0043	0.339176\\
10.0233	0.338128\\
10.0425	0.321388\\
10.0676	0.327003\\
10.1114	0.346303\\
10.1864	0.326917\\
10.2372	0.336053\\
10.2949	0.322329\\
10.3442	0.339896\\
10.3632	0.339928\\
10.3821	0.339668\\
10.4007	0.333172\\
10.4257	0.327193\\
10.4509	0.339497\\
10.4824	0.350945\\
10.5133	0.351097\\
10.5446	0.307401\\
10.5627	0.305448\\
10.6311	0.286876\\
10.65	0.286844\\
10.6688	0.286704\\
10.6876	0.286872\\
10.7068	0.281116\\
10.7315	0.280576\\
10.7566	0.27424\\
10.7884	0.286056\\
10.8195	0.274852\\
10.8379	0.275188\\
10.9065	0.27542\\
10.9257	0.275652\\
10.9446	0.275792\\
10.9632	0.27562\\
10.9822	0.275452\\
11.0012	0.269664\\
11.0447	0.288076\\
11.1085	0.289044\\
11.1392	0.301332\\
11.2082	0.301732\\
11.2278	0.302101\\
11.2464	0.301728\\
11.265	0.301732\\
11.284	0.301728\\
11.3351	0.290445\\
11.3534	0.29866\\
11.3844	0.319747\\
11.4105	0.30196\\
11.4417	0.302529\\
11.5297	0.30193\\
11.5548	0.30846\\
11.5796	0.314588\\
11.5985	0.314559\\
11.6175	0.314588\\
11.6361	0.313204\\
11.6553	0.301932\\
11.674	0.310684\\
11.7312	0.310688\\
11.8257	0.296001\\
11.8504	0.310741\\
11.9066	0.32704\\
11.9317	0.333201\\
11.9509	0.333372\\
11.9883	0.332972\\
12.0074	0.333372\\
12.0577	0.326499\\
12.165	0.314588\\
12.1901	0.3197\\
12.2403	0.33374\\
12.2718	0.340064\\
12.2906	0.339728\\
12.3095	0.340063\\
12.3283	0.34006\\
12.3471	0.339724\\
12.3661	0.338675\\
12.4037	0.345992\\
12.4357	0.321865\\
12.5047	0.339696\\
12.5799	0.339588\\
12.6052	0.333464\\
12.6304	0.33976\\
12.6493	0.339789\\
12.6681	0.339792\\
12.6869	0.33976\\
12.7059	0.339789\\
12.7247	0.338573\\
12.7688	0.333112\\
12.7999	0.344152\\
12.9186	0.318681\\
12.9553	0.249401\\
12.9737	0.249001\\
12.9925	0.223609\\
13.0108	0.210949\\
13.0539	0.185196\\
13.0849	0.166272\\
13.1215	0.1662\\
13.146	0.172328\\
13.1768	0.184352\\
13.1952	0.184356\\
13.2138	0.182972\\
13.2462	0.161327\\
13.3342	0.18824\\
13.3526	0.175804\\
13.3963	0.201084\\
13.4212	0.207412\\
13.4402	0.193989\\
13.5284	0.194148\\
13.5598	0.207184\\
13.5784	0.2058\\
13.6106	0.189428\\
13.6355	0.204308\\
13.7178	0.189404\\
13.7366	0.198227\\
13.7928	0.214348\\
13.824	0.225556\\
13.8549	0.219364\\
13.8679	0.222332\\
13.944	0.207405\\
13.9624	0.215789\\
14.0194	0.226603\\
14.0498	0.225876\\
14.0626	0.22868\\
14.0949	0.227003\\
14.1134	0.225576\\
14.1457	0.211223\\
14.1771	0.214719\\
14.2016	0.22922\\
14.259	0.209132\\
14.3086	0.245948\\
14.3403	0.245412\\
14.3717	0.258279\\
14.3903	0.244545\\
14.4223	0.245216\\
14.453	0.25134\\
14.5417	0.251439\\
14.5795	0.270932\\
14.5979	0.257629\\
14.6485	0.258236\\
14.6798	0.257429\\
14.6989	0.245984\\
14.8002	0.25848\\
14.8686	0.27066\\
14.8874	0.258036\\
14.9311	0.28258\\
14.9508	0.271067\\
14.9823	0.28336\\
15.0638	0.263553\\
15.0891	0.258784\\
15.1517	0.283151\\
15.1705	0.283116\\
15.2212	0.290083\\
15.2397	0.288897\\
15.2718	0.289472\\
15.2903	0.267964\\
15.3532	0.289444\\
15.3718	0.282748\\
15.441	0.289301\\
15.4602	0.289672\\
15.491	0.300544\\
15.5292	0.289483\\
15.5482	0.288428\\
15.5854	0.295056\\
15.6045	0.304156\\
15.6354	0.297348\\
15.6734	0.301588\\
15.6862	0.304388\\
15.7173	0.298064\\
15.7368	0.295812\\
15.7682	0.308063\\
15.8001	0.309035\\
15.8371	0.307853\\
15.856	0.307892\\
15.9071	0.311668\\
15.951	0.305539\\
16.0077	0.321487\\
16.0326	0.2958\\
16.051	0.31418\\
16.0706	0.302329\\
16.12	0.319941\\
16.1708	0.314784\\
16.202	0.325992\\
16.2402	0.302855\\
16.2711	0.320109\\
16.3593	0.32664\\
16.3784	0.327412\\
16.4034	0.3328\\
16.4283	0.337712\\
16.5044	0.333579\\
16.5226	0.341756\\
16.5734	0.333165\\
16.5924	0.32128\\
16.6683	0.30906\\
16.6994	0.330096\\
16.7314	0.328037\\
16.7493	0.345908\\
16.8499	0.345497\\
16.9067	0.333346\\
16.9259	0.333549\\
16.989	0.320661\\
17.0137	0.335373\\
17.0643	0.33292\\
17.0831	0.333833\\
17.1019	0.327705\\
17.1264	0.332217\\
17.1523	0.270664\\
17.1832	0.266508\\
17.232	0.167681\\
17.2505	0.167449\\
17.2688	0.166881\\
17.2873	0.160949\\
17.3243	0.17194\\
17.4123	0.170308\\
17.4313	0.18074\\
17.4502	0.181476\\
17.469	0.18168\\
17.525	0.200728\\
17.5885	0.207217\\
17.6258	0.19456\\
17.6507	0.20944\\
17.6702	0.201164\\
17.6881	0.2091\\
17.8404	0.189664\\
17.8588	0.198044\\
17.891	0.202928\\
17.9149	0.2168\\
17.946	0.216599\\
17.9778	0.222833\\
};

\addplot [color=mycolor2]
  table[row sep=crcr]{%
0.33764	0.12045\\
0.35345	0.126257\\
0.45865	0.12101\\
0.477	0.139354\\
0.49928	0.14583\\
0.50833	0.144699\\
0.55173	0.173278\\
0.63897	0.180316\\
0.65769	0.180696\\
0.67588	0.19888\\
0.69458	0.20758\\
0.71311	0.213833\\
0.75004	0.241715\\
0.78081	0.252483\\
0.97973	0.34076\\
1.02334	0.365649\\
1.04209	0.366215\\
1.06053	0.384655\\
1.57124	0.435713\\
1.66747	0.450821\\
1.91758	0.485963\\
1.96646	0.507113\\
1.98358	0.495804\\
2.02558	0.506952\\
2.04473	0.494984\\
2.16209	0.474616\\
2.29062	0.324165\\
2.30841	0.341957\\
2.35218	0.326608\\
2.37056	0.34498\\
2.38893	0.344196\\
2.51504	0.342955\\
2.54636	0.364276\\
2.67515	0.366733\\
2.69984	0.381429\\
2.74843	0.367872\\
2.82148	0.286441\\
2.90845	0.208604\\
2.9278	0.179372\\
2.94635	0.197921\\
2.96456	0.216137\\
3.12828	0.219837\\
3.14724	0.219437\\
3.16605	0.2197\\
3.18483	0.220268\\
3.21031	0.214896\\
3.37286	0.244576\\
3.39184	0.244608\\
3.4108	0.244748\\
3.4486	0.238284\\
3.4798	0.249491\\
3.64297	0.251124\\
3.65598	0.254136\\
3.675	0.245416\\
3.69358	0.253996\\
3.71313	0.233325\\
3.73134	0.251541\\
3.92044	0.264459\\
3.94506	0.27006\\
3.96997	0.276392\\
3.98878	0.275656\\
4.00773	0.276391\\
4.02652	0.275172\\
4.07726	0.282604\\
4.22168	0.276623\\
4.24679	0.27682\\
4.28459	0.276853\\
4.30354	0.277024\\
4.32828	0.281764\\
4.4924	0.270721\\
4.51748	0.275801\\
4.54262	0.277016\\
4.5614	0.276812\\
4.58055	0.277012\\
4.59897	0.285428\\
4.63106	0.302784\\
4.7874	0.295004\\
4.80676	0.289276\\
4.83207	0.289448\\
4.85681	0.295412\\
4.88822	0.307667\\
4.907	0.308032\\
4.93768	0.30662\\
4.96318	0.31212\\
5.12057	0.313816\\
5.14545	0.318696\\
5.1706	0.313784\\
5.19571	0.318896\\
5.21526	0.30826\\
5.23367	0.316672\\
5.25938	0.296196\\
5.29023	0.327048\\
5.45441	0.308955\\
5.47925	0.3138\\
5.51066	0.314952\\
5.52968	0.314416\\
5.54846	0.314788\\
5.56721	0.32354\\
5.64254	0.332312\\
5.69201	0.35094\\
5.81202	0.332772\\
5.83101	0.331756\\
5.85669	0.327012\\
5.8753	0.32684\\
5.89428	0.327072\\
5.91304	0.335824\\
6.06955	0.36754\\
6.18294	0.331937\\
6.20826	0.332957\\
6.22724	0.332953\\
6.24602	0.332985\\
6.2646	0.341569\\
6.37652	0.327361\\
6.43187	0.228921\\
6.45025	0.165645\\
6.49342	0.166235\\
6.5304	0.183219\\
6.59819	0.166323\\
6.61694	0.166691\\
6.64814	0.197891\\
6.67821	0.184796\\
6.77256	0.174372\\
6.80993	0.192988\\
6.84206	0.193915\\
6.86055	0.182336\\
6.91672	0.207465\\
6.96712	0.194564\\
7.03632	0.194269\\
7.0614	0.209349\\
7.13118	0.214459\\
7.1613	0.22458\\
7.24466	0.208339\\
7.26906	0.207652\\
7.29991	0.20762\\
7.35684	0.215664\\
7.43832	0.213189\\
7.46343	0.218767\\
7.47664	0.221973\\
7.50214	0.202231\\
7.53302	0.223103\\
7.57058	0.235012\\
7.69001	0.241695\\
7.70956	0.232928\\
7.7404	0.243765\\
7.76574	0.232728\\
7.79062	0.247607\\
7.90395	0.263256\\
7.94224	0.252227\\
7.96663	0.257065\\
7.99191	0.251508\\
8.0169	0.251108\\
8.0485	0.257869\\
8.0736	0.262949\\
8.1671	0.281136\\
8.2369	0.270264\\
8.262	0.270128\\
8.2875	0.270663\\
8.3188	0.27026\\
8.3376	0.269076\\
8.4007	0.277424\\
8.4393	0.262161\\
8.4884	0.301263\\
8.5388	0.276787\\
8.5578	0.275768\\
8.5831	0.264396\\
8.6148	0.277187\\
8.6708	0.270088\\
8.696	0.285379\\
8.7273	0.288003\\
8.7976	0.299192\\
8.8414	0.283548\\
8.8598	0.281964\\
8.904	0.289277\\
8.9358	0.290381\\
9.0041	0.298064\\
9.1496	0.28984\\
9.1678	0.298056\\
9.2253	0.28944\\
9.2432	0.307337\\
9.2803	0.324523\\
9.3062	0.302128\\
9.3689	0.32764\\
9.4263	0.290451\\
9.4575	0.307859\\
9.4764	0.308624\\
9.5324	0.307187\\
9.5514	0.316153\\
9.5763	0.333136\\
9.6089	0.328527\\
9.6268	0.320516\\
9.6963	0.317395\\
9.7395	0.35054\\
9.7782	0.320721\\
9.797	0.320552\\
9.853	0.32052\\
9.8727	0.321284\\
9.8974	0.321144\\
9.9164	0.320128\\
9.954	0.32724\\
9.9787	0.341952\\
10.1993	0.326612\\
10.2621	0.335697\\
10.3448	0.34608\\
10.4319	0.325871\\
10.4816	0.338923\\
10.5125	0.344512\\
10.5434	0.355377\\
10.5933	0.289648\\
10.6123	0.279566\\
10.7877	0.275188\\
10.8563	0.263023\\
10.8883	0.275967\\
11.0259	0.282291\\
11.0762	0.288476\\
11.1701	0.281859\\
11.1833	0.285053\\
11.3341	0.308113\\
11.3785	0.302331\\
11.4094	0.313225\\
11.4735	0.290195\\
11.4976	0.30422\\
11.5668	0.320071\\
11.6988	0.320317\\
11.7872	0.313669\\
11.8062	0.308656\\
11.8817	0.314952\\
11.9377	0.350961\\
12.0391	0.309231\\
12.1148	0.327591\\
12.1394	0.333172\\
12.2026	0.320916\\
12.2276	0.32586\\
12.2724	0.339759\\
12.3907	0.341548\\
12.4609	0.321479\\
12.4794	0.33006\\
12.5493	0.32176\\
12.6111	0.363497\\
12.7431	0.342395\\
12.825	0.345604\\
12.8812	0.300671\\
12.912	0.300959\\
12.9367	0.290555\\
12.9679	0.311682\\
13.0351	0.190097\\
13.0909	0.168837\\
13.1275	0.159602\\
13.1828	0.194985\\
13.2706	0.17974\\
13.3334	0.205984\\
13.3772	0.184325\\
13.4715	0.200884\\
13.5098	0.176404\\
13.5468	0.203403\\
13.6676	0.196116\\
13.7611	0.204285\\
13.8614	0.232081\\
13.9875	0.226396\\
14.0182	0.237099\\
14.0563	0.237969\\
14.2264	0.238852\\
14.2895	0.233156\\
14.3391	0.262811\\
14.5109	0.22144\\
14.5475	0.258036\\
14.63	0.259373\\
14.768	0.25704\\
14.9062	0.26622\\
14.9815	0.281132\\
15.0381	0.270124\\
15.1328	0.271065\\
15.1891	0.282948\\
15.22	0.293851\\
15.271	0.301449\\
15.3155	0.277384\\
15.3404	0.282328\\
15.4221	0.289268\\
15.4529	0.320139\\
15.497	0.307888\\
15.5731	0.292164\\
15.6364	0.295964\\
15.6611	0.30072\\
15.7184	0.296336\\
15.7688	0.315897\\
15.8187	0.290613\\
15.8749	0.301732\\
15.9063	0.313136\\
15.9753	0.31418\\
16.089	0.310145\\
16.145	0.326272\\
16.2078	0.332968\\
16.3024	0.327044\\
16.3336	0.338315\\
16.3599	0.333489\\
16.4534	0.3328\\
16.4726	0.327612\\
16.5548	0.315925\\
16.6114	0.30906\\
16.6426	0.330267\\
16.812	0.339372\\
16.8307	0.338156\\
16.9698	0.327218\\
17.0459	0.315776\\
17.1451	0.323161\\
17.1769	0.207036\\
17.1951	0.215252\\
17.2257	0.179857\\
17.3058	0.160648\\
17.3683	0.173172\\
17.3871	0.171956\\
17.4875	0.181744\\
17.5184	0.20262\\
17.5573	0.188983\\
17.5877	0.200688\\
17.6133	0.185124\\
17.669	0.230912\\
17.7137	0.19532\\
17.7392	0.185716\\
17.7771	0.1894\\
17.808	0.200272\\
17.8902	0.2112\\
17.947	0.207794\\
17.9713	0.222124\\
};

\end{axis}

\end{tikzpicture}%

%% file: wns3_2019_quic_arxiv.bbl

\newcommand{\SortNoop}[1]{}
\begin{thebibliography}{21}


\ifx \showCODEN    \undefined \def \showCODEN     #1{\unskip}     \fi
\ifx \showDOI      \undefined \def \showDOI       #1{#1}\fi
\ifx \showISBNx    \undefined \def \showISBNx     #1{\unskip}     \fi
\ifx \showISBNxiii \undefined \def \showISBNxiii  #1{\unskip}     \fi
\ifx \showISSN     \undefined \def \showISSN      #1{\unskip}     \fi
\ifx \showLCCN     \undefined \def \showLCCN      #1{\unskip}     \fi
\ifx \shownote     \undefined \def \shownote      #1{#1}          \fi
\ifx \showarticletitle \undefined \def \showarticletitle #1{#1}   \fi
\ifx \showURL      \undefined \def \showURL       {\relax}        \fi
\providecommand\bibfield[2]{#2}
\providecommand\bibinfo[2]{#2}
\providecommand\natexlab[1]{#1}
\providecommand\showeprint[2][]{arXiv:#2}

\bibitem[\protect\citeauthoryear{Bishop}{Bishop}{2018}]%
        {draftquichttp}
\bibfield{author}{\bibinfo{person}{M. Bishop}.}
  \bibinfo{year}{2018}\natexlab{}.
\newblock \bibinfo{booktitle}{\emph{{Hypertext Transfer Protocol (HTTP) over
  QUIC}}}.
\newblock \bibinfo{type}{{draft-ietf-quic-http-13}}.
  \bibinfo{institution}{IETF}.
\newblock
\urldef\tempurl%
\url{https://tools.ietf.org/html/draft-ietf-quic-http-13}
\showURL{%
\tempurl}


\bibitem[\protect\citeauthoryear{Carlucci, Cicco, and Mascolo}{Carlucci
  et~al\mbox{.}}{2015}]%
        {Carlucci-2015-HOU}
\bibfield{author}{\bibinfo{person}{G. Carlucci}, \bibinfo{person}{L.~De Cicco},
  {and} \bibinfo{person}{S. Mascolo}.} \bibinfo{year}{2015}\natexlab{}.
\newblock \showarticletitle{{HTTP over UDP: An Experimental Investigation of
  QUIC}}. In \bibinfo{booktitle}{\emph{Proceedings of the 30th Annual ACM
  Symposium on Applied Computing}} \emph{(\bibinfo{series}{SAC '15})}.
  \bibinfo{publisher}{ACM}, \bibinfo{address}{New York, NY, USA},
  \bibinfo{pages}{609--614}.
\newblock
\showISBNx{978-1-4503-3196-8}
\urldef\tempurl%
\url{https://doi.org/10.1145/2695664.2695706}
\showDOI{\tempurl}


\bibitem[\protect\citeauthoryear{Casoni and Patriciello}{Casoni and
  Patriciello}{2016}]%
        {CASONI201681}
\bibfield{author}{\bibinfo{person}{Maurizio Casoni} {and}
  \bibinfo{person}{Natale Patriciello}.} \bibinfo{year}{2016}\natexlab{}.
\newblock \showarticletitle{Next-generation TCP for ns-3 simulator}.
\newblock \bibinfo{journal}{\emph{Simulation Modelling Practice and Theory}}
  \bibinfo{volume}{66} (\bibinfo{year}{2016}), \bibinfo{pages}{81 -- 93}.
\newblock
\showISSN{1569-190X}
\urldef\tempurl%
\url{https://doi.org/10.1016/j.simpat.2016.03.005}
\showDOI{\tempurl}


\bibitem[\protect\citeauthoryear{Cisco}{Cisco}{2017}]%
        {ciscoVni}
\bibfield{author}{\bibinfo{person}{Cisco}.} \bibinfo{year}{2017}\natexlab{}.
\newblock \bibinfo{booktitle}{\emph{{Cisco Visual Networking Index, Forecast
  and Methodology, 2016–2021}}}.
\newblock \bibinfo{type}{{T}echnical {R}eport}.
\newblock


\bibitem[\protect\citeauthoryear{Iyengar and Swett}{Iyengar and Swett}{2018}]%
        {draftquickrec}
\bibfield{author}{\bibinfo{person}{J. Iyengar} {and} \bibinfo{person}{I.
  Swett}.} \bibinfo{year}{2018}\natexlab{}.
\newblock \bibinfo{booktitle}{\emph{{QUIC Loss Detection and Congestion
  Control}}}.
\newblock \bibinfo{type}{{draft-ietf-quic-recovery-13}}.
  \bibinfo{institution}{IETF}.
\newblock
\urldef\tempurl%
\url{https://tools.ietf.org/id/draft-ietf-quic-recovery-13.txt}
\showURL{%
\tempurl}


\bibitem[\protect\citeauthoryear{Iyengar and Thomson}{Iyengar and
  Thomson}{2018}]%
        {draftquicktr}
\bibfield{author}{\bibinfo{person}{J. Iyengar} {and} \bibinfo{person}{M.
  Thomson}.} \bibinfo{year}{2018}\natexlab{}.
\newblock \bibinfo{booktitle}{\emph{{QUIC: A UDP-Based Multiplexed and Secure
  Transport}}}.
\newblock \bibinfo{type}{{draft-ietf-quic-transport-13}}.
  \bibinfo{institution}{IETF}.
\newblock
\urldef\tempurl%
\url{https://tools.ietf.org/id/draft-ietf-quic-transport-13.txt}
\showURL{%
\tempurl}


\bibitem[\protect\citeauthoryear{Kakhki, Jero, Choffnes, Nita-Rotaru, and
  Mislove}{Kakhki et~al\mbox{.}}{2017}]%
        {kakhki2017taking}
\bibfield{author}{\bibinfo{person}{Arash~Molavi Kakhki},
  \bibinfo{person}{Samuel Jero}, \bibinfo{person}{David Choffnes},
  \bibinfo{person}{Cristina Nita-Rotaru}, {and} \bibinfo{person}{Alan
  Mislove}.} \bibinfo{year}{2017}\natexlab{}.
\newblock \showarticletitle{{Taking a Long Look at QUIC: An Approach for
  Rigorous Evaluation of Rapidly Evolving Transport Protocols}}. In
  \bibinfo{booktitle}{\emph{Proceedings of the 2017 Internet Measurement
  Conference}} \emph{(\bibinfo{series}{IMC '17})}. \bibinfo{publisher}{ACM},
  \bibinfo{address}{New York, NY, USA}, \bibinfo{pages}{290--303}.
\newblock
\showISBNx{978-1-4503-5118-8}
\urldef\tempurl%
\url{https://doi.org/10.1145/3131365.3131368}
\showDOI{\tempurl}


\bibitem[\protect\citeauthoryear{Langley, Riddoch, Wilk, Vicente, Krasic,
  Zhang, Yang, Kouranov, Swett, Iyengar, Bailey, Dorfman, Roskind, Kulik,
  Westin, Tenneti, Shade, Hamilton, Vasiliev, Chang, and Shi}{Langley
  et~al\mbox{.}}{2017}]%
        {Langley:2017:QTP:3098822.3098842}
\bibfield{author}{\bibinfo{person}{Adam Langley}, \bibinfo{person}{Alistair
  Riddoch}, \bibinfo{person}{Alyssa Wilk}, \bibinfo{person}{Antonio Vicente},
  \bibinfo{person}{Charles Krasic}, \bibinfo{person}{Dan Zhang},
  \bibinfo{person}{Fan Yang}, \bibinfo{person}{Fedor Kouranov},
  \bibinfo{person}{Ian Swett}, \bibinfo{person}{Janardhan Iyengar},
  \bibinfo{person}{Jeff Bailey}, \bibinfo{person}{Jeremy Dorfman},
  \bibinfo{person}{Jim Roskind}, \bibinfo{person}{Joanna Kulik},
  \bibinfo{person}{Patrik Westin}, \bibinfo{person}{Raman Tenneti},
  \bibinfo{person}{Robbie Shade}, \bibinfo{person}{Ryan Hamilton},
  \bibinfo{person}{Victor Vasiliev}, \bibinfo{person}{Wan-Teh Chang}, {and}
  \bibinfo{person}{Zhongyi Shi}.} \bibinfo{year}{2017}\natexlab{}.
\newblock \showarticletitle{{The QUIC Transport Protocol: Design and
  Internet-Scale Deployment}}. In \bibinfo{booktitle}{\emph{Proceedings of the
  Conference of the ACM Special Interest Group on Data Communication}}
  \emph{(\bibinfo{series}{SIGCOMM '17})}. \bibinfo{publisher}{ACM},
  \bibinfo{address}{New York, NY, USA}, \bibinfo{pages}{183--196}.
\newblock
\showISBNx{978-1-4503-4653-5}
\urldef\tempurl%
\url{https://doi.org/10.1145/3098822.3098842}
\showDOI{\tempurl}


\bibitem[\protect\citeauthoryear{Lee, Hong, et~al\mbox{.}}{Lee
  et~al\mbox{.}}{2016}]%
        {libquic}
\bibfield{author}{\bibinfo{person}{Junseong Lee}, \bibinfo{person}{Brian Hong},
  {et~al\mbox{.}}} \bibinfo{year}{2016}\natexlab{}.
\newblock \bibinfo{title}{libquic}.
\newblock   (\bibinfo{year}{2016}).
\newblock
\urldef\tempurl%
\url{github.com/devsisters/libquic}
\showURL{%
\tempurl}


\bibitem[\protect\citeauthoryear{Mishra, Vankar, and Tahiliani}{Mishra
  et~al\mbox{.}}{2016}]%
        {mishra2016tcp}
\bibfield{author}{\bibinfo{person}{Dharmendra~Kumar Mishra},
  \bibinfo{person}{Pranav Vankar}, {and} \bibinfo{person}{Mohit~P. Tahiliani}.}
  \bibinfo{year}{2016}\natexlab{}.
\newblock \showarticletitle{{TCP Evaluation Suite for ns-3}}. In
  \bibinfo{booktitle}{\emph{Proceedings of the Workshop on Ns-3}}
  \emph{(\bibinfo{series}{WNS3 '16})}. \bibinfo{publisher}{ACM},
  \bibinfo{address}{New York, NY, USA}, \bibinfo{pages}{25--32}.
\newblock
\showISBNx{978-1-4503-4216-2}
\urldef\tempurl%
\url{https://doi.org/10.1145/2915371.2915388}
\showDOI{\tempurl}


\bibitem[\protect\citeauthoryear{Nguyen, Gangadhar, Rahman, and
  Sterbenz}{Nguyen et~al\mbox{.}}{2016}]%
        {Nguyen:2016:ISV:2915371.2915386}
\bibfield{author}{\bibinfo{person}{Truc Anh~N. Nguyen},
  \bibinfo{person}{Siddharth Gangadhar}, \bibinfo{person}{Md~Moshfequr Rahman},
  {and} \bibinfo{person}{James~P.G. Sterbenz}.}
  \bibinfo{year}{2016}\natexlab{}.
\newblock \showarticletitle{{An Implementation of Scalable, Vegas, Veno, and
  YeAH Congestion Control Algorithms in Ns-3}}. In
  \bibinfo{booktitle}{\emph{Proceedings of the Workshop on Ns-3}}
  \emph{(\bibinfo{series}{WNS3 '16})}. \bibinfo{publisher}{ACM},
  \bibinfo{address}{New York, NY, USA}, \bibinfo{pages}{17--24}.
\newblock
\showISBNx{978-1-4503-4216-2}
\urldef\tempurl%
\url{https://doi.org/10.1145/2915371.2915386}
\showDOI{\tempurl}


\bibitem[\protect\citeauthoryear{Papastergiou, Fairhurst, Ros, Brunstrom,
  Grinnemo, Hurtig, Khademi, Tüxen, Welzl, Damjanovic, and
  Mangiante}{Papastergiou et~al\mbox{.}}{2017}]%
        {7738442}
\bibfield{author}{\bibinfo{person}{G. Papastergiou}, \bibinfo{person}{G.
  Fairhurst}, \bibinfo{person}{D. Ros}, \bibinfo{person}{A. Brunstrom},
  \bibinfo{person}{K. Grinnemo}, \bibinfo{person}{P. Hurtig},
  \bibinfo{person}{N. Khademi}, \bibinfo{person}{M. Tüxen},
  \bibinfo{person}{M. Welzl}, \bibinfo{person}{D. Damjanovic}, {and}
  \bibinfo{person}{S. Mangiante}.} \bibinfo{year}{2017}\natexlab{}.
\newblock \showarticletitle{{De-Ossifying the Internet Transport Layer: A
  Survey and Future Perspectives}}.
\newblock \bibinfo{journal}{\emph{IEEE Communications Surveys Tutorials}}
  \bibinfo{volume}{19}, \bibinfo{number}{1} (\bibinfo{date}{Firstquarter}
  \bibinfo{year}{2017}), \bibinfo{pages}{619--639}.
\newblock
\showISSN{1553-877X}
\urldef\tempurl%
\url{https://doi.org/10.1109/COMST.2016.2626780}
\showDOI{\tempurl}


\bibitem[\protect\citeauthoryear{Patriciello}{Patriciello}{2017}]%
        {Patriciello:2017:SCL:3067665.3067666}
\bibfield{author}{\bibinfo{person}{Natale Patriciello}.}
  \bibinfo{year}{2017}\natexlab{}.
\newblock \showarticletitle{{A SACK-based Conservative Loss Recovery Algorithm
  for Ns-3 TCP: A Linux-inspired Proposal}}. In
  \bibinfo{booktitle}{\emph{Proceedings of the Workshop on Ns-3}}
  \emph{(\bibinfo{series}{WNS3 '17})}. \bibinfo{publisher}{ACM},
  \bibinfo{address}{New York, NY, USA}, \bibinfo{pages}{1--8}.
\newblock
\showISBNx{978-1-4503-5219-2}
\urldef\tempurl%
\url{https://doi.org/10.1145/3067665.3067666}
\showDOI{\tempurl}


\bibitem[\protect\citeauthoryear{Polese, Chiariotti, Bonetto, Rigotto, Zanella,
  and Zorzi}{Polese et~al\mbox{.}}{2018}]%
        {polese2019survey}
\bibfield{author}{\bibinfo{person}{Michele Polese}, \bibinfo{person}{Federico
  Chiariotti}, \bibinfo{person}{Elia Bonetto}, \bibinfo{person}{Filippo
  Rigotto}, \bibinfo{person}{Andrea Zanella}, {and} \bibinfo{person}{Michele
  Zorzi}.} \bibinfo{year}{2018}\natexlab{}.
\newblock \showarticletitle{A Survey on Recent Advances in Transport Layer
  Protocols}.
\newblock \bibinfo{journal}{\emph{Submitted to IEEE Communications Surveys and
  Tutorials}} (\bibinfo{year}{2018}).
\newblock
\urldef\tempurl%
\url{https://arxiv.org/abs/1810.03884}
\showURL{%
\tempurl}


\bibitem[\protect\citeauthoryear{R{\"u}th, Poese, Dietzel, and
  Hohlfeld}{R{\"u}th et~al\mbox{.}}{2018}]%
        {DBLP:journals/corr/abs-1801-05168}
\bibfield{author}{\bibinfo{person}{Jan R{\"u}th}, \bibinfo{person}{Ingmar
  Poese}, \bibinfo{person}{Christoph Dietzel}, {and} \bibinfo{person}{Oliver
  Hohlfeld}.} \bibinfo{year}{2018}\natexlab{}.
\newblock \showarticletitle{{A First Look at QUIC in the Wild}}. In
  \bibinfo{booktitle}{\emph{Passive and Active Measurement}},
  \bibfield{editor}{\bibinfo{person}{Robert Beverly}, \bibinfo{person}{Georgios
  Smaragdakis}, {and} \bibinfo{person}{Anja Feldmann}} (Eds.).
  \bibinfo{publisher}{Springer International Publishing},
  \bibinfo{address}{Cham}, \bibinfo{pages}{255--268}.
\newblock
\showISBNx{978-3-319-76481-8}


\bibitem[\protect\citeauthoryear{Seemann and Clemente}{Seemann and
  Clemente}{2018}]%
        {QGo}
\bibfield{author}{\bibinfo{person}{Marten Seemann} {and} \bibinfo{person}{Lucas
  Clemente}.} \bibinfo{year}{2018}\natexlab{}.
\newblock \bibinfo{title}{Quic implementation in Go}.
\newblock   (\bibinfo{year}{2018}).
\newblock
\urldef\tempurl%
\url{github.com/lucas-clemente/quic-go}
\showURL{%
\tempurl}


\bibitem[\protect\citeauthoryear{Shade et~al\mbox{.}}{Shade
  et~al\mbox{.}}{2016}]%
        {proto-quic}
\bibfield{author}{\bibinfo{person}{Robbie Shade} {et~al\mbox{.}}}
  \bibinfo{year}{2016}\natexlab{}.
\newblock \bibinfo{title}{proto-quic}.
\newblock   (\bibinfo{year}{2016}).
\newblock
\urldef\tempurl%
\url{github.com/google/proto-quic}
\showURL{%
\tempurl}


\bibitem[\protect\citeauthoryear{Thomson and Turner}{Thomson and
  Turner}{2018}]%
        {draftquictls}
\bibfield{author}{\bibinfo{person}{M. Thomson} {and} \bibinfo{person}{S.
  Turner}.} \bibinfo{year}{2018}\natexlab{}.
\newblock \bibinfo{booktitle}{\emph{{Using TLS to Secure QUIC}}}.
\newblock \bibinfo{type}{{draft-ietf-quic-tls-13}}.
  \bibinfo{institution}{IETF}.
\newblock
\urldef\tempurl%
\url{https://tools.ietf.org/html/draft-ietf-quic-tls-13}
\showURL{%
\tempurl}


\bibitem[\protect\citeauthoryear{Tikhonov and Prodoehl}{Tikhonov and
  Prodoehl}{2017}]%
        {LSQ}
\bibfield{author}{\bibinfo{person}{Dmitri Tikhonov} {and}
  \bibinfo{person}{Brian Prodoehl}.} \bibinfo{year}{2017}\natexlab{}.
\newblock \bibinfo{title}{LiteSpeed QUIC}.
\newblock   (\bibinfo{year}{2017}).
\newblock
\urldef\tempurl%
\url{github.com/litespeedtech/lsquic-client}
\showURL{%
\tempurl}


\bibitem[\protect\citeauthoryear{Yan Soares~Couto}{Yan Soares~Couto}{2018}]%
        {nsquic}
\bibfield{author}{\bibinfo{person}{Daniel Macedo~Batista Yan Soares~Couto,
  Diego~Camarinha}.} \bibinfo{year}{2018}\natexlab{}.
\newblock \bibinfo{booktitle}{\emph{{nsQUIC: Uma Extensao para Simulacao do
  Protocolo QUIC no NS-3}}}.
\newblock \bibinfo{type}{{T}echnical {R}eport}.
\newblock
\urldef\tempurl%
\url{http://www.sbrc2018.ufscar.br/wp-content/uploads/2018/04/180520_1.pdf}
\showURL{%
\tempurl}


\bibitem[\protect\citeauthoryear{Zhang, Polese, Mezzavilla, Zhu, Rangan,
  Panwar, and Zorzi}{Zhang et~al\mbox{.}}{2019}]%
        {8613277}
\bibfield{author}{\bibinfo{person}{M. Zhang}, \bibinfo{person}{M. Polese},
  \bibinfo{person}{M. Mezzavilla}, \bibinfo{person}{J. Zhu},
  \bibinfo{person}{S. Rangan}, \bibinfo{person}{S. Panwar}, {and}
  \bibinfo{person}{M. Zorzi}.} \bibinfo{year}{2019}\natexlab{}.
\newblock \showarticletitle{{Will TCP Work in mmWave 5G Cellular Networks?}}
\newblock \bibinfo{journal}{\emph{IEEE Communications Magazine}}
  \bibinfo{volume}{57}, \bibinfo{number}{1} (\bibinfo{date}{January}
  \bibinfo{year}{2019}), \bibinfo{pages}{65--71}.
\newblock
\showISSN{0163-6804}
\urldef\tempurl%
\url{https://doi.org/10.1109/MCOM.2018.1701370}
\showDOI{\tempurl}


\end{thebibliography}
